\documentclass[a4paper,12pt]{article}

\usepackage{latexstyle}

\title{The spatial evolution of economic activities:\\ from theory to estimation}

\author{Davide Fiaschi \thanks{University of Pisa, Via Ridolfi, 10, 56124 Pisa, Italy (e-mail: davide.fiaschi@unipi.it).}
 \and Angela Parenti\thanks{University of Pisa, Via Ridolfi, 10, 56124 Pisa, Italy (e-mail: angela.parenti@unipi.it).}
	\and Cristiano Ricci\thanks{University of Pisa, Via Ridolfi, 10, 56124 Pisa, Italy (e-mail: cristiano.ricci@ec.unipi.it).}}

\date{\today}

\begin{document}

\maketitle

\begin{abstract}
	
This paper studies the evolution of economic activities using a continuous time-space aggregation-diffusion model, which encompasses competing effects of agglomeration and congestion. To bring the model to the real data, a novel discretization technique over time and space is introduced. This technique effectively disentangles spatial effects into pure topography, agglomeration, repulsion, and diffusion forces, which is crucial for developing robust econometric methods in spatial economics. Our empirical analysis of personal income across Italian municipalities from 2008 to 2019 validates the model's primary predictions and demonstrates superior performance compared to the most common spatial econometric models in the literature. 
\end{abstract}



\section{Introduction\label{sec:Introduction}}
The investigation of the causes of the spatial distribution of economic activities is a very active field of analysis \citep{allen2014trade, redding2017quantitative, desmet2018geography}. Several models have been introduced to explain the emergence of spatial patterns characterized by geographical aggregations of economic activities, e.g. cities, and specific relationships among the locations of these aggregations, i.e. city size distribution, and their spatial organization \citep{krugman1991increasing}, as well as 
to quantify the effect of spatial agglomeration on aggregate income \citep{duranton2023urban}.
Despite this increasing literature, a definitive explanation of how aggregation economies emerge from the behaviour of individual agents is still needed \citep{rossi2019geography}, and, such micro-foundations appear still more crucial in the light of the increasing availability of accurate data at fine geographical scale \citep{allen2014trade, ahlfeldt2015economics,desmet2018geography}.
Furthermore, the complex nature of the spatial distribution of cross-sectional economic activities, together with its nonlinear evolution through time, needs specific quantitative techniques. Currently, neither spatial econometrics, which aims at capturing the spatial correlation in the \textit{equilibrium outcome} from models of social/spatial interactions (e.g., \citealp{anselin2002under}; \citealp{brueckner2003strategic}; \citealp{ertur2007growth}; \citealp{combes2015empirics}, and \citealp{xu2019theoretical}), nor geo-statistical models, which focus on the specific spatial processes underlying the observed spatial distributions abstracting from the spatial interactions (e.g., \citealp{cressie2015statistics}), are able to provide such techniques.
  
In this paper, we study the dynamics over time and space of economic activities by a micro-founded continuous time-space aggregation-diffusion model, where the collective macroscopic behaviour emerges from the dynamic of interacting and locally optimizing agents, in the mean-field limit as the number of agents becomes infinite. The macroeconomic dynamic is expressed by a Partial Differential Equation (PDE), which belongs to the class of \textit{Aggregation-Diffusion Equations} (ADEs), which over the past 20 years have been employed in several biological applications and stimulated many mathematical works (see \citealp{carrillo2019aggregation} for a review). The competing effects of aggregation, repulsion and diffusion lead to several interesting properties, such as metastability, symmetrization, and non-uniqueness in the equilibrium solutions.\footnote{A complete characterization of equilibria is not still available, and so far the only remarkable result is by \cite{carrillo2019nonlinear}, who proved that equilibria are radially symmetric in the absence of exogenous factors; for a specific class of interactions, the authors also proved the uniqueness of the equilibrium as well as the convergence to the unique equilibrium independent of the initial distribution.} The estimation of ADEs is primarily addressed in two key works: \citet{huang2019learning} tackle the challenge by observing individuals rather than aggregate variables, while \citet{carrillo2024sparse} focus specifically on estimating the interaction kernel, neglecting the broader parameter set.
We propose a new discretization technique over time and space \citep{donaghy2001solution,ait2002maximum,piras2007nonlinear,oud2012continuous,wymer2012continuous}, which complements the literature on quantitative spatial economics \citep{redding2017quantitative}, whose outcome is a new class of spatial econometric models, denoted Spatial Aggregation Repulsion Diffusion (SARD) models. In particular, the use of \textit{generalised finite difference methods} to transpose the system of PDEs defined in the continuum into a set of discrete locations (\citealp{jensen1972finite}) allows for dealing with any non-uniform spatial organization of economic activities (and other socio-economic and cross-sectional variables), which are in general not arranged on a regular Cartesian grid but into irregular units, such as administrative regions.\footnote{This constitutes one of the main differences with respect to spatial econometrics, where locations are assumed to be discrete by the beginning of analysis \citep{anselin2002under}.} 
Moreover, the formalism of PDEs in continuous space provides an innovative tool to introduce new elements in the theory of spatial econometrics. In particular, as described in more detail in Section \ref{sec:theoreticalModel}, it allows to construct regressors having the remarkable feature of being cross-sectional zero-sum, which naturally leads to a disentangle between growth over time and spatial reallocation of economic activities. Here spatial reallocation is intended as a purely redistributive process by which if a location loses (gains) a certain amount, the same amount is distributed to (taken from) some other locations. Moreover, instead of simply measuring the effect of spatial correlation through local averages, by exploiting the differentials in the variables of interest across neighbouring locations, our approach is able to distinguish among the \textit{aggregative} dynamics towards more appealing locations (a centripetal force, for example arising from the higher wages in specific labour markets), the \textit{repulsive} dynamics (a centrifugal force, for example caused by the higher cost of living in a congested location), and the \textit{diffusive} dynamics (an equalizing force, for example, caused by the random preferences of the underlying individuals).\footnote{Here and in the rest of the paper the term \emph{diffusion} is intended as in the literature of PDEs, that is a pure reallocation force which has the effect of driving the spatial process to a uniform distribution.}
Finally, the linear econometric model resulting from the discretization allows to directly study the process in the transient regime, i.e. not assuming to be in a neighbour of the equilibrium, therefore without relying on a (log-)linearization around the steady state \citep{ertur2007growth}. 

To summarise, relative to the existing literature on spatial econometrics and quantitative spatial economics, we propose a model which is able to disentangle growth over time from spatial reallocation and, for the latter, to separately identify topographical, aggregative, repulsive and diffusive forces. Numerical investigations show the capacity of our proposed procedure to correctly estimate the model's parameters when discretization over time and space is sufficiently fine. Maximum likelihood estimation appears to be the best method, although the simplest OLS produces in some circumstances acceptable outcomes. Moreover, our procedure is able to deal with the spatial correlation in the errors induced, among other factors, by the discretization.

An application to the spatio-temporal dynamics of Italian municipal income over the period 2008-2019 finds evidence of the presence of agglomerative, repulsive and diffusive forces, jointly with the key role of geographical characteristics. SARD model outperforms the most common spatial econometric models used in the literature, as spatial LAG and DURBIN model \citep{lesage2009introduction}. We also show the capacity of the SARD model to quantify the contribution of each reallocation force to the average growth rate of each municipality, as well as to the overall (di)convergence in local incomes.

There are two important limitations in our analysis. First, the SARD model can be thought as a reduced form of a structural spatial model of income (see \citealp{fiaschiRicci2023spatialDynamicsPopulationABA}), which by its nature limits the possibility to be used to explore the dynamics of other key economic variables such as local wages, prices, rents, etc. It remains that it can be used both for forecast as well as policy purposes in the same vein as the VAR approach. Second, our methodology requires a sufficiently high level of geographical resolution in order to prevent significant discretization bias in the estimation.

The paper is organized as follows: Section \ref{sec:theoreticalModel} presents a prototype model of spatial growth; Section \ref{sec:methodology} discusses the methodology to estimate such a model and explore its properties by numerical investigations; Section \ref{sec:empiricalApplication} discusses an empirical application of the proposed methodology to Italian municipalities; and, finally, Section \ref{sec:concludingRemarks} concludes. Technical details are gathered in the Appendix.

\section{A prototype model of spatial growth \label{sec:theoreticalModel}}

This section introduces a class of spatial growth models encompassing the main features present in the literature, i.e. i) the spatial non-uniformity caused by geographical and socio-economic factors; ii) the spatial aggregation of economic activities driven by positive spatial spillovers; iii) the centrifugal dynamics driven by congestion and excessive crowding; and, finally, iv) the existence of randomness in the individuals' preferences \citep{fujita_thisse_2002}.

Let $y(t,z)$ be the variable of interest of our model in location $z$ at time $t$, e.g. municipal total income per square kilometre, where $\Omega \subseteq \mathbb{R}^2$. Each point $z \in \Omega$ is identified by two components $z = (z_1,z_2)$, i.e. latitude and longitude. The dynamics of $y(t,z)$ is assumed to obey the following partial differential equation (PDE):
\begin{eqnarray} \nonumber
	\partial_{t} y(t,z) &=&
	a\left(t,z\right) + \phi y(t,z) +   \\ \nonumber
	&+& \gamma_S \div_z \left( y(t,z) \nabla_z S(z)\right) +  \\ \nonumber
	&+& \gamma_A \div_z \left(y(t,z) \nabla_z \left(K_{h_A}  * y\right) (t,z)\right)  +  \\ \nonumber
	&+& \gamma_R \div_z \left(y(t,z) \nabla_z \left(K_{h_R} * y\right) (t,z)\right) +  \\
	&+& \gamma_D \Delta_z y(t,z),
	\label{eq:dataGenerationProcessCellTotalIncome}
\end{eqnarray}
with $\gamma_D\geq 0$.\footnote{Eq. \eqref{eq:dataGenerationProcessCellTotalIncome}   is general not well-posed, i.e solutions often grow unbounded in finite time or even fail to exist, in the case $\gamma_D < 0$, see \citet{taylor1975reflection}.} 
Eq. \eqref{eq:dataGenerationProcessCellTotalIncome} makes wide use of concepts used in PDE literature, which will be discussed in detail below.  In particular, the variable on the left-hand side $\partial_t y(t,z)$ represents the total variation of the quantity $y(t,z)$ at time $t$ in location $z$, expressed through the use of the partial derivative $\partial_t$ with respect to time. 
In particular, $\div_z$ is the \textit{divergence operator}, i.e. $\div_z F \equiv \partial_{z_1} F_{z_1} + \partial_{z_2} F_{z_2}$, where $F = (F_{z_1}, F_{z_2})$ is a vector field $F:\Omega \to \RR^2$ and $\partial_{z_i}$ are the partial derivatives with respect to each of the components of $z = (z_1,z_2)$. All the partial derivatives expressed above are to be intended in the spatial sense, i.e. they measure how the spatial profile varies along one of the two components, either latitude or longitude. When $f:\Omega \to \RR$ is a scalar function the term $\nabla_z f$ stands for the \textit{gradient operator,} i.e. $\nabla_z f \equiv \left(\partial_{z_1} f,\partial_{z_2} f\right)$ while finally the term $\Delta_z f \equiv \partial^2_{z_1z_1} f + \partial^2_{z_2 z_2} f$ is called the \textit{Laplace operator} and involves the second order pure partial derivatives $\partial^2_{z_1z_1}$ and $\partial^2_{z_2z_2}$.

Neglecting boundary effects one has
\begin{eqnarray*}
	\int_\Omega \div_z \left( y(t,z) \nabla_z S(z)\right) dz &=& 0; \\
	\int_\Omega \div_z \left(y(t,z) \nabla_z \left(K_{h_A} * y\right) (t,z)\right) dz &=& 0; \\
	 \int_\Omega \div_z \left(y(t,z) \nabla_z \left(K_{h_R} * y\right) (t,z)\right) dz &=&0; \text{ and } \\
	 \int_\Omega \Delta_z y(t,z) dz &=& 0,
\end{eqnarray*}
which expresses that the last four terms of Eq. \eqref{eq:dataGenerationProcessCellTotalIncome} are purely reallocative.

\cite{fiaschiRicci2023spatialDynamicsPopulationABA} shows that Eq. \eqref{eq:dataGenerationProcessCellTotalIncome} can be derived in the limit of an infinity-agent economy, where agents can be understood as units of production/consumers, as the outcome of the agents' mobility driven by locally maximizing behaviour (Appendix \ref{sec:microfoundation} summarise the key steps in the derivation). 

The first term on the right-hand side of Eq. \eqref{eq:dataGenerationProcessCellTotalIncome} $a\left(t,z\right)$ accounts for the \textit{exogenous local component} of the dynamics, e.g. migration shocks, as well as \textit{local effects} (amenities) and \textit{time effects} (e.g., exogenous technological progress, see \citealp{ertur2007growth}). 

Parameter $\phi$ instead represents the change in $y(t,z)$ due to the \textit{local specific process of accumulation}, i.e. the impact of variable $y(t,z)$ on the time change of the variable itself, independent of any spatial interaction. Taking as reference the Solovian framework, $\phi$ should reflect the \textit{shape of production function}, \textit{saving behaviour}, \textit{depreciation rates of factors} and \textit{employment growth}. The dynamic at the aggregate level is given by:
\begin{equation}
	\int_\Omega \partial_{t} y(t,z) dz = \dot{Y}(t)= \int_\Omega a(t,z) dz + \phi Y(t),
	\label{eq:aggregateRateOfGrowth}
\end{equation}
which points out as $\phi$ represents the part of the \textit{aggregate long-run growth rate of the economy} independent of
the spatial distribution of activities, while the first term in Eq. \eqref{eq:aggregateRateOfGrowth} represents the aggregate effect of exogenous growth components.

The main reference behind the terms from the second to the fifth line of the right-hand side of Eq. \eqref{eq:dataGenerationProcessCellTotalIncome} is the so-called \textit{Fokker-Planck Equation}. In particular, they contain \textit{differential operators}, which belong to the language of PDEs, and are particularly effective for describing the various sources of change in the spatial distribution of $y(t,z)$. 
Since partial derivatives express a measure of differentials across different locations, the sign of their coefficients reflects the direction driving the reallocation. For example, as the second line, if the coefficient $\gamma_S$ is positive, then the term $\gamma_S \div_z \left( y(t,z) \nabla_z S(z)\right)$ expresses the tendency of $y(t,z)$ to increase in those locations where $S(z)$ is lower, and to decrease of the same amount where $S(z)$ is higher.\footnote{A simple example in one dimension showing the interpretation of $\gamma_S >0$ is the following. Assume that the distribution $y(z)$ is uniform in the space, i.e $y(z)=\overline{y} \; \forall z$, and the function $S(z)$ is a sigmoid of the form $S(z) = \frac{1}{1+e^{-z}}$, i.e. increasing in $z$ and with $S^{''}(z)<0$ for $z >  0$. In this case, the second term of Eq. \eqref{eq:dataGenerationProcessCellTotalIncome} becomes $\gamma_S\overline{y}S^{''}(z)$ and, therefore: i) for $z>0$, $\gamma_S\overline{y}S^{''}(z)$ is negative, implying that the values of $y(z)$ are decreasing; while, ii) for $z < 0$, $\gamma_S\overline{y}S^{''}(z)$ is positive, implying that the values of $y(z)$ are increasing.} The definition of higher and lower is to be intended as a relative comparison between locations and is provided by the use of $\nabla_z S(z)$ which measures the \textit{steepness of the transition} moving from one location to the other. This increase-decrease effect is made in such a way that the total amount of mass in the distribution of $y$ which is gained where it increases, is lost in the other locations. Hence, the total variation accounted for by the second term, when one takes into account all the locations, is zero. For this reason, we refer to all the terms in Eq. \eqref{eq:dataGenerationProcessCellTotalIncome} except for the first two, as purely \textit{reallocation} terms, since they don't affect the total amount of $y$.

The second line of Eq. \eqref{eq:dataGenerationProcessCellTotalIncome} $\gamma_S \div_z \left( y(t,z) \nabla_z S(z)\right)$ is introduced to take into account the \textit{topography} of $ \Omega$. In particular, $\nabla_z S(z)$ indicates the \textit{pure geographical and exogenous} (dis)advantages to move from location $z$ to neighbouring locations. In this regard, the coefficient $\gamma_S$ is expected to be positive reflecting a reallocation towards places where the exogenous variable $S(z)$ reaches its lowest levels (e.g. from the mountains towards the plains). 

The third line on the right-hand side of Eq. \eqref{eq:dataGenerationProcessCellTotalIncome} $\gamma_A \div_z \left(y(t,z) \nabla_z \left(K_{h_A}  * y\right) (t,z)\right)$ represents the effect of \textit{aggregation} and of cultural features such as roads, land boundaries, and buildings, i.e. the tendency of $y$ to concentrate in specific locations. \cite{fiaschiRicci2023spatialDynamicsPopulationABA} shows that this term, and the next two, of Eq. \eqref{eq:dataGenerationProcessCellTotalIncome} are the outcome of the agents' mobility driven by locally maximizing behaviour. The intensity of this process is measured by $\gamma_A<0$. $K_{h_A}$ is a kernel function with bandwidth $h_{A} \geq 0$, and "$*$" is the \textit{convolution operator}, i.e. $\left(K_{h_A}  * y\right)\left(t,z\right) \equiv \int_{\Omega} K_{h_A}\left(k-z\right)y(t,k)dk$. In other words, $\left(K_{h_A}  * y\right) (t,z)$ is the weighted sum of all $y$ around location $z$ at period $t$, where the weights are defined by kernel $K_{h_A} $; in particular, the shape of $K_{h_A} $ and the value of ${h_A}$ decide how these weights change with the distance from location $z$. Since $\nabla_z \left(K_{h_A}  * y\right) (t,z)$ points to the directions where the level of $y(t,z)$ is higher when averaged among neighbours, it plays a similar role of $\nabla_z S(z)$ as the term on the second line of Eq. \eqref{eq:dataGenerationProcessCellTotalIncome}. An important difference is that, instead of being purely exogenous as for $S(z)$, the direction which drives the reallocation is now endogenous since it depends on the spatial distribution of $y$. In particular, we assume $\gamma_A < 0$, so that the reallocation is driven towards those areas where the local average of $y$, $\left(K_{h_A}  * y\right)\left(t,z\right)$ is higher, providing the intuition on why this term will tend to concentrate $y$ over space.
A standard explanation in economics is the observed process of aggregation of workers, i.e. the emergence of cities based on the positive externalities generated by working in places where other activities and/or skilled workers are already present  (\citealp{fujita1996economics,krugman1998space,Moretti2004,ahlfeldt2015economics}).

The fourth line on the right-hand side of Eq. \eqref{eq:dataGenerationProcessCellTotalIncome} $\gamma_R\, \div_z \left(y(t,z) \nabla_z \left(K_{h_R} * y\right) (t,z)\right)$ represents the effect of \textit{repulsion} of $y$ across different locations, that is the tendency of $y$ to flow away from locations with higher levels. The intensity of this process is measured by $\gamma_R>0$. This term is exactly analogous to the one related to aggregation, except that the Kernel function $K_{h_R}$ can be different from $K_{h_A}$ (for example, in the speed of decay as a function of the distance), and that the sign $\gamma_R$ is assumed to be positive, so the effect instead of being centripetal is centrifugal. The population and income density, increasing the price of lands and living, operates as a centrifugal force \citep{duranton2020economics}. For Kernels $K_{h_A}$ and $K_{h_R}$ we adopt distance-decay functions as:
\begin{eqnarray}
		&K_{h_A}(z) \equiv \left\{ \begin{array}{ll}
		\left(\frac{c_K}{h_A}\right)^2\frac{1}{\left(\norm{\frac{z}{h_A}}+1\right)^2} & \textrm{if $\norm{z} \leq h_A$};\\
		0 & \textrm{otherwise}\\
	\end{array} \right., \nonumber \\
	&K_{h_R}(z) \equiv \left\{ \begin{array}{ll}
			\left(\frac{c_K}{h_R}\right)^2\frac{1}{\left(\norm{\frac{z}{h_R}}+1\right)^2} & \textrm{if $\norm{z} \leq h_R$};\\
		0 & \textrm{otherwise}\\
	\end{array} \right., 
	\label{eq:kernelAandR}
\end{eqnarray}
where $c_K =  1/\sqrt{2 \pi \left(\log 2 - 1/2\right) }$, $h_A$ and $h_R$ are the distance within which there exists a non-null (aggregative and repulsive) spatial interaction.

Finally, the last line on the right-hand side of Eq. \eqref{eq:dataGenerationProcessCellTotalIncome} $\gamma_D \Delta_z y(t,z)$, represents the effect of \textit{diffusion} across different locations of $y$, which tends to uniformly spread $y$ over space. The intensity of this diffusion process in location $z$ at time $t$ depends on the parameter $\gamma_D>0$ and on the sign and magnitude of second derivatives of $y(t,z)$.\footnote{The use of the second derivative can be understood intuitively if one thinks of the one-dimensional example of a bell-shaped distribution of $y(t,z)$ (e.g. a Gaussian distribution). In this case, the spatial second derivative measures the convexity/concavity of the distribution, being negative in the centre of the peak (concave) and positive in the area outside the peak (convex). Therefore, when $\gamma_D > 0$ the effect is to decrease where the distribution is convex (i.e. on the local peaks) and to increase in all other regions. \citet[p. 12]{farlow1993partial} provides an intuition of why the second derivative is crucial for describing a diffusion process, which tends to uniformly spread the variable of interest over space.} In economics, the diffusion process can be justified by the \textit{idiosyncratic random component}  in the agents' choice, determined by their hidden characteristics/preferences \citep{Wozniak2010}. At the aggregate level, this idiosyncratic component drives the distribution of the variable of interest in space towards a uniform distribution. 

While is it true that both \textit{repulsion} and \textit{diffusion} represent centrifugal forces, they are very different in nature. The repulsion effect only takes place in the presence of overcrowding; the diffusion, instead, is always affecting the dynamics. In particular, diffusion always tends to equalize all the levels of $y(t,z)$ across locations, while repulsion only lowers the level of $y$ with high density. In other words, in the same way that aggregation expresses the tendency of individuals to relocate to be closer to each other, repulsion can be seen as an effect due to \textit{congestion} and \textit{crowding}. A possible source of the observed outflows of individuals from very crowded locations can arise from the higher housing prices and moving costs and, in general, from the higher cost of living in locations with high population density (\citealp{krugman1998space}). Moreover, this tendency is generally justified by the factor-return equalization across different locations in the presence of decreasing marginal returns to factors.\footnote{Suppose to consider two locations, 1 and 2, with a different endowment of capital $k_1>k_2$ but with the same production function; then $f^\prime(k_1)<f^\prime(k_2)$ under the hypothesis of $f^{\prime \prime}(\cdot)<0$; with free movement of capital we should observe a flow of capital from location 1 to location 2. The second derivative with respect to the distribution over space of capital is a proxy for the difference in the level of $k_1$ and $k_2$, which, in turn, is reflected in the difference in factor returns and, hence, in the intensity of reallocation.}
Finally, theory suggests that some signs restrictions have to be tested in the estimation; in particular, in Eq. \eqref{eq:dataGenerationProcessCellTotalIncome} $\gamma_S \leq 0,\gamma_A \leq 0$, to reflect the reallocation due to topography and the aggregation effect respectively, while $\gamma_R \geq  0, \gamma_D \geq 0$ to reflect the repulsive effect and the presence of diffusion.

\section{From theory to empirics\label{sec:methodology}}

This section discusses a methodology to estimate the parameters of Eq. \eqref{eq:dataGenerationProcessCellTotalIncome} when only a sample of $N$ spatial units observed at times $0$ and $\tau$ is available.
In particular, Section \ref{sec:methodologyEstimate} reports the main steps of the estimate; Section \ref{sec:derivationSARD} describes in detail how the econometric model used in the estimate is derived by an appropriate approximation over time and space of Eq. \eqref{eq:dataGenerationProcessCellTotalIncome}; Section \ref{sec:remarksSARD} contains some remarks on the derivation of the econometric model and its possible extensions; Section \ref{subsec:caveats} discusses some possible issues in the estimate; finally, Section \ref{sec:MCsimulations} investigates the properties of the proposed methodology by numerical simulations.

\subsection{The econometric model \label{sec:methodologyEstimate}}
The econometric model associated to Eq. \eqref{eq:dataGenerationProcessCellTotalIncome}, denoted  \textit{Spatial Aggregation Repulsion Diffusion} (SARD), is given by:
\begin{eqnarray}
	\label{eq:SARD}
	\Delta_\tau{\mathbf{y}} &=&  \tilde{\alpha}  \mathbf{1} + \tilde{\phi} \mathbf{y} + \tilde{\gamma}_{S} \mathbf{x}_S + \tilde{\gamma}_A \mathbf{x}_A +\tilde{\gamma}_R \mathbf{x}_R + \tilde{\gamma}_D \mathbf{x}_D +  \\ \nonumber
	&+& \tilde{\rho}_{S} M_S \Delta_\tau {\mathbf{y}} + \tilde{\rho}_{A} M_A \Delta_\tau{\mathbf{y}}+ \tilde{\rho}_{R} M_R \Delta_\tau{\mathbf{y}}+ \tilde{\rho}_{D} M_D \Delta_\tau{\mathbf{y}} +
	\\ \nonumber
	&+& \boldsymbol{\epsilon},
\end{eqnarray}
where $\Delta_\tau{\mathbf{y}}$ is the vector of the time changes of $ \mathbf{y}$ of length $N$ (the number of the units of observations), $ \mathbf{1}$, $ \mathbf{y}$, $\mathbf{x}_S$, $\mathbf{x}_A$, $\mathbf{x}_R$, $\mathbf{x}_D$ are the vectors of regressors of length $N$, $M_S$, $M_A$, $M_R$, $M_D$, are $N \times N$ matrices whose calculation is described in Section \ref{sec:derivationSARD}, and $\boldsymbol{\epsilon}$ is an error component whose characteristics are discussed below.

Section \ref{sec:derivationSARD} shows that Model \eqref{eq:SARD} represents a (corrected) first order Taylor approximation over time and a second order approximation over space of Eq. \eqref{eq:dataGenerationProcessCellTotalIncome}.
However, the parameters $\tilde{\alpha}$, $\tilde{\phi}$, $\tilde{\gamma}_S$, $\tilde{\gamma}_A$, $\tilde{\gamma}_R$, $\tilde{\gamma}_D$ correspond to those of  the theoretical model up to a scale factor, i.e:
\begin{equation}
 	\tilde{\alpha} = \dfrac{\mathbf{\alpha} }{1 - \tau \rho_\phi/2}, \quad
	\tilde{\phi} = \dfrac{\phi}{1 - \tau \rho_\phi/2}, \quad \tilde{\gamma}_j = \dfrac{\gamma_j}{1 - \tau \rho_\phi/2}, \quad \tilde{\rho}_j = \dfrac{ \tau \rho_j / 2}{1 - \tau \rho_\phi/2},
	\label{eq:fromSARDtoModelsParameters}
\end{equation}
with $j \in \{S,A,R,D\}$. 
To identify parameters of Eq. \eqref{eq:dataGenerationProcessCellTotalIncome} we use the dynamics of aggregate variable of interest $Y$ from Eq. \eqref{eq:aggregateRateOfGrowth}, i.e.:
\begin{equation}
	Y\left( \tau \right) = \left( Y\left(0\right) + \dfrac{{\alpha}}{\phi}\right)
	\exp\left(\phi \tau \right) - \dfrac{{\alpha}}{\phi},
	\label{eq:aggregateDynamicsVariableOfInterest}
\end{equation}
to get an estimate of $\rho_\phi$, i.e.\footnote{Substituting the expression for $\alpha$ and $\phi$ from Eq. \eqref{eq:fromSARDtoModelsParameters} into Eq. \eqref{eq:aggregateDynamicsVariableOfInterest} we get an equation for $\rho_\phi$.}
\begin{equation}
	\rho_\phi = \left(\dfrac{2}{\tau}\right)\left[ 1- \left(\dfrac{1}{\tilde{\phi} \tau }\right) \log \left( \dfrac{ Y\left(\tau\right)+ \tilde{\alpha}/\tilde{\phi} }{ Y\left(0\right) + \tilde{\alpha}/\tilde{\phi} } \right) \right].
	\label{eq:estimateRhoPhi}
\end{equation}


Model \eqref{eq:SARD} admits a panel structure with individual and time fixed effects as shown in Section \ref{sec:derivationSARD}. In particular, assuming that $a\left(t,z\right) = \alpha\left(z\right) +  d\left(t\right)$, then:
\begin{equation}
	\tilde{\boldsymbol{\alpha}} \equiv \dfrac{\boldsymbol{\alpha}}{1 - \tau \rho_\phi/2} \text{ and } \quad
	\tilde{\mathbf{d}} \equiv \dfrac{\mathbf{d} + \tau \Delta_\tau \mathbf{d}/2 }{1 - \tau \rho_\phi/2},
	\label{eq:spatialTimeFectedEffects}
\end{equation}
where $\tilde{\boldsymbol{\alpha}}$ is the vector of location fixed effects of length $N$ and $\tilde{\mathbf{d}}$ is the vector of time fixed effect of length equal to the number of time periods. 
The identification of $\tilde{\boldsymbol{\alpha}}$ and $\tilde{\mathbf{d}}$ however requires a sample of more than two periods and to specify a functional form for $d(t)$ (e.g. a  linear trend) in order to identify the parameters by exploiting the aggregate dynamics of $Y$ given by:
\begin{equation}
	Y\left( t \right) = \left( Y\left(t_0\right) + \dfrac{{\bar{\alpha}}}{\phi}\right)
	\exp\left(\phi \left(t-t_0\right)\right)  - \dfrac{{\bar{\alpha}}}{\phi} + \exp\left(\phi \left(t-t_0\right)\right) \int_{t_0}^t \exp\left(-\phi \left(s-t_0\right)\right)  d(s)ds,
	\label{eq:aggregateDynamicsVariableOfInterestWithspatialTimeFectedEffects}
\end{equation}
for $t_0 \in \{0, \tau, 2\tau, \cdots\}$ and $t \in \{\tau, 2\tau, \cdots\}$, where $\bar{\alpha} \equiv \int_\Omega \alpha\left(z\right)\,dz$.

The error term $\boldsymbol{\epsilon}$  in the Model \eqref{eq:SARD} includes, in addition to the classical i.i.d stochastic component, the randomness deriving from the finite number of individuals and firms and the potential presence of measurement errors. In particular, assuming $y^{OBS}_{it}=y_{it}+ \xi_{it}$ for $t \in \{0, \tau\}$, where  $y^{OBS}_{it}$ is the \textit{observed} value of the variable of interest and $\xi_{it}$ is a classical error (i.e. a random variable i.i.d. over time and space), $\partial_t y^{OBS}_{it}= \partial_t  y_{it}+ d\xi_{it}$, i.e. the dynamics of the observed variable is the sum of two components: the first one systematic, which induces a structure of spatial correlation (see Section \ref{sec:fluctuationsFiniteWorld}), while the second one purely random.  Inspired by the spatial econometric literature \citep[p. 27]{lesage2009introduction}, we consider:
\begin{eqnarray}\nonumber
\boldsymbol{\epsilon} &=& \lambda {W}_\epsilon \boldsymbol{\epsilon} + \boldsymbol{\eta}, \text{ and}\\ 
\boldsymbol{\eta} & \sim & \mathcal{N}(0,\sigma^2 I),
\label{eq:SARDspatialError}
\end{eqnarray}
where $\lambda$ is a parameter measuring the intensity of spatial spillovers in the error component, ${W}_\epsilon$ is a $N \times N $ spatial weights matrix, and $\sigma^2$ is the variance of the i.i.d. random component. The choice of ${W}_\epsilon$ is discussed in Section \ref{sec:derivationSARD} and Appendix \ref{app:spatialMatrixErrors}. It is worth noticing that the specification of the error in Eq. \eqref{eq:SARDspatialError} as a spatial error only affects the efficiency of estimates but not their consistency \citep[p. 28]{lesage2009introduction}.

From the theoretical model, we expected the following signs for the estimated parameters: i) $\phi (\tilde{\phi}) > 0$, which should reflect the positive impact of initial conditions on the variation of the variable of interest over time;  ii) $ {\gamma}_{S} (\tilde{\gamma}_{S}) < 0$ representing the spatial effect of reallocating towards more spatially appealing locations;  iii) ${\gamma}_A (\tilde{\gamma}_A ) < 0$ indicating evidence of aggregation effect over space; iv) $\gamma_R (\tilde{\gamma}_R ) > 0$ showing evidence of repulsion effect; finally v) $\gamma_D (\tilde{\gamma}_D ) > 0$, highlighting a diffusion effect over space.



\subsection{The derivation of SARD model \label{sec:derivationSARD}}

In this section, we describe in detail the procedure to derive Model \eqref{eq:SARD} in the presence of $N$ units of observation  $z^i$ for $i = 1,\dots N$  on $T+1$ periods $t=0, \dots, T$. 
Let 
\begin{equation*}
\int_{t-1}^{t} y(s,z^{i}) \,ds 
\end{equation*}
be the total cumulative value of the variable of interest in \emph{period} $t$ in unit $z^{i}$. Since the variable of interest is observed only on periods $t=0, \dots, T$ we assume that the flow $y(s,z^{i})$ is homogeneous for $s \in [t-1,t]$, and therefore we consider 
\begin{equation}\label{eq:defyti}
y_{ti} \equiv y(t,z^{i}),
\end{equation}
the discrete measure of our variable of interest at period $t$ in spatial unit $z^{i }$.\footnote{It is also possible to consider the case where the process is observed at dates $0,\Delta, 2\Delta,\dots,T\Delta$ where $\Delta > 0$ is fixed. This amounts to consider the cumulative variation as $\int_{(t-1)\Delta}^{t\Delta}y(s,z^{i})\,ds$, reducing Eq. \eqref{eq:defyti} to $y_{ti} \equiv y(t\Delta,z^{i})\cdot \Delta$. In the text, we have assumed $\Delta \equiv 1$ (i.e. 1 year in Section \ref{sec:empiricalApplication}).}  With $y_{i} \equiv y(0,z^i)$ we denote its measure at period $0$.

\subsubsection{Discretization over space}

As the discretization over space, starting from the second line of Eq. \eqref{eq:dataGenerationProcessCellTotalIncome}, $\gamma_S \div_z \left( y(t,z) \nabla_z S(z)\right)$, making partial derivatives with respect to $(z_1,z_2)$ explicit yields: 
\begin{equation}\label{eq:discretization0}
	\gamma_S \div_z \left( y(t,z) \nabla_z S(z)\right) = \gamma_S\left[ \partial_{z_1}\left(y(t,z)\partial_{z_1} S(z) \right) + \partial_{z_2}\left(y(t,z)\partial_{z_2} S(z) \right)\right].
\end{equation}
Using the notation introduced in Appendix \ref{sec:computationGradientLaplacian}, Matrices $M_{z_1}$ and $M_{z_2}$ respectively denote the (approximate) derivative operators of the first order partial derivatives $\partial_{z_1}$ and $\partial_{z_2}$, i.e., for unit $z^i$:
\begin{equation}\label{eq:discretization1}
	\partial_{z_1}y(t,z)\big\vert_{z = z^i} \approx (M_{z_1}\mathbf{y}_t)_{i} \text{ and } \partial_{z_2}y(t,z)\big\vert_{z = z^i} \approx (M_{z_2}\mathbf{y}_t)_{i},
\end{equation}
where $\mathbf{y}_t=(y_{t1},...,y_{tN})$.

The smaller the distance among contiguous spatial units the more accurate the approximation in Eq. (\ref{eq:discretization1}) (see \citealp{thomas2013numerical} for a reference in the case of a uniform grid).
From Eqq. \eqref{eq:discretization0} and \eqref{eq:discretization1} the discrete counterpart of the term in the second line in Eq. \eqref{eq:dataGenerationProcessCellTotalIncome} is given by:
\begin{equation*}
	\gamma_S \div_z \left( y(t,z) \nabla_z S(z)\right) \approx \gamma_{S} \left[ M_{z_1} \left(\mathbf{y}_t\odot M_{z_1} \mathbf{s} \right)  +  M_{z_2} \left(\mathbf{y}_t \odot M_{z_2} \mathbf{s} \right)  \right],
\end{equation*}
with $s_i \equiv S(z^i)$,  $\mathbf{s}=(s_{1},...,s_{N})$ and where $\odot$ denotes the element-wise product between vectors. The approximation of the third and fourth line of Eq. \eqref{eq:dataGenerationProcessCellTotalIncome} follows in the same manner, but the discretization of the convolution terms $\left(K_{h_A}*y\right)(t,z)$ and $\left(K_{h_R}*y\right)(t,z)$. In particular, taking
\begin{equation}
	\left(K_{h_A}  * y\right)\left(t,z\right) \equiv \int_{\Omega} K_{h_A}\left(k-z\right)y(t,k)dk \approx W_{h_A} \mathbf{y}_t, 
\end{equation}
with $W_{h_A}$ a $\left( N \times N \right)$ matrix representing the kernel $K_{h_A}$ defined as:
\begin{equation}
	\left(W_{h_A}\right)_{i,j} = K_{h_A}\left({z^i - z^j} \right)A_j, 
\label{eq:WhA}
\end{equation}
and $A_j$ the area of the $j$-th spatial unit, then the fourth term in Eq. \eqref{eq:dataGenerationProcessCellTotalIncome} can be approximated as:
\begin{equation}
	\gamma_A \div_z \left( y(t,z) \nabla_z \left(K_{h_A}*y\right)(t,z)\right) \approx \gamma_{A} \left[ M_{z_1} \left(\mathbf{y}_t\odot M_{z_1}  W_{h_A} \mathbf{y}_t \right)  +  M_{z_2} \left(\mathbf{y}_t \odot M_{z_2}  W_{h_A} \mathbf{y}_t \right)  \right].
\end{equation}
The same applies to the fifth term in Eq. \eqref{eq:dataGenerationProcessCellTotalIncome}, with $W_{h_R}$ a $\left( N \times N \right)$ matrix representing the kernel $K_{h_R}$. ${W}_{h_A}$ and ${W}_{h_R}$ have a clear similarity with the spatial weight matrices used in the spatial econometric literature \citep{lesage2009introduction}; however, since the matrices ${W}_{h_A}$ and ${W}_{h_R}$ approximate in discrete space integrals in continuous space, the value of the function in the centre of the neighbour where integration takes place, i.e. their diagonal elements, are non-zero.

The (approximate) derivative operators of the second order partial derivatives $\partial^2_{z_1z_1}$ and $\partial^2_{z_2z_2}$ for unit $z^i$ are
\begin{equation}\label{eq:discretization2}
	\partial^2_{z_1z_1}y(t,z)\big\vert_{z = z^i} \approx (M_{z_1z_1}\mathbf{y}_t)_{i} \text{ and } \partial^2_{z_2z_2}y(t,z)\big\vert_{z = z^i} \approx (M_{z_2z_2}\mathbf{y}_t)_{i},
\end{equation}
which can be used to approximate the sixth term of Eq. \eqref{eq:dataGenerationProcessCellTotalIncome}.
The overall outcome of the discretization over space is therefore given by: 
\begin{eqnarray} \nonumber
	\partial_t {y(t,z^i)} & = &   a\left(t,z^i\right) + \phi y_{ti} +\\ \nonumber
	&+& \gamma_{S} \left[ M_{z_1} \left(\mathbf{y}_t \odot M_{z_1} \mathbf{s} \right)  +  M_{z_2} \left(\mathbf{y}_t \odot M_{z_2} \mathbf{s} \right)  \right]_i + \\ \nonumber
	&+&\gamma_A \left[ M_{z_1} \left(\mathbf{y}_t \odot M_{z_1} W_{h_A} \mathbf{y}_t \right)  +  M_{z_2} \left(\mathbf{y}_t \odot M_{z_2} W_{h_A} \mathbf{y}_t \right)  \right]_i + \\  \nonumber
	&+& \gamma_R \left[ M_{z_1} \left(\mathbf{y}_t \odot M_{z_1} W_{h_R} \mathbf{y}_t \right)  +  M_{z_2} \left(\mathbf{y}_t \odot M_{z_2} W_{h_R} \mathbf{y}_t \right)  \right]_i + \\  \nonumber
	&+& \gamma_D \left[(M_{z_1 z_1} + M_{z_2 z_2})\mathbf{y}_t\right]_{i}\\
	&+& r^{ti}_{s},
	\label{eq:dataGenerationProcessCellTotalIncomeDiscritizationSpace}
\end{eqnarray}
where $r^{ti}_{s}$ is a reminder of the discretization over space such that $\norm {\mathbf{r}^{t}_{s}}$ goes to zero as the space  discretization gets finer. The statistical properties of this reminder will be explored in Section \ref{sec:MCsimulations}.

\subsubsection{Discretization over time}\label{subsec:discretizationTime}
Assume $\tau \in \NN $ and $ \tau > 1$. As the discretization in time, Taylor's theorem in Lagrange form states that there exists a $t' \in (0,\tau)$ such that:\footnote{See Theorem 5.15 in \cite{rudin1964principles}.}
\begin{equation}
	\partial_{t} y(0,z^i) = \Delta_\tau y_i   - \frac{1}{2}\tau\partial_{tt}^{2}y(t',z^i),
\end{equation}
where
\begin{equation}
\Delta_\tau y_i	\equiv \dfrac{y(\tau,z^i)-y(0,z^i)}{\tau}
\end{equation}
is the time variation of $y$ in the period $[0,\tau]$, from which:
\begin{equation}
	\Delta_\tau y_i  =  \partial_{t} y(0,z^i) + \frac{1}{2}\tau\partial_{tt}^{2}y(t',z^i),
	\label{eq:timediscretizationBaselineEquation}
\end{equation}
where from the previous, under the assumption that $\partial_{tt}^{2}y$ is bounded, one can see that the discrete derivative is a good approximation of the ``true'' derivative as $\tau$ approaches zero.
A ``naive'' discretization over time would consider only the first term on the right-hand side in Eq. \eqref{eq:timediscretizationBaselineEquation}, which can be directly taken from Eq. \eqref{eq:dataGenerationProcessCellTotalIncome} evaluated in $t=0$ using Eq. \eqref{eq:dataGenerationProcessCellTotalIncomeDiscritizationSpace}. The second term on the right-hand side of Eq. \eqref{eq:timediscretizationBaselineEquation} measures the error in the discretization of considering the first term only. In our case, this error has a non negligible magnitude due to the time length of discretization, the exponential shape in the dynamics, and the presence of spatial spillovers.
To illustrate the point, assume that $\gamma_{S},\gamma_{A},\gamma_{R} = 0$; from Eq. \eqref{eq:dataGenerationProcessCellTotalIncome} for $t = t'$, it follows: 
\begin{multline}\label{eq:partialtty} 
	\partial_{tt}^{2}y(t',z^i) = 	\partial_{t}\left[\partial_{t}y(t,z^i)\right]\big\vert_{t=t'}=\\
	 = \partial_t a\left(t,z^i\right)\big\vert_{t=t'} + \phi\partial_{t}y(t,z^i)\big\vert_{t=t'} + \gamma_D \left[\partial_{z_1 z_1} \partial_t y(t',z)\big\vert_{z=z^i} + \partial_{z_2 z_2} \partial_t y(t',z)\big\vert_{z=z^i}  \right].
\end{multline}
Since $t'$ is not determined, a natural choice is to approximate the first order time derivative in unit $z^i$ at period $t'$ by the time variation in the whole period, i.e. $\partial_t y(t',z^i) \approx \Delta_\tau y_i$ and, at the same time, to run the space discretization as in Eq. \eqref{eq:dataGenerationProcessCellTotalIncomeDiscritizationSpace}. This approximation is more accurate as $t'$ goes to $t$, i.e. $\tau$ goes to zero.
Consequently, the coefficients $\phi$  and $\gamma_D$ appearing in Eq. \eqref{eq:partialtty} will be affected by this approximation. Therefore:  
\begin{equation}\label{eq:partialttyMatrix}
\partial_{tt}^{2}y(t',z^i) \approx \partial_t a\left(t',z^i\right) + {\rho_\phi} \Delta_\tau y_i + {\rho}_D \left[( M_{z_1 z_1} +  M_{z_2 z_2}) \Delta_\tau \mathbf{y}  \right]_i,
\end{equation}
for some ${\rho}_{\phi}\approx \phi$ and ${\rho}_D \approx \gamma_D$.
Hence, from Eq. \eqref{eq:timediscretizationBaselineEquation}:
\begin{eqnarray} \nonumber
\Delta_\tau y_i & \approx & a\left(0,z^i\right) + \phi y_{i} +  \gamma_D \left[(M_{z_1 z_1} + M_{z_2 z_2})\mathbf{y}\right]_{i} + \\
&+& \frac{\tau}{2}\left\{ \partial_t a\left(t,z^i\right)\big\vert_{t=0} + {\rho_\phi} \Delta_\tau y_i + {\rho}_D \left[( M_{z_1 z_1} +  M_{z_2 z_2}) \Delta_\tau \mathbf{y}  \right]_i\right\}, \nonumber
\end{eqnarray}
or, taking $a_i \equiv a\left(0,z^i\right)$ and $\Delta_\tau a_i \approx \partial_t a\left(t,z^i\right)\big\vert_{t=0} $, in matrix form:
\begin{eqnarray} \nonumber
	\Delta_\tau \mathbf{y} & \approx & \mathbf{a} + \phi \mathbf{y} +  \gamma_D (M_{z_1 z_1} + M_{z_2 z_2})\mathbf{y} + \\
	&+& \frac{\tau}{2}\left[ \Delta_\tau \mathbf{a} + {\rho_\phi} \Delta_\tau \mathbf{y} + {\rho}_D ( M_{z_1 z_1} +  M_{z_2 z_2}) \Delta_\tau \mathbf{y} \right], \nonumber
\end{eqnarray}
from which:
\begin{multline}\label{eq:discretizationSpaceTimeSimplifiedModel}
\Delta_\tau \mathbf{y} \approx \left[ \left(1 - \frac{\tau{\rho_\phi}}{2}\right)\mathbf{I} - \frac{\tau{\rho}_D}{2}( M_{z_1 z_1} +  M_{z_2 z_2}) \right]^{-1}\times  \\ \times  \left[ \mathbf{a} +  \frac{\tau}{2}\Delta_\tau \mathbf{a} + \phi \mathbf{y} + \gamma_D (M_{z_1 z_1} + M_{z_2 z_2})\mathbf{y}\right].
\end{multline}
Eq. \eqref{eq:discretizationSpaceTimeSimplifiedModel} makes clear the importance of considering the second term in Eq. \eqref{eq:timediscretizationBaselineEquation}, which is represented by the expression within the square bracket to be inverted. The magnitude of this correction is directly related to $\tau$, i.e. the length of time discretization, and to the extent of spatial spillovers measured by matrices $M_{z_1 z_1}$ and $M_{z_2 z_2}$. Moreover, Eq. \eqref{eq:discretizationSpaceTimeSimplifiedModel} shows that all the parameters are identifiable up to the scale factor $ \left(1 - {\tau{\rho_\phi}}/{2}\right)$.
Appendix \ref{sec:appendixMatricesdiscretizationOverTime} deals with the general case, and it shows that:
\begin{eqnarray}	\label{eq:modelDiscretizedOvertimeWithoutRandom}
 \nonumber
	\Delta_\tau \mathbf{y} & = &  \tilde{\mathbf{a}} + \tilde{\phi} \mathbf{y} + \\ \nonumber
	&+& \tilde{\gamma}_{S} \left[ M_{z_1} \left(\mathbf{y} \odot M_{z_1} \mathbf{s} \right)  +  M_{z_2} \left(\mathbf{y} \odot M_{z_2} \mathbf{s} \right)  \right] +  \tilde{\rho}_S M_S \Delta_\tau \mathbf{y} \\ \nonumber
	&+&\tilde{\gamma}_A \left[ M_{z_1} \left(\mathbf{y} \odot M_{z_1} W_{h_A} \mathbf{y} \right)  +  M_{z_2} \left(\mathbf{y} \odot M_{z_2} W_{h_A} \mathbf{y} \right)  \right] +  \tilde{\rho}_A M_A \Delta_\tau \mathbf{y}\\  \nonumber
	&+&\tilde{\gamma}_R \left[ M_{z_1} \left(\mathbf{y} \odot M_{z_1} W_{h_R} \mathbf{y} \right)  +  M_{z_2} \left(\mathbf{y} \odot M_{z_2} W_{h_R} \mathbf{y} \right)  \right] + \tilde{\rho}_R M_R \Delta_\tau \mathbf{y}\\ \nonumber 
	&+& \tilde{\gamma}_D (M_{z_1 z_1} + M_{z_2 z_2})\mathbf{y} + \tilde{\rho}_D M_D \Delta_\tau \mathbf{y}\\ 
	&+& \mathbf{r}_{s,t},
\end{eqnarray}
where $M_S$, $M_A$, $M_R$ and $M_D$ are defined in Appendix \ref{sec:appendixMatricesdiscretizationOverTime} and $\mathbf{r}_{s,t}$ is the reminder of the process of discretization over space and time and it is such that $\norm {\mathbf{r}_{s,t}}$ goes to zero as the space and time discretization gets finer. The statistical properties of this reminder will be explored in Section \ref{sec:MCsimulations}.

\subsubsection{The reminder of the discretization as an error term \label{sec:fluctuationsFiniteWorld}}

Some of the properties of the reminder $\mathbf{r}_{s,t}$ of Model \eqref{eq:modelDiscretizedOvertimeWithoutRandom} are still unknown. This suggests to treat it as an error term in the estimate.
Our guess is that such a term can show some spatial dependence induced by the spatial discretization.
Moreover,  Eq. \eqref{eq:dataGenerationProcessCellTotalIncome} is derived by considering an infinite number of agents in every location $z \in \Omega$  in the spirit of the laws of large numbers (see Appendix \ref{sec:microfoundation}). However, in the real world the units of observation can include many agents but not infinite and, therefore, some fluctuations from the mean behaviour should be present at the unit level. The characteristics of these fluctuations are still under investigation on the theoretical ground \citep{graham1996asymptotic}. Heuristically, the presence of spatial interactions between agents suggests that fluctuations should contain some degree of spatial correlation already in the continuous time-space framework. This spatial correlation should be further magnified in the discretization over time and space by the presence of spatial spillovers in the dynamics (represented by matrices $M_S$, $M_A$, $M_R$ and $ M_D$, and $W_{h_R}$ and $W_{h_A}$).

To account for the reminders of the approximation over time and space and of the possible finite-world fluctuations, Model \eqref{eq:modelDiscretizedOvertimeWithoutRandom} is therefore amended as it follows:
\begin{eqnarray} \nonumber
	\Delta_\tau \mathbf{y} &=&\tilde{\mathbf{a}}  + \tilde{\phi} \mathbf{y} +  \\ \nonumber
	&+& \tilde{\gamma}_{S} \underbrace{\left[ M_{z_1} \left(\mathbf{y} \odot M_{z_1} \mathbf{s} \right)  +  M_{z_2} \left(\mathbf{y} \odot M_{z_2} \mathbf{s} \right)  \right]}_{\mathbf{x}_S} +  \\ \nonumber
	&+&\tilde{\gamma}_A \underbrace{\left[ M_{z_1} \left(\mathbf{y} \odot M_{z_1} W_{h_A} \mathbf{y} \right)  +  M_{z_2} \left(\mathbf{y} \odot M_{z_2} W_{h_A} \mathbf{y} \right)  \right]}_{\mathbf{x}_A} +  \\  \nonumber
	&+&\tilde{\gamma}_R \underbrace{\left[ M_{z_1} \left(\mathbf{y} \odot M_{z_1} W_{h_R} \mathbf{y} \right)  +  M_{z_2} \left(\mathbf{y} \odot M_{z_2} W_{h_R} \mathbf{y} \right)  \right] }_{\mathbf{x}_R}+ \\  \nonumber
	&+& \tilde{\gamma}_D \underbrace{(M_{z_1 z_1} + M_{z_2 z_2})\mathbf{y}}_{\mathbf{x}_D} + \\  \nonumber
    &+& \tilde{\rho}_S M_S \Delta_\tau \mathbf{y} + \tilde{\rho}_A M_A \Delta_\tau \mathbf{y} + \tilde{\rho}_R M_R \Delta_\tau \mathbf{y} + \tilde{\rho}_D M_D \Delta_\tau \mathbf{y} + \\
    &+& \boldsymbol{\epsilon},	
	\label{eq:modelDiscretizedOvertime}
\end{eqnarray}
where $\boldsymbol{\epsilon}$ is a spatially correlated error.
As already said, a feasible way to model $\boldsymbol{\epsilon}$ inspired by the spatial econometric literature is:
\begin{equation}
	\boldsymbol{\epsilon} = \lambda {W}_\epsilon \boldsymbol{\epsilon} + \boldsymbol{\eta},
	\label{eq:spatialErrors}
\end{equation}
where the matrix ${W}_\epsilon$ should reflect the structure of the spatial dependence of $\boldsymbol{\epsilon}$, while $\lambda$ its intensity, and, finally, $\boldsymbol{\eta}$ is a vector of well-behaved i.i.d. stochastic components. Unfortunately, the theory is silent on the shape of ${W}_\epsilon$; therefore, its specification will be data driven and such that it ensures the following properties: i) ${W}_\epsilon$ has zeros on its main diagonal; ii) ${W}_\epsilon$ is symmetric; iii) the spatial correlation is allowed to be nonlinear with respect to the distance  (see Appendix \ref{app:spatialMatrixErrors}). Numerical simulations in Section \ref{sec:MCsimulations} confirm the presence of spatial dependence in $\boldsymbol{\epsilon} $ and also the goodness of our proposed procedure for the specification of ${W}_\epsilon$.

%

\subsection{Some remarks on SARD model \label{sec:remarksSARD}}

The methodology used to estimate the parameters of Eq. \eqref{eq:dataGenerationProcessCellTotalIncome} strongly departs from the typical approach used in growth empirics, i.e. a log-linear approximation around the long-run equilibrium to obtain a reduced-form model, that is linear with respect to the model parameters \citep{durlauf2005growth}. In particular, both the difficulty of characterizing the \textit{spatial equilibrium} and the joint presence of salient \textit{non-linear dynamics} and \textit{spatial spillovers} are the key properties of Eq.  \eqref{eq:dataGenerationProcessCellTotalIncome}.  Therefore, the SARD econometric Model \eqref{eq:SARD} has been derived with no reference to the equilibrium spatial distribution, but it results linear in the parameters since the discretization over space and time preserves the linearity. Finally, the relationship between the local growth and the initial local level of income $\mathbf{y}_0$ is strongly non linear as the outcome of aggregative, repulsive and diffusive forces.

As regards the space, it can be modelled either in a continuous two-dimensional plane (longitude and latitude) or as a set of discrete spatial units; in the latter case, the \textit{spatial structure} can be represented by connecting arcs, reflecting the interaction between couples of spatial units (\citealp{mendes2015parametric}). Spatial econometrics generally adopts the discrete approach to match the only availability of discrete spatial data, and models the spatial structure by the \textit{(spatial) weight matrix}. In particular, the weight matrix identifies if two spatial units are neighbours and, also, the intensity of their interaction; the proximity might be specified in several ways, for example using geographical distance and/or spatial contiguity  \citep{lesage2009introduction}.\footnote{This choice should be driven by the ``problem being modelled, and perhaps particular additional non-sample information which may be available'' (\citealp{mendes2015parametric}).}
The spatial structure implied by Eq. \eqref{eq:dataGenerationProcessCellTotalIncome} is, however, more complex since it is expressed in continuous space, which prevents the possibility of specifying the pairwise interactions by a simple weight matrix. Instead, the spatial structure is specified by a function depending on the relative position between two spatial units, and by another function describing the geographical landscape. The latter specifies the intrinsic characteristics of the territory (e.g. the presence of rivers, mountains, etc.), that may affect the strength of the connection between two spatial units. The former, instead, describes how the intensity of the interaction between two units depends on their reciprocal distance, keeping into account also the direction to move from one unit to the other. This effect of directionality is measured by \textit{spatial gradients}, i.e. spatial partial derivatives. 
The discrepancy between the continuous locations of Eq. \eqref{eq:dataGenerationProcessCellTotalIncome} and the actual discrete data imposes an intermediate step of discretization to approximate partial derivatives to arrive at the econometric model. In particular, we employ the \textit{generalized finite difference method}, which is based on a Taylor approximation to compute partial derivatives on irregular grids \citep{jensen1972finite}.\footnote{Appendix \ref{sec:computationGradientLaplacian} contains a self-contained overview of the method, which amounts to constructing a set of matrices that approximate the derivative operators up to the appropriate order in space.}

As regards time, continuous time and spatial spillovers induce a contemporaneous effect between two locations. In the discrete time approximation, the final outcome is a model specification which strongly resembles the class of spatial lag models \citep{lesage2009introduction}. This implies that endogeneity is a key feature in the estimate.


Finally, Model \eqref{eq:SARD} resembles typical spatial econometric models, such as the spatial autoregressive model (SAR), the spatial error model (SEM), and also a flavour of an SLX model (\citealp{lesage2009introduction}). However, the calculation of the matrices $M$s, i.e. the counterparts of the spatial weights matrix, is strictly driven from the discretization of Eq. \eqref{eq:dataGenerationProcessCellTotalIncome} and not a priori assumed. In particular, the model in its parts of spatial lag and error is not derived by strategic decision of rational agents as in \cite{brueckner2003strategic} or other types of strategic social interaction as in \cite{brock2001discrete, brock2001interactions}, but as results of the discretization over time and space of a reallocation process affected by exogenous factors (topography), purely random component (diffusion) or the endogenous forces of aggregation and repulsion.
This feature is crucial for the identification of the model parameters. As discussed in \cite{gibbons2012mostly}, the latter is crucially based on the use of the ``true'' spatial matrix in the estimate, which in our case is driven by the theoretical model and exogenous. Moreover, since Model \eqref{eq:SARD} falls back into the \cite{blume2015linear}'s case of ``known sociomatrices'', its identification seems to be guaranteed from the shape of the matrices $M$s.

\subsection{Caveats on SARD estimation}\label{subsec:caveats}

The proposed methodology has some challenges and intrinsic limitations discussed below. 
Firstly, the estimation of Model \eqref{eq:SARD} requires a Maximum Likelihood (ML) or Instrumental Variable (IV) procedure given the presence of endogenous regressors $M_S \Delta_\tau \mathbf{y}$, $M_A \Delta_\tau \mathbf{y}$, $M_R \Delta_\tau \mathbf{y}$ and $M_D \Delta_\tau \mathbf{y}$.  On the one hand, ML can be computationally intensive given that one needs to optimize over four parameters (even five in the presence of spatial dependence in the error term) and the normality of the error term is not guaranteed given the presence of nonlinear spatial dependence. In this regard, the simple OLS estimate provides a benchmark robust to non-normality of residuals (see Section \ref{sec:MCsimulations}). On the other hand, the IV requires the specification of not weak and valid instruments. Unfortunately, in Section \ref{sec:MCsimulations} we will show that the validity of the usual instruments used in spatial econometrics, based on spatially lagged variables of higher order, is not always guaranteed in our framework. 

Secondly, the discretization procedure can introduce a bias in the model, which we investigate in detail in Section \ref{sec:MCsimulations}. However, discretization errors are proportional to the distance between the evaluation points and, therefore, bias can be overlooked with a sufficiently fine discretization. In this respect, if the geographical resolution of data is low, i.e. the distance between spatial locations is too wide, then a substantial bias cannot be avoided.
As a concrete example, the proposed methodology would most likely fail (or produce meaningless results) if applied to countries' GDP per capita spatial distribution.
Also, the proposed framework cannot provide any diagnostic on the minimum resolution needed to get reliable estimates. However, a mismatch between the expected sign of parameters and its estimate could be a signal that the spatial resolution of the data used in the analysis is too low (or that the model is misspecified). In this respect, attention must be given to the fact that parameters are identifiable up to the scale factor $\left(1-\tau \rho_\phi/2\right)$. Given that $\rho_\phi \approx \phi$ and $\phi$ is approximately the instantaneous growth rate of $y$, one should keep into consideration the expected instantaneous growth rate to identify possible changes in the expected signs. 
The magnitude of the spatial correlation among the units of observation is another important feature to evaluate the bias from discretization. If they are highly correlated, that is neighbouring locations don't differ too much among themselves, then the bias is limited even if locations are spread far apart. On the contrary, the approximation may be very poor even when the units of observations are of small size if the spatial correlation is very low.

Thirdly, the theoretical model also imposes limitations on the possible choice of the (kernel) matrices $W_{h_A}$ and $W_{h_R}$. The latter derives from the space discretization of the kernel functions $K_{h_A}$ and $K_{h_R}$ in Eq. \eqref{eq:dataGenerationProcessCellTotalIncome}. Accordingly, they require to be defined in terms of \textit{relative distance} between spatial units and, therefore, they are not allowed to be based on contiguity between areas of observations (as, e.g., $N$-order contiguity). 


Fourthly, since the ML estimation of the SARD model is computationally challenging, an educated choice of the initial conditions for the numerical optimization is advised. In this respect, the initial conditions will be set based on an instrumental variable (IV) estimate, which is easy to compute and does not rely on the assumption of normality of the disturbances, although it can be less efficient than the ML estimator \citep{kelejian1998generalized}. To instrument the spatially lagged dependent variable, \cite{kelejian2004instrumental} suggests to use $\left[{X} \; W{X} \dots W^g {X}\right]$, where $g$ is usually taken equal to 1 or 2 according to the type of model estimated. In the same vein, we use as instruments $\left[{M}_S^2 {X} \; {M}_A^2{X} \; {M}_R^2 {X} \; {M}_D^2 {X}\right]$.

Finally, since the econometric model is obtained by the discretization of a continuous space model, all variables must share the same scale with respect to the unit of observation; for example, the total income of a region must be rescaled by its area.

\subsection{Numerical investigations \label{sec:MCsimulations}}

In this section, we examine the properties of the proposed estimation method of spatial PDEs by numerical investigations. We focus only on the errors induced by discretization and neglect other sources of randomness in $\epsilon$, such as the finite number of individuals and firms and the presence of measurement errors in $y$.

\subsubsection{The baseline setting}

\begin{figure}[!htbp]
	\begin{subfigure}[b]{0.32\textwidth}
		\includegraphics[width=\textwidth]{y0_continuosSpace.pdf}
		\caption{The initial spatial distribution $y_0$ in continuos space.}
		\label{fig:y0_continuosSpace}
	\end{subfigure}
	\begin{subfigure}[b]{0.32\textwidth}
		\includegraphics[width=\textwidth]{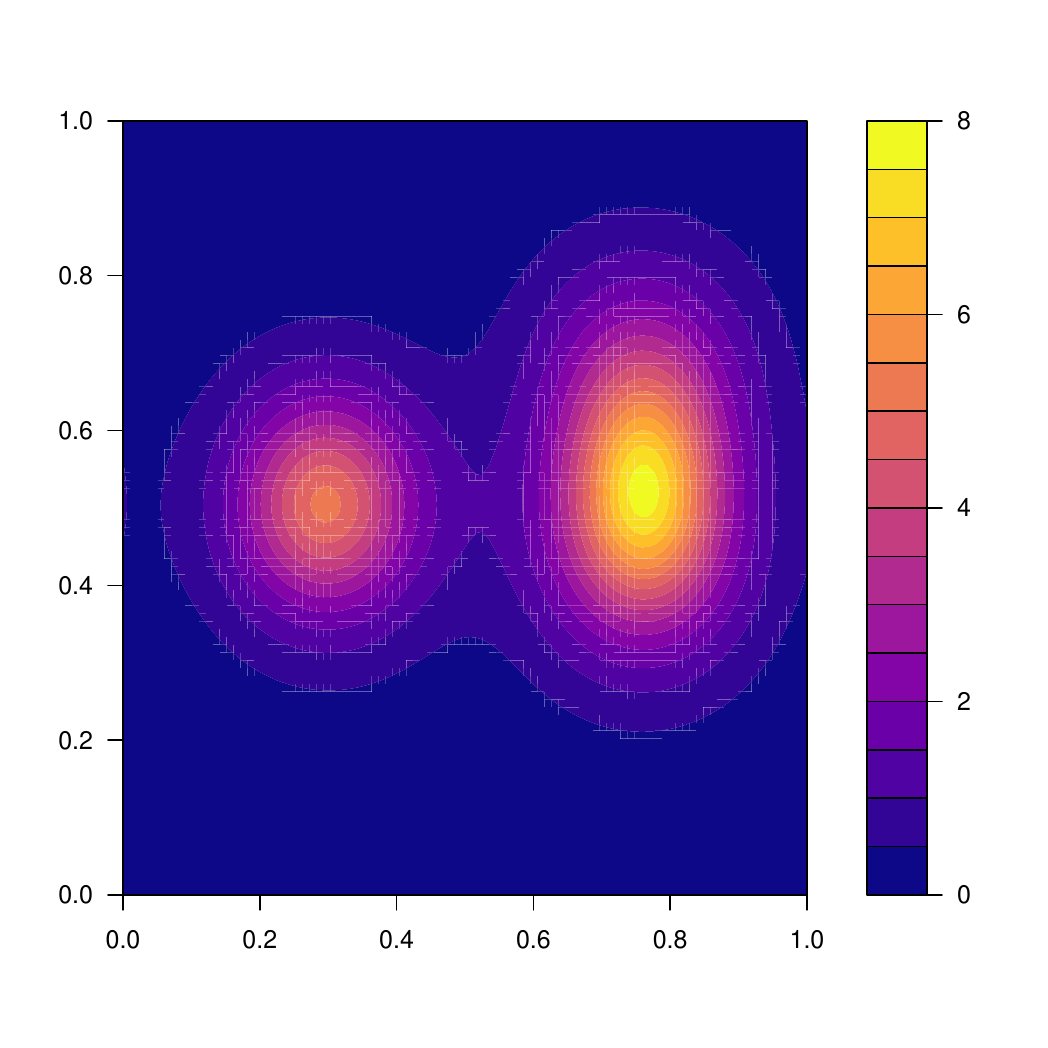}
		\caption{The final spatial distribution $y_1$ in continuos space.}
		\label{fig:yT_continuosSpace_tau1}
	\end{subfigure}
	\begin{subfigure}[b]{0.32\textwidth}
		\includegraphics[width=\textwidth]{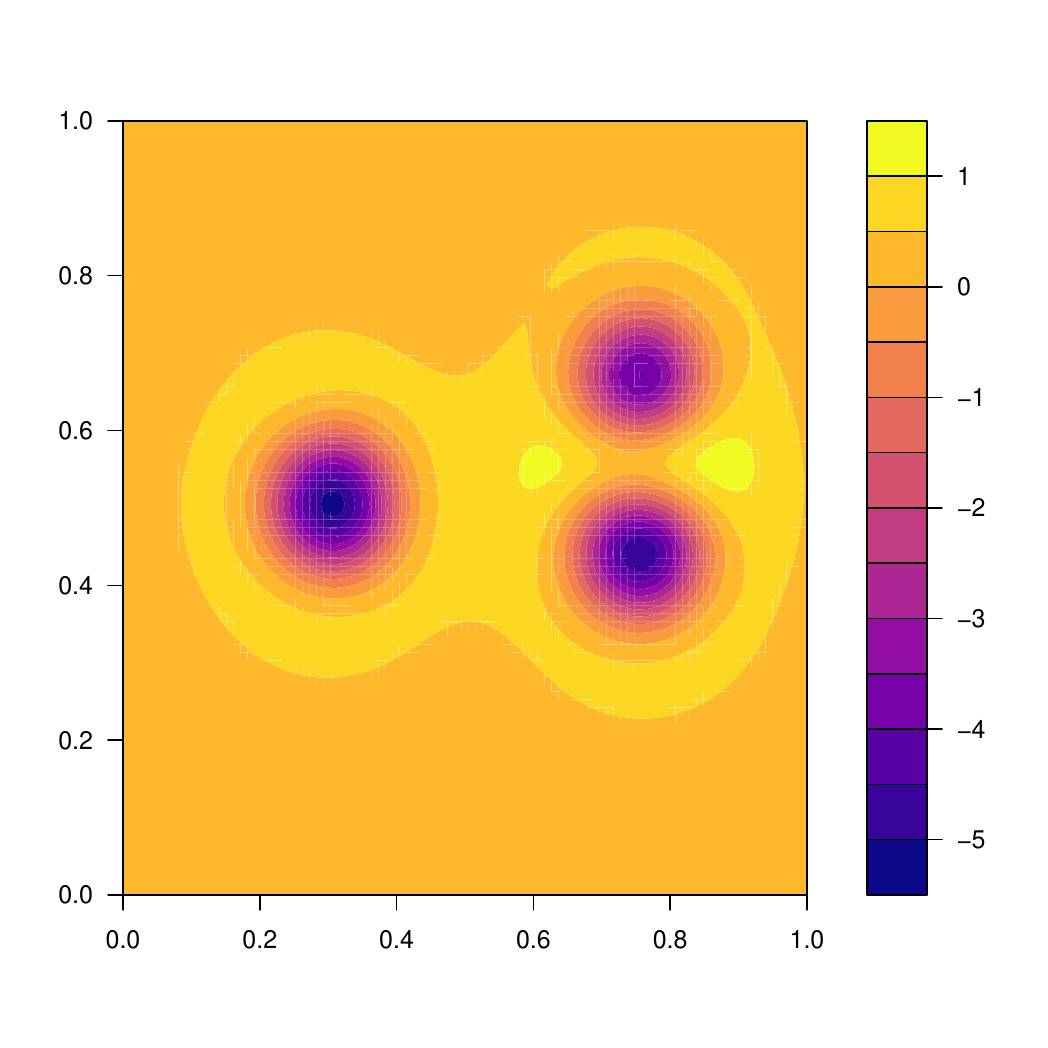}
		\caption{The spatial distribution of $\Delta_1 y$ in continuos space.\\\phantom{cricci}}
		\label{fig:delta_continuosSpace_tau1}
	\end{subfigure}
\caption{Panel (a) reports the initial spatial distribution; Panel (b) the final spatial distribution; and Panel (c) the distribution of time change  of $y$ with $\tau=1$ in the 2-dimensional torus of area one. The colour reflects the value of the variable in the location (the blue corresponds to the lowest one).}
\label{fig:MonteCarloBasicVariablestau1}
\end{figure}

As baseline setting, consider a two-dimensional torus of area one, the kernels in Eq. \eqref{eq:kernelAandR} with $h_A=0.15$ and $h_R=0.4$, an initial distribution $y_{0}$ with three peaks, $\gamma_S=0$ (no effect from spatial exogenous heterogeneity), $\alpha=0.01$, $\phi=0.01$, $\gamma_A=-0.00175$, $\gamma_R = 0.0025$, and $\gamma_D=0.00525$. The parameters and the initial distribution are chosen such that the time-space evolution of $y$ simulated by Eq. \eqref{eq:dataGenerationProcessCellTotalIncome} displays the three spatial behaviours of aggregation, repulsion, and diffusion.
\begin{figure}[!htbp]
	\begin{subfigure}[b]{0.32\textwidth}
		\includegraphics[width=\textwidth]{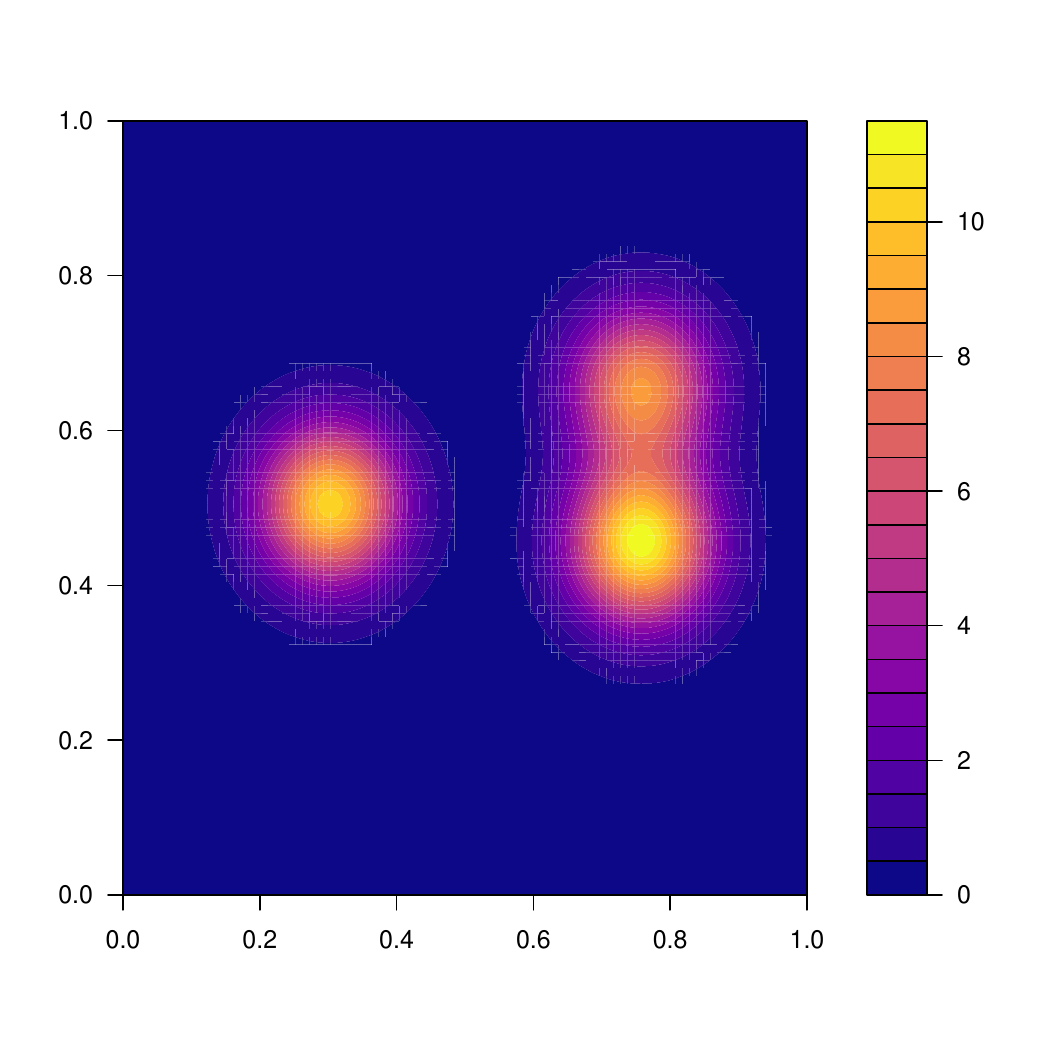}
		\caption{The initial spatial distribution $y_0$ in continuos space.}
		\label{fig:y0_continuosSpace_tau1}
	\end{subfigure}
	\begin{subfigure}[b]{0.32\textwidth}
		\includegraphics[width=\textwidth]{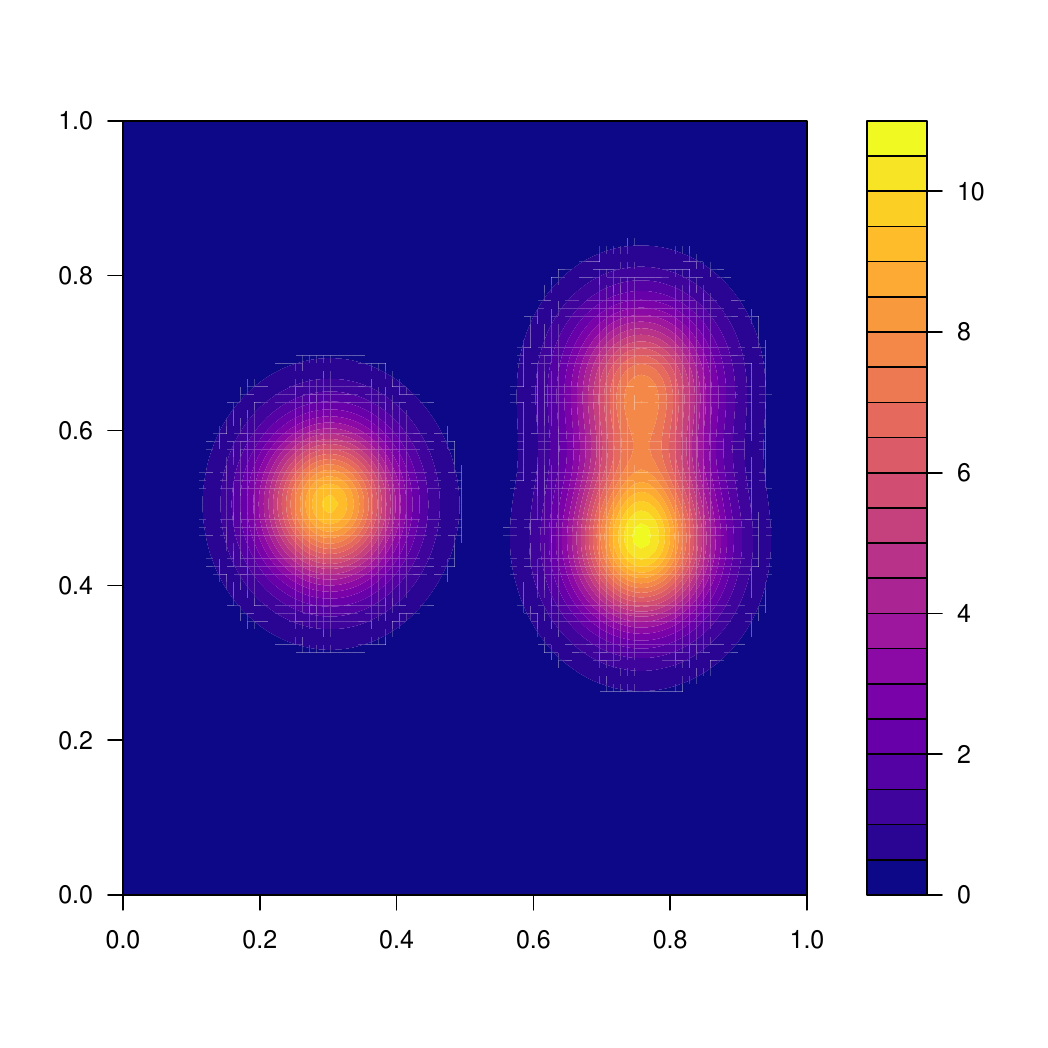}
		\caption{The final spatial distribution $y_{0.1}$ in continuos space.}
		\label{fig:yT_continuosSpace_tau01}
	\end{subfigure}
	\begin{subfigure}[b]{0.32\textwidth}
		\includegraphics[width=\textwidth]{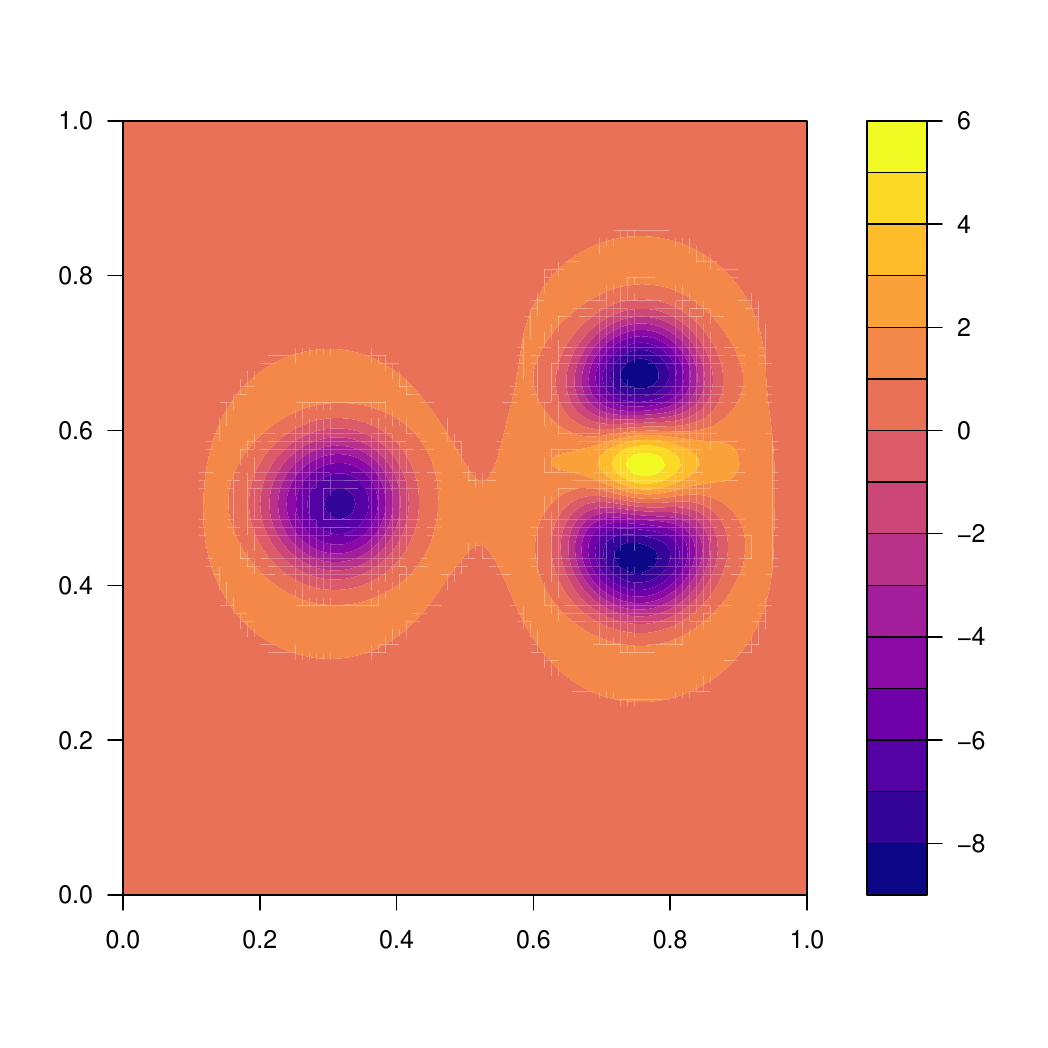}
		\caption{The spatial distribution of $\Delta_{0.1} y$ in continuos space.\\\phantom{cricci}}
		\label{fig:delta_continuosSpace_tau01}
	\end{subfigure}
\caption{Panel (a) reports the initial spatial distribution; Panel (b) the final spatial distribution; and Panel (c) the distribution of time change  of $y$ with $\tau=0.1$ in the 2-dimensional torus of area one. The colour reflects the value of the variable in the location (the blue corresponds to the lowest one).}
\label{fig:MonteCarloBasicVariablestau01}
\end{figure}
For testing the importance of the time discretization, we consider two-time intervals, one with  $\tau=0.1$, reported in Figure \ref{fig:MonteCarloBasicVariablestau01}, and one with $\tau=1$, reported in Figure \ref{fig:MonteCarloBasicVariablestau1}.\footnote{With a little abuse of notation we will denote by $\tau = 0.1$ the case where $\tau = 1$ and $\Delta = 0.1$.} After 0.1 periods, two of the three peaks on the right in $y_{0}$ are merging as the prevalence of the aggregative force; the third peak on the left instead becomes flatter, signalling the prevalence of repulsive and diffusive forces (see Figures \ref{fig:yT_continuosSpace_tau01} and \ref{fig:delta_continuosSpace_tau01}). However, after one period the aggregative force led to a complete merge of the two peaks on the right, while the one on the left spread even more due to repulsive and diffusive forces (see Figure \ref{fig:yT_continuosSpace_tau1} and \ref{fig:delta_continuosSpace_tau1}).

\begin{figure}[!htpb]
	\begin{subfigure}[b]{0.32\textwidth}
		\includegraphics[width=\textwidth]{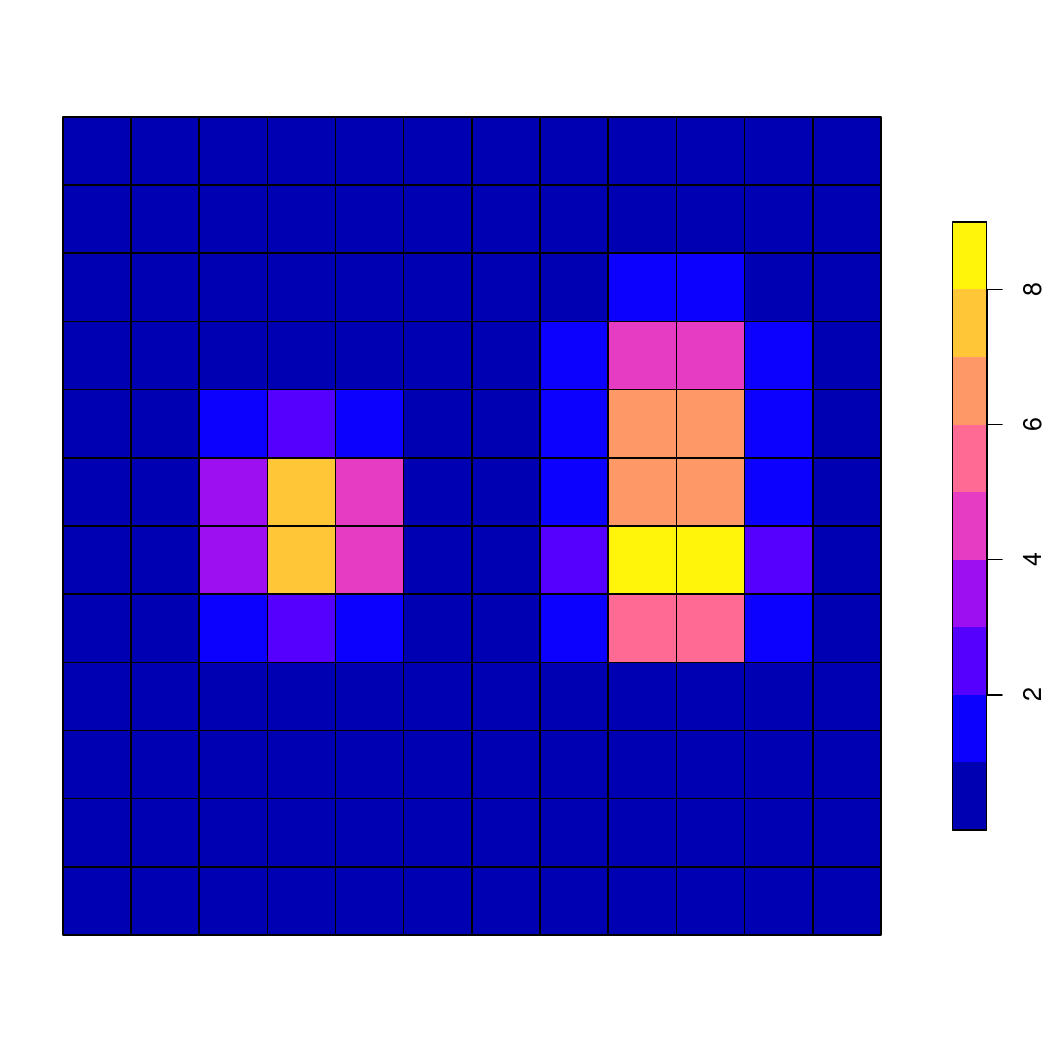}
		\caption{The initial spatial distribution $y_0$ partitioned into 144 cells.}
		\label{fig:y0144}
	\end{subfigure}
\begin{subfigure}[b]{0.32\textwidth}
	\includegraphics[width=\textwidth]{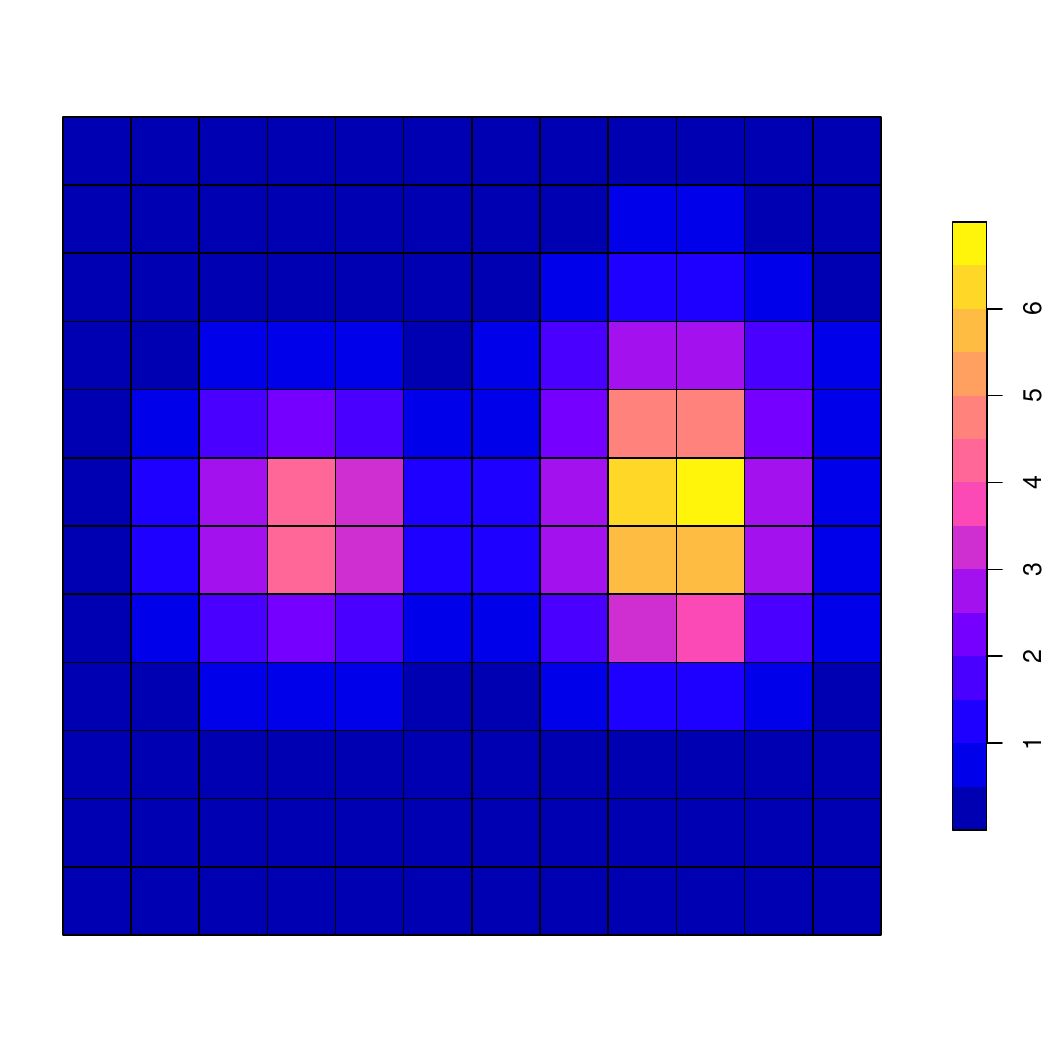}
	\caption{The final spatial distribution $y_1$ partitioned into 144 cells.}
	\label{fig:yT144}
\end{subfigure}
	\begin{subfigure}[b]{0.32\textwidth}
	\includegraphics[width=\textwidth]{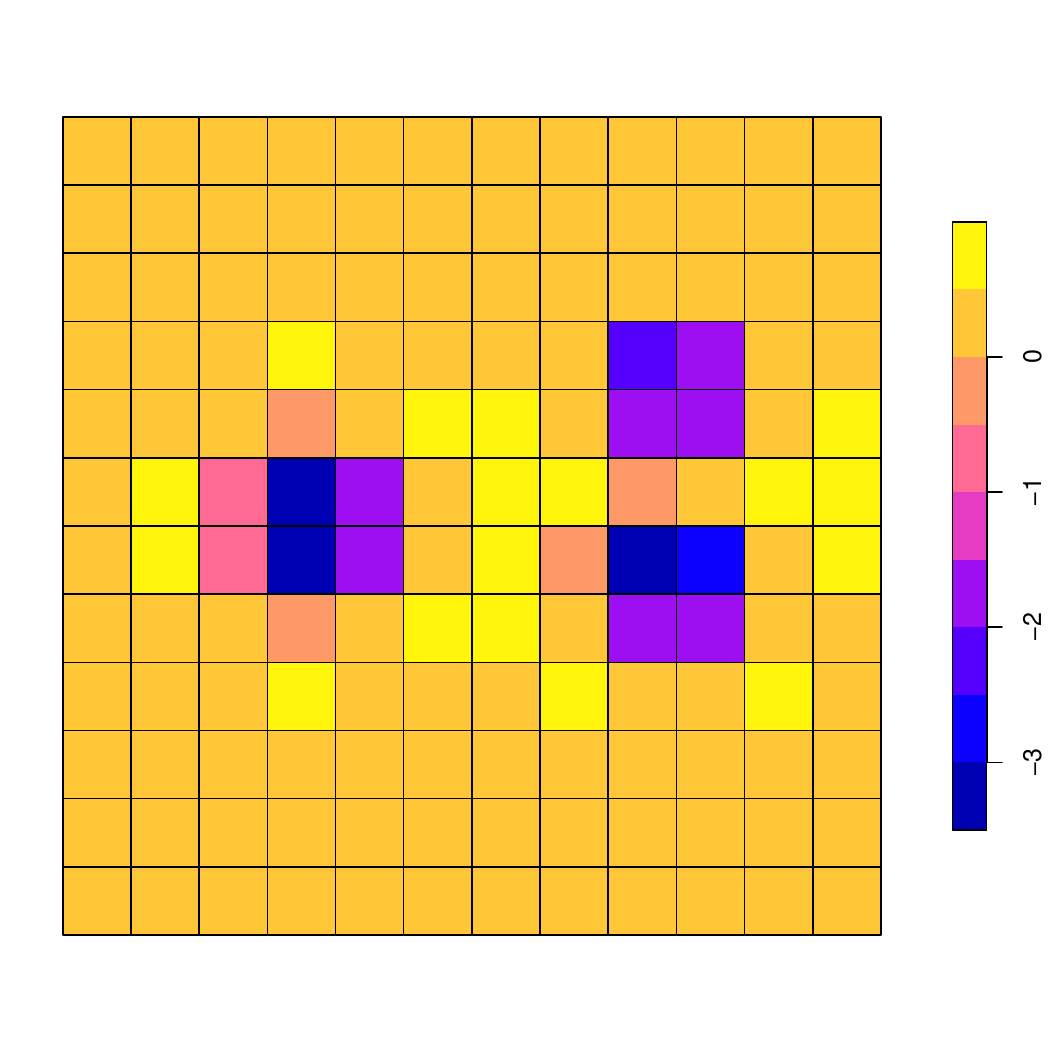}
	\caption{The spatial distribution of $\Delta_1 y$ partitioned into 144 cells.\\\phantom{parenti}}
	\label{fig:delta144}
\end{subfigure}
	\begin{subfigure}[b]{0.32\textwidth}
	\includegraphics[width=\textwidth]{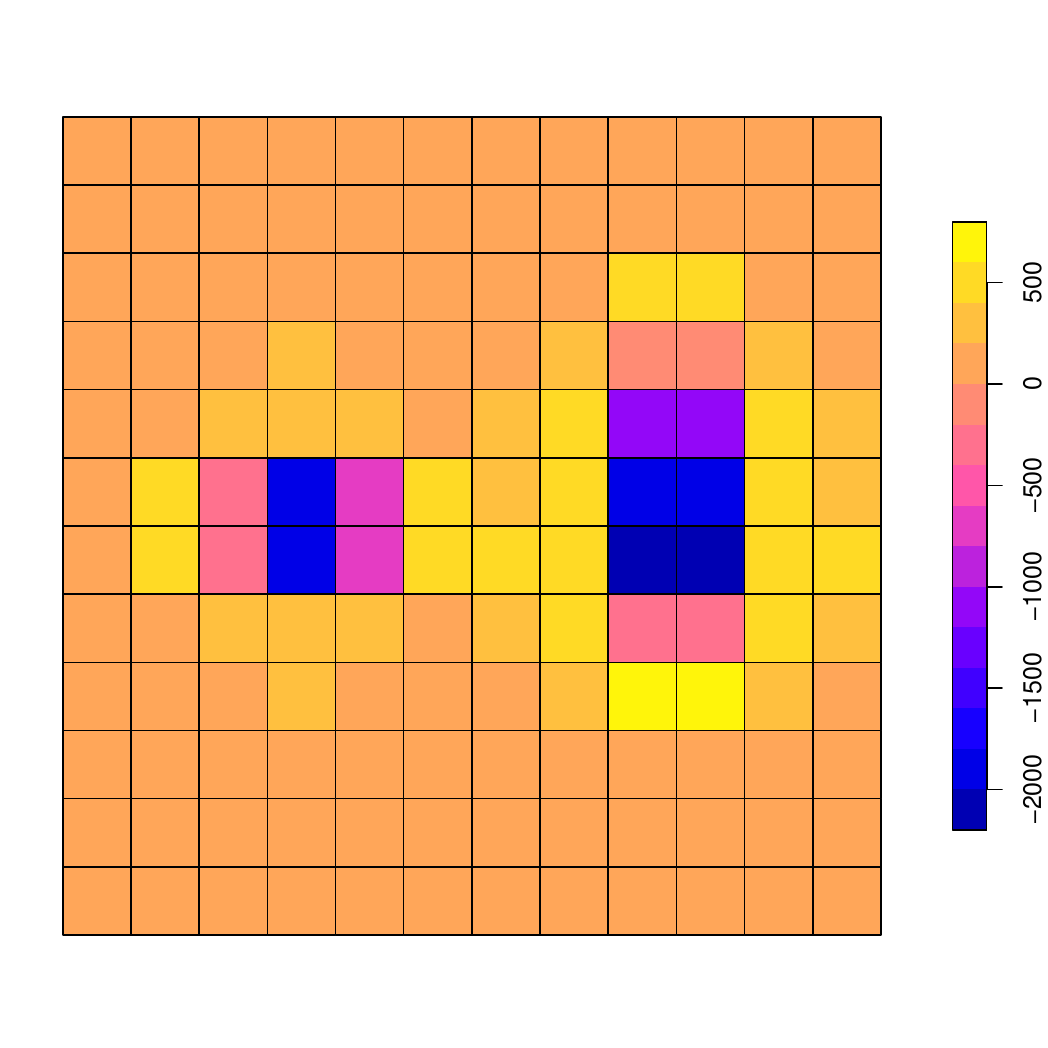}
	\caption{The spatial distribution of $x_A$ partitioned into 144 cells.}
	\label{fig:xA144}
\end{subfigure}
	\begin{subfigure}[b]{0.32\textwidth}
	\includegraphics[width=\textwidth]{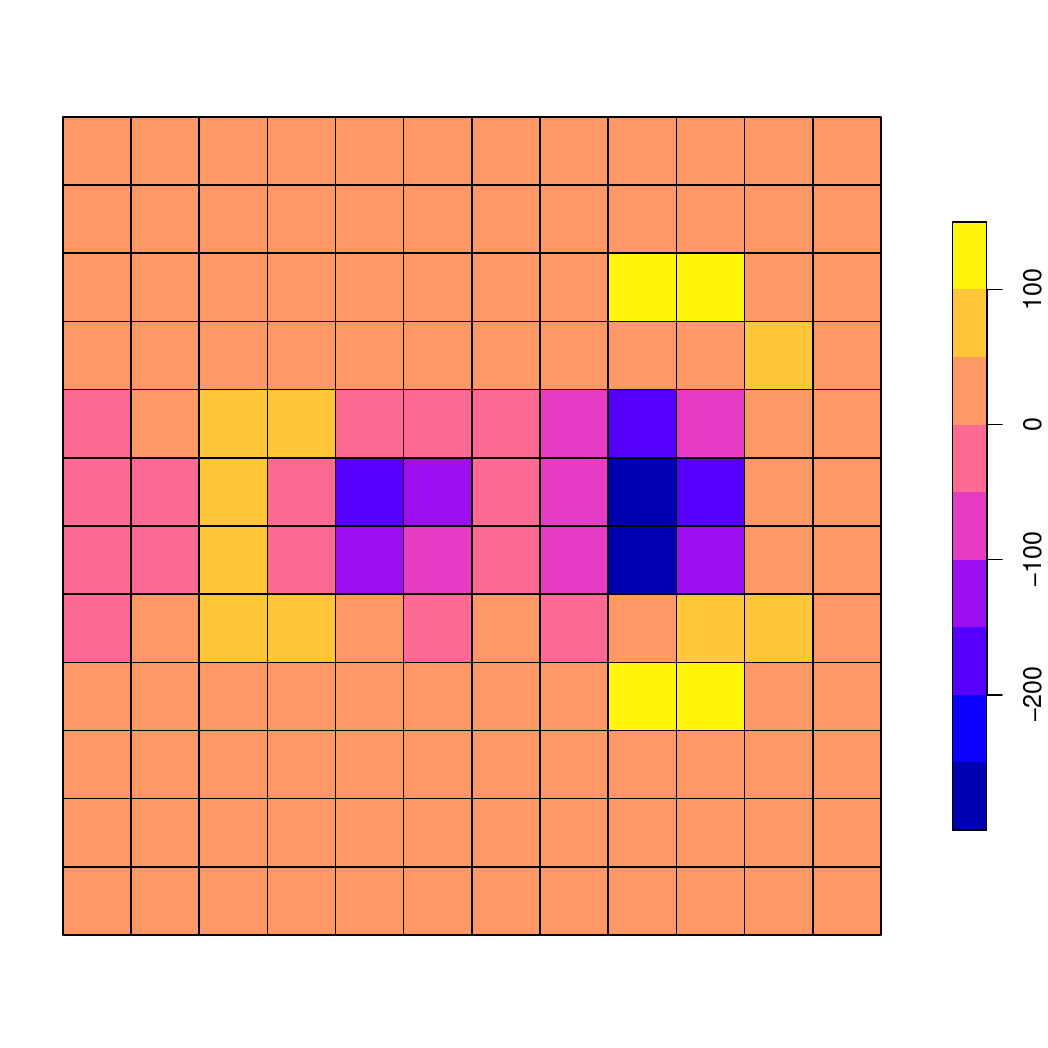}
	\caption{The spatial distribution of $x_R$ partitioned into 144 cells.}
	\label{fig:xR144}
\end{subfigure}
	\begin{subfigure}[b]{0.32\textwidth}
	\includegraphics[width=\textwidth]{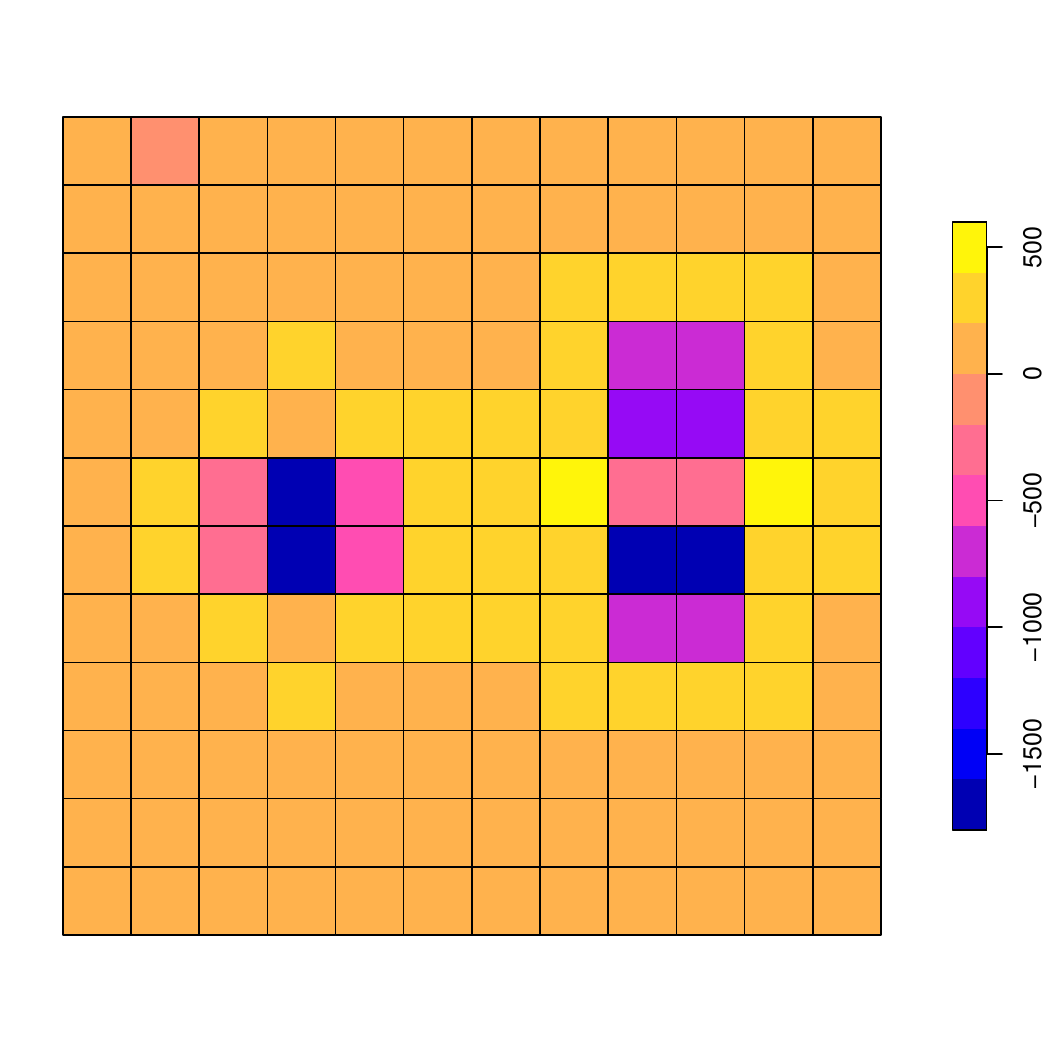}
	\caption{The spatial distribution of $x_D$ partitioned into 144 cells.}
	\label{fig:xD144}
\end{subfigure}
\caption{For a partition of space into $N=144$ regular cells, Panel (a) reports the initial spatial distribution; Panel (b) the final spatial distribution; Panel (c) the distribution of time change of $y$ with $\tau=1$; 
Panel (d) the spatial distribution of $x_A$; Panel (e) the spatial distribution of $x_R$; and Panel (f) the spatial distribution of $x_D$ of Model (\ref{eq:SARD}). The colour reflects the value of the variable in the units of observation (the blue corresponds to the lowest one).}
\label{fig:MonteCarloBasicVariablesSpacediscretization}
\end{figure}

To appreciate the impact of the space discretization, Figures \ref{fig:y0144}, \ref{fig:yT144} and \ref{fig:delta144} depict the case of a partition of space into $N=144$ regular cells for $\tau=1$. Given this discretization, Figures \ref{fig:xA144}, \ref{fig:xR144} and \ref{fig:xD144} report the maps of regressors $x_A$, $x_R$ and $x_D$ respectively. Given that $\gamma_{A}<0$ while $\gamma_R > 0$, the attractive and repulsive forces show a similar spatial pattern but an opposite effect on the dynamics of $y$: the first force induces the emergence of one peak on the right, while the second works to flatten the peak on the left. Finally, the diffusive force has a particularly strong (negative) magnitude in correspondence of three peaks.

\begin{figure}[!htpb]
	\centering
	\includegraphics[width=0.4\textwidth]{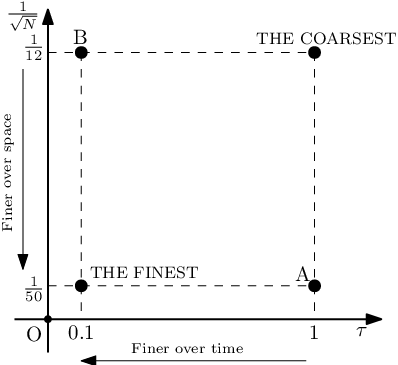}
	\caption{The possible combinations in the discretization over time and space when $N \in \left\{144,400,900,1600,2500 \right\}$ and $\tau \in \left\{0.1,0.25,0.5,0.75,1\right\}$.}
	\label{fig:discretizationOverTimeSpace}
\end{figure}

\subsubsection{The parameters' estimate}

To test the capacity of our methodology to identify the parameters of Eq. \eqref{eq:dataGenerationProcessCellTotalIncome} under different geographic granularity of data, consider five discretizations of space, i.e $N \in \left\{144,400,900,1600,2500 \right\}$ and five possible time intervals, i.e. $\tau \in \left\{0.1,0.25,0.5,0.75,1\right\}$. Figure \ref{fig:discretizationOverTimeSpace} is a graphical illustration of all the combinations of discretization over time and space, where the origin of the axes $O$ represents the case of continuous time and space, the combination $N=2500$ and $\tau=0.1$ the finest discretization, while the one with $N=144$ and $\tau=1$ the coarsest.

We consider four types of estimate: i) OLS NAIVE, i.e. the ordinary least squares estimate of Model \eqref{eq:SARD} without any control for the bias due to discretization (i.e. imposing $\rho_S=\rho_A=\rho_R=\rho_D=0$); ii) OLS, i.e. the ordinary least squares estimate of Model \eqref{eq:SARD}, which means including the controls for the bias due to discretization without dealing with their endogeneity; iii) IV, i.e. the instrumental variables estimate of Model \eqref{eq:SARD}, which means including the controls for the bias due to discretization and dealing with their endogeneity, using as instruments the second order spatially lagged exogenous regressors; and finally, iv) ML, i.e. the maximum likelihood estimate of Model \eqref{eq:SARD}, taking errors specified as in Eq. \eqref{eq:SARDspatialError}, which means including the controls for the bias due to discretization and dealing with their endogeneity and, in addition, accounting for the presence of spatial correlation in the residuals. 

\begin{table}[!htbp]
	\hspace{-1.5cm} \footnotesize
	\begin{tabular}{cccccccccc}  
\hline\hline
		\multicolumn{3}{c}{Parameters}   & $\alpha=0.01$ & $\phi=0.01$ & $\gamma_{A}=-0.00175$  &$\gamma_{R}=0.0025$ & $\gamma_{D}=0.00525$ & AICc & MSE \\ \cline{1-10}
		&&&&&&&\\
		$N$&$\tau$ &\multicolumn{7}{c}{\textbf{OLS NAIVE}} \\
		\cline{1-10}
		\\[-1.8ex]
      $144$& $1$ & \textit{Mean}& 0.16797 & -0.14647 & -0.00097 & 0.00054 & 0.00327 & -228.7 & 0.01096 \\
    & & \textit{Bias}& 0.15797 & -0.15647 & 0.00078 &  -0.00196&  -0.00198&  &  \\
	&&\textit{SE} &0.01185 &0.00784 &0.00005&0.00023&0.00006& & \\ 
	&&\textit{$SE^B$} &0.69501&0.58907&0.00437&0.00279&0.01396& & \\ 
   & & \textit{RMSE}&  0.69265& 0.59059& 0.00863 & 0.00681& 0.01419 &  &  \\ \\[-1.8ex]
 $2500$&$0.10$ & \textit{Mean} &0.03659 & -0.01804 & -0.00136 & 0.00175 & 0.00426 & -6290.2 & 0.00471 \\
   & & \textit{Bias}& 0.02659 &  -0.02804& 0.00039 & -0.00075 &  -0.00099&  &  \\
	&&\textit{SE} &0.00168 &0.00097&0.0000&0.00002&0.00001& & \\ 
	&&\textit{$SE^B$}	&0.05900&0.02909&0.00222&0.00285&0.00693& & \\
   & & \textit{RMSE}& 0.06700 & 0.03747&  0.00825& 0.00637& 0.00836 &  &  \\
	&&&&&&&\\
		&&\multicolumn{7}{c}{\textbf{OLS}} \\ 
		\cline{1-10}
		\\[-1.8ex]
      $144$& $1$ & \textit{Mean}  &0.04803 &-0.03236&-0.00119&0.00139&0.00406&-556.1 &0.00589\\
    & & \textit{Bias}&  0.03803&  -0.04236&  0.00056&  -0.00111& -0.00119 &  &  \\
	&&\textit{SE} &0.01467 &0.01311&0.00019&0.00054&0.00036& & \\ 
&&	\textit{$SE^B$}&0.19821&0.14384&0.01107&0.01083&0.02988& & \\ 
   & & \textit{RMSE}& 0.20220 & 0.14835& 0.01359 & 0.01225& 0.03007 &  &  \\ \\[-1.8ex]
 $2500$& $0.10$  & \textit{Mean}& 0.00855 & 0.01111 & -0.00175 & 0.00246 & 0.00521 & -19023.4 & 0.00025 \\
   & & \textit{Bias}& -0.00145 & 0.00111 & 0.00000 & -0.00004 & -0.00004 &  &  \\
	&&\textit{SE} &0.00044&0.00031&0.00000&0.00001&0.00000& & \\ 
 &&\textit{$SE^B$}	&0.01347 &0.01727&0.00273&0.00383&0.00812& & \\
   & & \textit{RMSE}& 0.01464 &0.01884 & 0.00870 & 0.00656&  0.00934&  &  \\ 
	&&&&&&&\\
		&&\multicolumn{7}{c}{\textbf{IV}} \\ 
		\cline{1-10}
		\\[-1.8ex]
      $144$& $1$ &  \textit{Mean}& 0.29173& -0.26550& -0.00247& -0.00167 & 0.00643& -408.9& 0.01740\\
  & & \textit{Bias}&  0.28173& -0.27550 & -0.00072 &-0.00417  & 0.00118 &  &  \\
	&&\textit{SE} &0.06446 &0.06346&0.00057&0.00173&0.00131& & \\ 
&&	\textit{$SE^B$}	&0.73241&0.65902&0.01480&0.01617&0.03611& & \\
   & & \textit{RMSE}& 0.75834 &0.68652 & 0.01664 &0.01731 &  0.03510&  &  \\ \\[-1.8ex]
 $2500$& $0.10$ & \textit{Mean}& 0.01757 & 0.00169 & -0.00174 & 0.00245 & 0.00510 & -18985.7 & 0.00032\\
    & & \textit{Bias}& 0.00757 & -0.00831 & 0.00001 & -0.00005 &-0.00015  &  &  \\
	&&\textit{SE} &0.00179&0.00176&0.00000&0.00001&0.00001& & \\ 
&&	\textit{$SE^B$}	&0.03353 &0.02036 &0.00275&0.00388&0.00808& & \\
  & & \textit{RMSE}&  0.03613& 0.02101&  0.00870& 0.00659& 0.00930 &  &  \\
	&&&&&&&\\
		&&\multicolumn{7}{c}{\textbf{ML}} \\ 
		\cline{1-10}
		\\[-1.8ex]
      $144$& $1$  & \textit{Mean}&  -0.06940 & 0.08946 & -0.00392& 0.00600& 0.00916 & -661.8 & 0.01056\\
  & & \textit{Bias}&  -0.07940& 0.07946 & -0.00217 & 0.00350 & 0.00391 &  &  \\
&&	\textit{SE}	&0.00684&0.00677&0.00007&0.00025&0.00016& & \\ 
  & & \textit{RMSE}& 0.07178 &0.08051 & 0.01003 & 0.00454& 0.00575 &  &  \\ \\[-1.8ex]
 $2500$& $0.10$  & \textit{Mean}& 0.00740 & 0.01256 & -0.00179 & 0.00252 & 0.00532 & -20611.7 & 0.00026\\
  & & \textit{Bias}&  -0.00260&  0.00256& -0.00004 &0.00002  & 0.00007 &  &  \\
&&	\textit{SE}	&0.00115 &0.00115&0.00015&0.00021&0.00045& & \\
  & & \textit{RMSE}& 0.00516 & 0.00874& 0.00833 &0.00524 & 0.00453 &  &  \\ 
	&&&&&&&\\
\hline\hline
\end{tabular}
\caption{\scriptsize The estimate of Model \eqref{eq:SARD} parameters for the discretization $N=2500$ and $\tau=0.1$ (the finest) and $N=144$ and $\tau=1$ (the coarsest), using four estimation methods: OLS NAIVE (ordinary least squares without controls for time discretization); OLS (ordinary least squares); IV (instrumental variables); and ML (maximum likelihood). $SE$ are standard errors; $SE^B$ non-parametric bootstrap standard errors; $RMSE$ root mean squared error; AICc (corrected) Akaike information criterion; MSE: Mean Squared Error.}
\label{tab:MCresults}
\end{table}

The comparison between the finest and the coarsest discretization models reported in Table \ref{tab:MCresults} highlights the improvement in the estimated parameters both in terms of bias and overall fitness as measured by the AICc and Mean Squared Error  (MSE). 
Across the four estimation methods, ML outperforms all other methods in terms of AICc although it has a modestly higher bias with respect to OLS and almost no difference in the MSE. Moreover,  in the ML estimation, the true values are always within the $95\%$ confidence intervals. The use of IV results in a poor estimation both in terms of AICc, MSE and bias. 
The closeness of the performance of ML and OLS highlights that for our setting, the endogeneity in the time discretization controls as well as the possible departure from the normality of the errors are not too strong.
While the ML estimated parameters are always statistically significant at the usual levels, the OLS and IV are no longer significant when we estimate the standard errors via non-parametric bootstrap (i.e. no assumptions on the error component).

\begin{figure}[!htbp]
	\begin{subfigure}[b]{0.49\textwidth}
		\includegraphics[width=\textwidth]{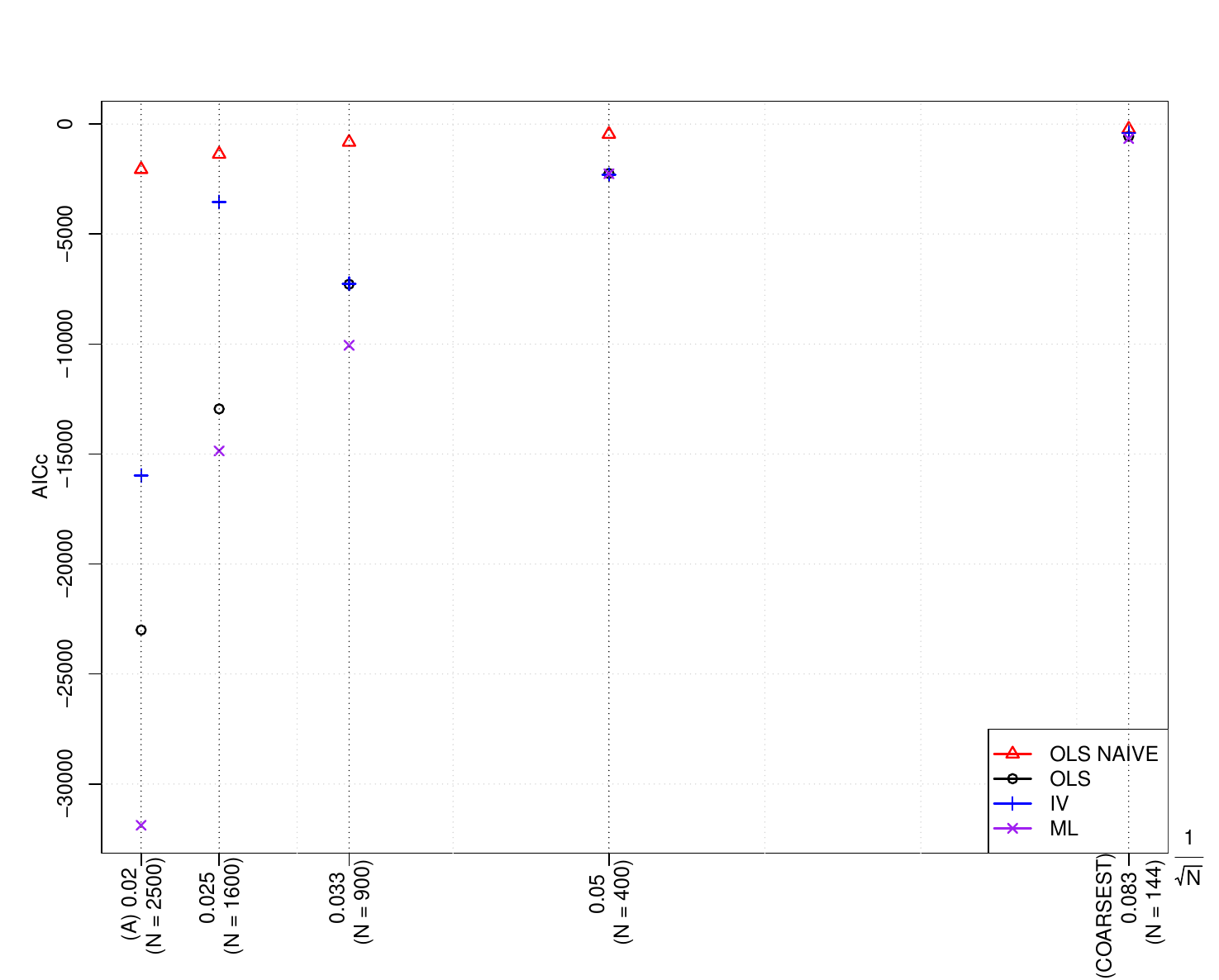}
		\caption{AICc for alternative spatial discretization and type of estimation starting from the coarsest case of Figure \ref{fig:discretizationOverTimeSpace} and increasing the finesse of spatial discretization in the case $\tau = 1$.}
		\label{fig:montecarloPDEAICc}
	\end{subfigure}
	\hspace{0.5cm}
	\begin{subfigure}[b]{0.49\textwidth}
		\includegraphics[width=\textwidth]{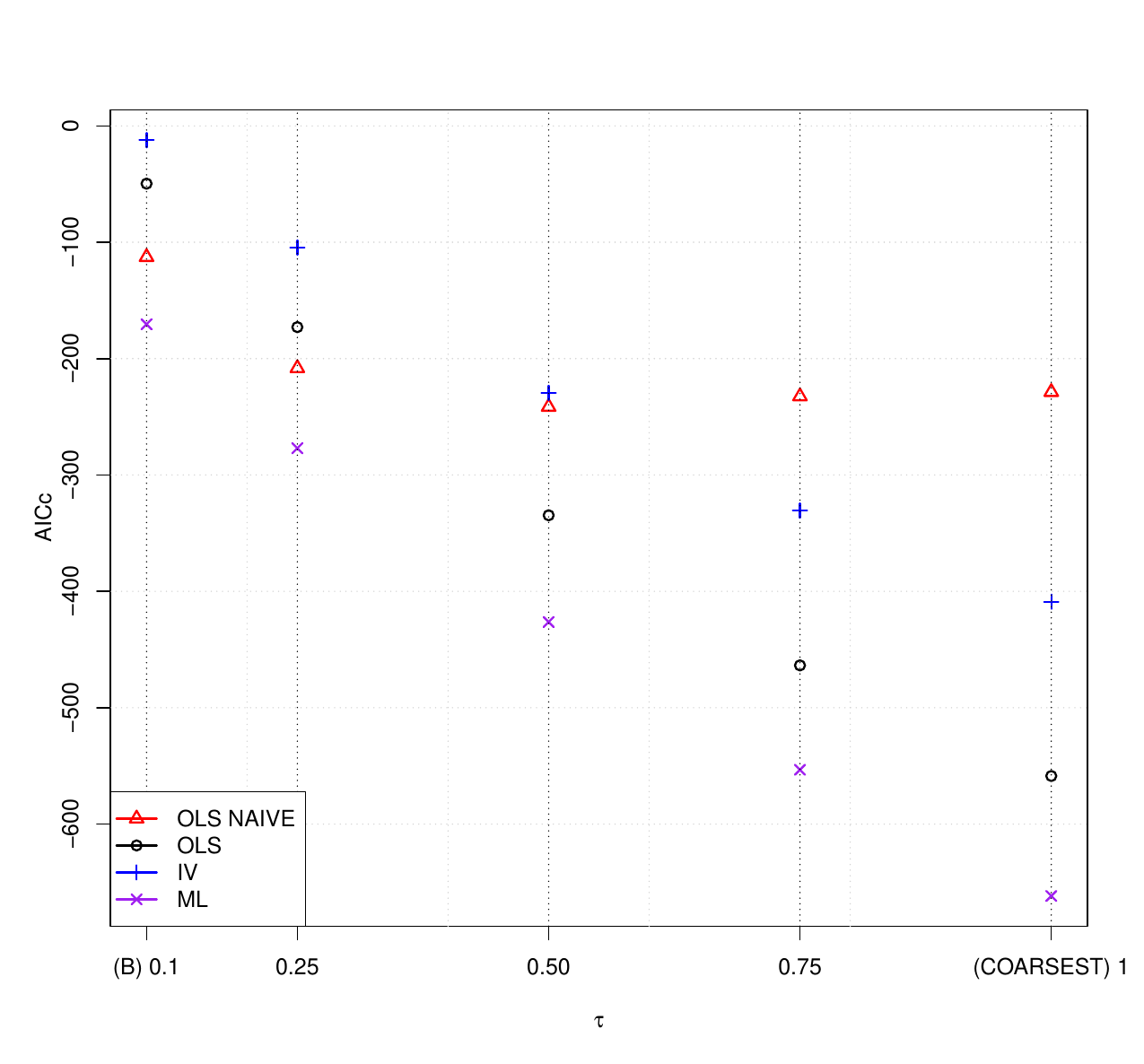}
		\caption{AICc for alternative time discretization and type of estimation starting from the coarsest case of Figure \ref{fig:discretizationOverTimeSpace} and increasing the finesse of time discretization in the case $N = 144$.}
		\label{fig:montecarloPDEAICcTau}
	\end{subfigure}
	\caption{AICc for different levels of time and spatial discretization for the four estimation methods: OLS NAIVE (ordinary least squares without any control for time discretization); OLS (ordinary least squares); IV (instrumental variables); and ML (maximum likelihood).}
	\label{fig:AICcDifferentModels}
\end{figure}

Finally, Figure \ref{fig:AICcDifferentModels} confirms that ML estimation outperforms all three other types of estimation in terms of AICc for alternative discretizations starting from the coarsest case of Figure \ref{fig:discretizationOverTimeSpace} (see Appendix \ref{app:estimatedParametersNumericalSimulation} for the point estimate of the model parameters). Overall, OLS and IV appear to be a valid alternative to ML in some circumstances.

\subsubsection{The spatial dependence in the reminder}
\begin{table}[!htbp]
	\scriptsize
	\hspace{-1cm}\begin{tabular}{ccccccccccccc} 
		\hline \hline \\[-1.8ex]
		$N$ & $\tau$ & $\hat{\lambda}$ & $\hat{\ell}_1$ & $\hat{\ell}_2$ & $\hat{\ell}_3$ & $\hat{\ell}_4$ & $\hat{\ell}_5$ & $\hat{\ell}_6$ & $\hat{\ell}_7$ & $\hat{\ell}_8$ & $\hat{\ell}_9$ & $\hat{\ell}_{10}$ \\  
		\hline \\[-1.8ex]
		144 & 1 & 0.622 & -2.743 &-2.494 &-2.151 &-1.861 &-1.537& -1.239 &-0.973& -0.691& -0.477& -0.190 \\
		& & (0.068)& (0.380) &(0.335) &(0.297) &(0.256) &(0.217) &(0.179) &(0.142) &(0.109) &(0.078) &(0.041)\\  \\[-1.8ex]
		2500 & 0.10 & 0.775 & 0.159 &-0.049 & 0.024& -0.007 & 0.011& -0.002  &0.003&  0.004& -0.001&  0.002 \\ 
		& & (0.011)& (0.005) &(0.004) &(0.003) &(0.003) &(0.003) &(0.002) &(0.002) &(0.002) &(0.001) &(0.001)\\  \\[-1.8ex]
		\hline
		\hline
	\end{tabular}
	\caption{The estimate of $\lambda$ and $\ell$s used for the calculation of ${W}_\epsilon$ in the ML estimate following the procedure described in Appendix \ref{app:spatialMatrixErrors}; $\hat{\ell}_i-\hat{\ell}_{i+1}$ measures the importance of spatial dependence among cells with order of contiguity $i$. Standard errors are reported between brackets.}
	\label{tab:estSpatialMatrixNumericalSimulations}                                  
\end{table}

The significant and positive value of $\lambda$ in the ML estimation reported in Table \ref{tab:estSpatialMatrixNumericalSimulations} shows evidence in favour of the presence of spatial dependence in the error term. 
The same table reports the estimated  $\ell$s used in the estimate of $W_{\epsilon}$ with the procedure described in Appendix \ref{app:spatialMatrixErrors}. While the estimate for the coarsest discretization is not satisfying, for the finest we have the expected outcome that the first order of contiguity is dominant with respect to the other orders, with a maximum significant order equal to 5 and an estimated $\lambda$ positive and close to 1 (see Appendix \ref{app:spatialMatrixErrors}).

\begin{figure}[!htbp]
	\begin{subfigure}[t]{0.30\textwidth}
		\includegraphics[width=1.3\textwidth]{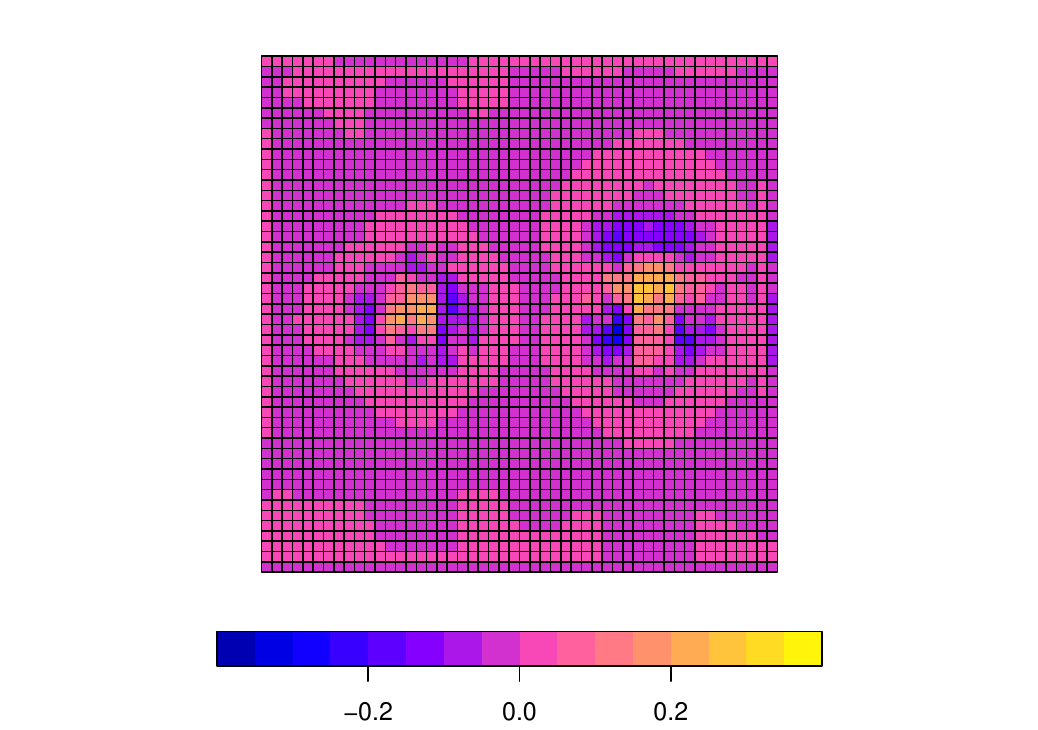}
		\caption{The spatial distribution of the reminder from the theoretical model}
		\label{fig:spatDistReminder}
	\end{subfigure}
	\begin{subfigure}[t]{0.30\textwidth}
		\includegraphics[width=1.3\textwidth]{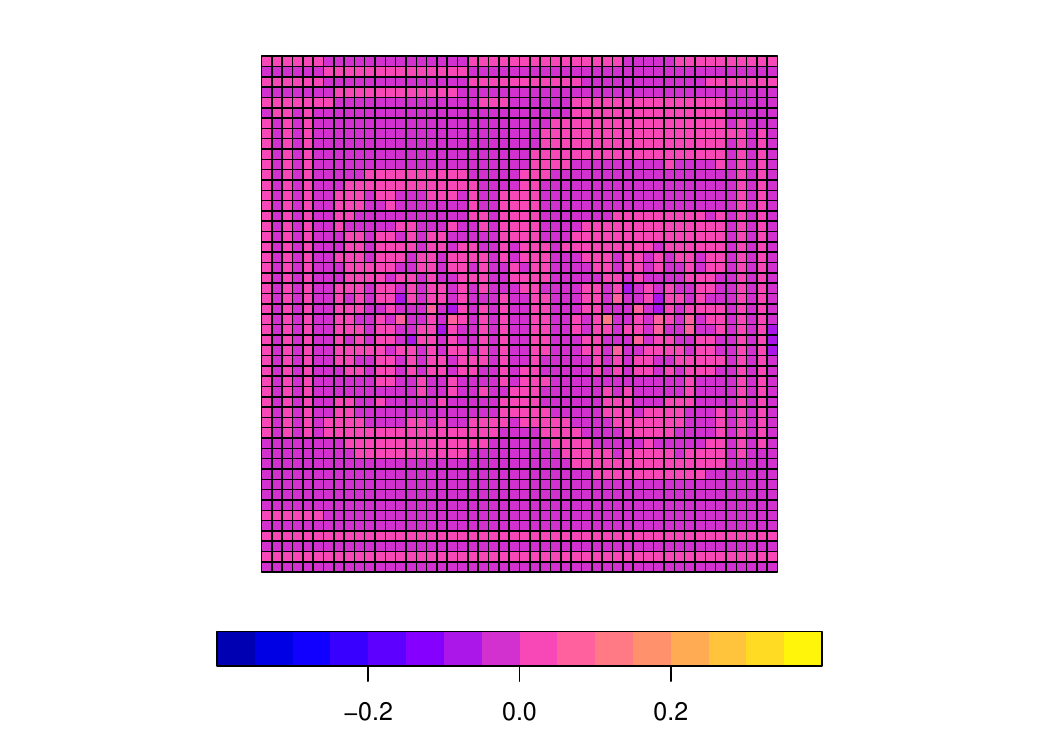}
		\caption{The spatial distribution of the remainder after the filtering by $W_\epsilon$.}
		\label{fig:spatDistReminderFiltered}
	\end{subfigure}
	\begin{subfigure}[t]{0.33\textwidth}
		\includegraphics[width=0.85\textwidth]{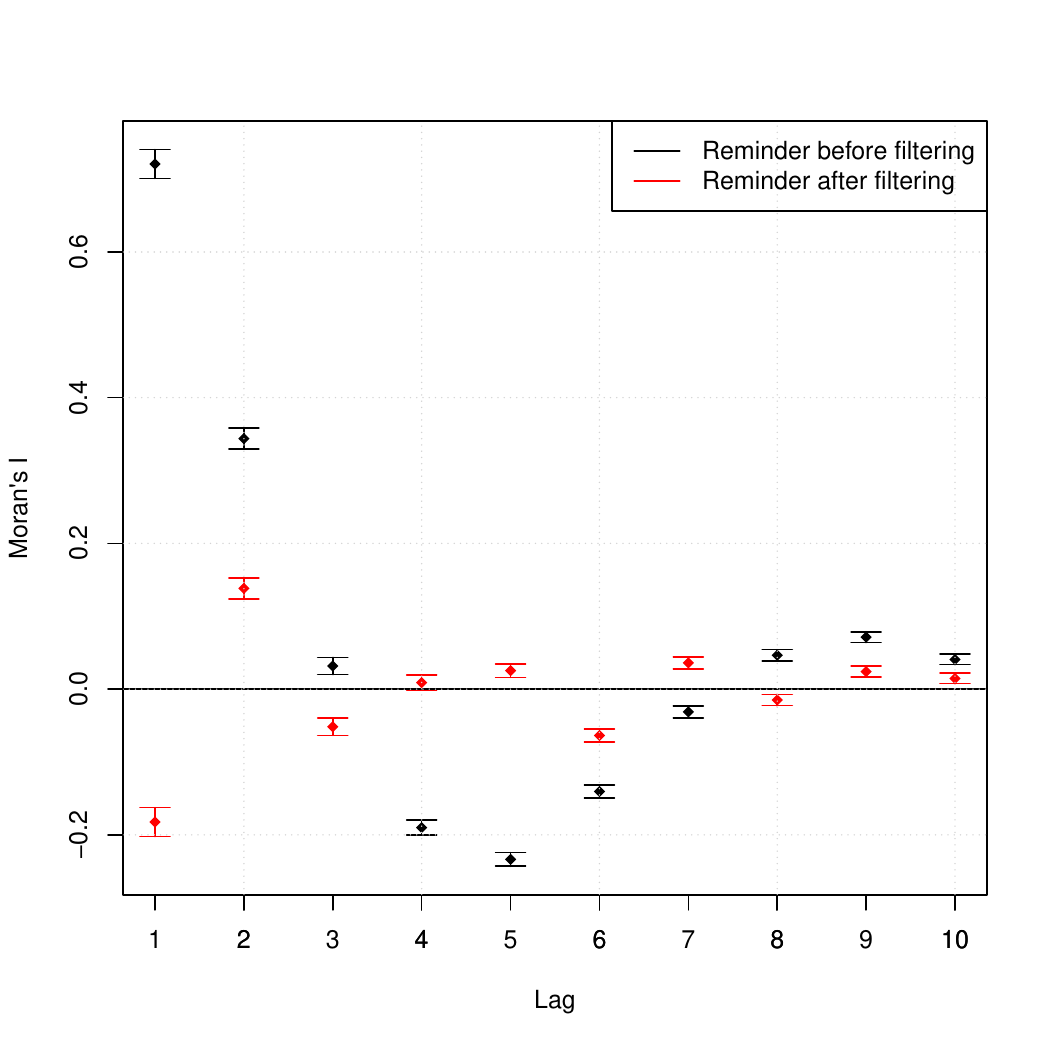}
		\caption{The estimated spatial correlogram (Moran's I) of the reminder for alternative orders of spatial contiguities (with their 95\% confidence bands).}
		\label{fig:montecarloCorrelogramm2500}
	\end{subfigure}  
	\caption{The reminder in the finest case ($N=2500, \tau=0.1$).}
	\label{fig:montecarloCorrelograms}
\end{figure}
Figure \ref{fig:spatDistReminder} reports the spatial distribution of the reminder in the finest case ($N=2500, \tau=0.1$) directly calculated from the theoretical model assuming that $\rho_{j} = \left(\tau/2\right)\gamma_{j}$ for $j \in \{A,R,D\}$. As expected, the stronger spatial dependence (both positive and negative) is present around the three peaks. Figure \ref{fig:spatDistReminderFiltered}, instead, reports the spatial distribution of filtered reminders, where filtering is made by $W_\epsilon$ estimated applying on the reminders the procedure described in Appendix \ref{app:spatialMatrixErrors} taking $\lambda = 1$.
A comparison between Figures \ref{fig:spatDistReminder} and \ref{fig:spatDistReminderFiltered} confirms the ability of the proposed procedure to remove the most of spatial dependence. Such ability is further confirmed by the estimate of the spatial correlogram (Moran's I) of the reminder (before and after the filtering) for different orders of spatial contiguities reported in Figure \ref{fig:montecarloCorrelogramm2500}.

\section{Empirical application \label{sec:empiricalApplication}}

In this section, we estimate Model \eqref{eq:SARD} for personal income of Italian municipalities in the period 2008-2019 and compare our estimate with the most common spatial  econometric models.

\subsection{Data on Italian municipalities \label{sec:dataItalianMunicipalities}}

Italian Ministry of Economy and Finance (Agenzia delle Entrate) releases information at the municipal level (about 8000 in Italy) on the \textit{nominal personal income} declared for tax purposes (IRPEF) for the period 2008-2020 from resident households.\footnote{\url{https://www1.finanze.gov.it/finanze/pagina_dichiarazioni/public/dichiarazioni.php}} 
During this period, Italy was hit by three main shocks: the subprime mortgage crisis coming from the US in 2007-2011, the sovereign debt crisis started in 2011-2013, and the COVID pandemic in 2020. Given the peculiarities of the COVID pandemic, we exclude 2020. In the analysis, personal income is not corrected for inflation because the latter is not available at the municipal level. However, in the period 2008-2019, the inflation was extremely low, with an annual average inflation rate at the national level of about 1.2\%, and with marginal regional heterogeneity, with the most of Italian region having a rate in the range [1\%,1.7\%].\footnote{Source: ISTAT (\textit{Italian National Institute of Statistics}).}    

\begin{figure}[!htbp]
	\centering
	\begin{subfigure}[b]{0.4\textwidth}
		\centering
		\includegraphics[width=\textwidth]{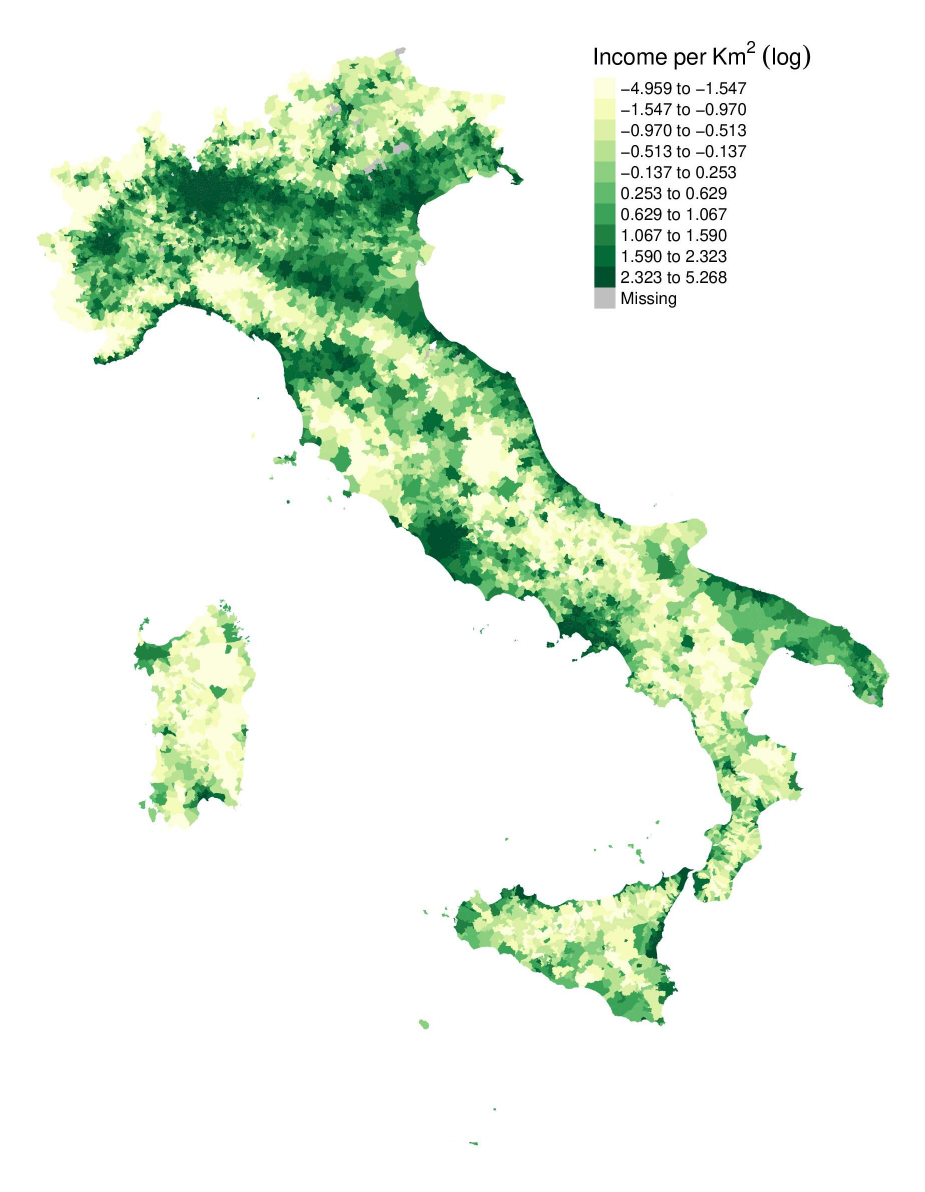}
		\caption{Income per Km$^2$ in 2008}
		\label{fig:incomeDistribution2008}
	\end{subfigure}
	\begin{subfigure}[b]{0.4\textwidth}
	\centering
	\includegraphics[width=\textwidth]{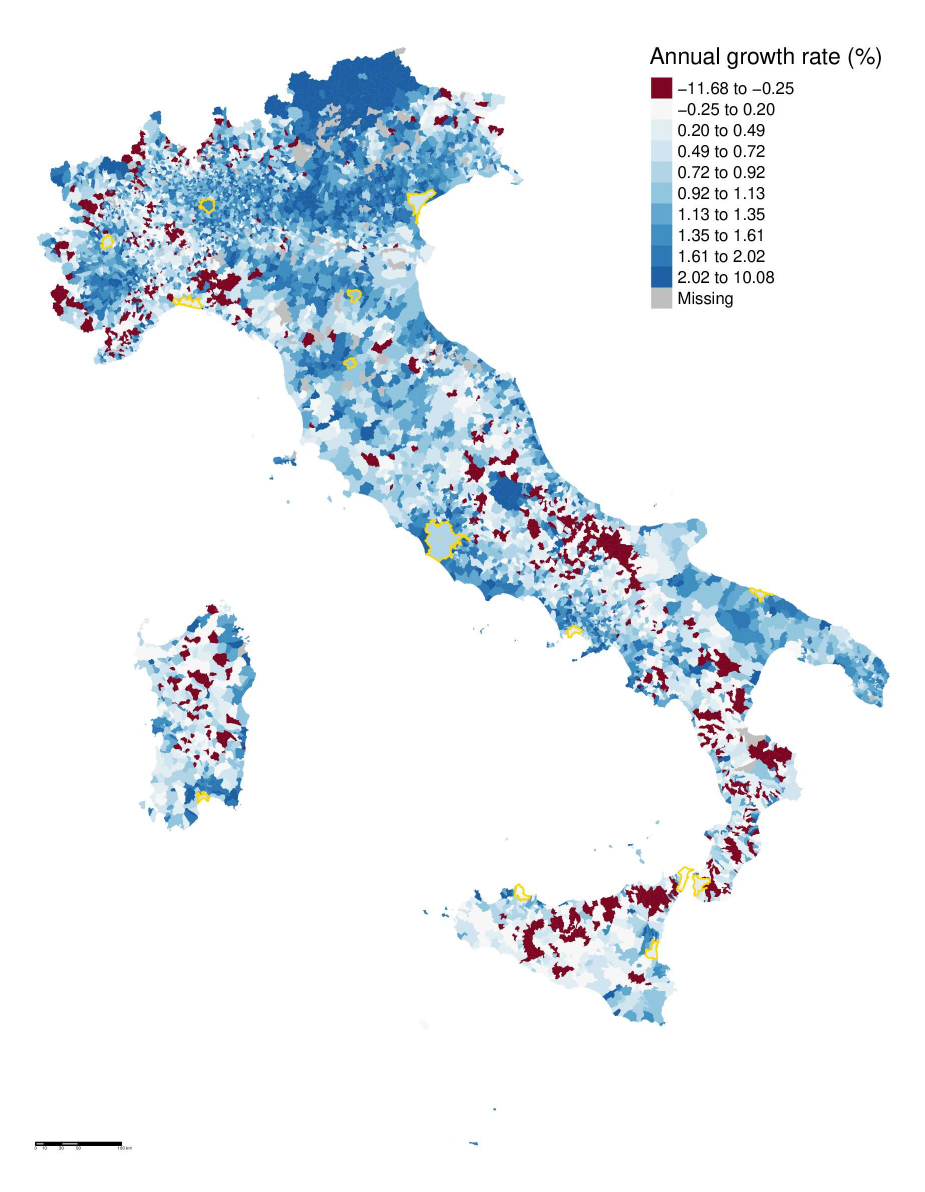}
	\caption{Average annual growth rate 2008-2019}
	\label{fig:differenceIncomeDistribution20082019}
\end{subfigure}
	\caption*{\textit{Source: Italian Ministry of Economy and Finance (Agenzia delle Entrate)} and ISTAT (\textit{Italian National Institute of Statistics}).}
	\caption{The map of the (log of) total personal income per $\text{Km}^2$ of Italian municipalities in 2008 and annual growth rates over the period 2008-2019. Yellow borders indicate the 14 Italian metropolitan municipalities.}
	\label{fig:mapsMunicipalIncome}
\end{figure}
To make comparable the income of municipalities with very different sizes and different types of territory, the total personal incomes are divided by the area of each municipality, measured in $\text{Km}^2$.
Figure \ref{fig:mapsMunicipalIncome} reports the map of total income per $\text{Km}^2$ of 7807 Italian municipalities for the year 2008 and its annual growth rate over the period 2008-2019.
The general impression is of a high level of agglomeration in the spatial income distribution, which seems to be persistent over time. There also exists a remarkable difference between the Northern and Southern parts of Italy, with the highest level of agglomeration in the North, between the inner and coastal areas and finally, the key impact of mountains which hurts the agglomeration process. In this respect, the estimate will include also the (average) altimetry of municipalities, gathered from ISTAT.


\subsection{The estimate of SARD model \label{sec:estimation}}

In the empirical application, we consider $\tau=11$ and several alternative values both for $h_A$ and $h_R$ of Eq. \eqref{eq:kernelAandR}. In particular, we probe all the combinations of $h_A \in \{5, 10, 15, 20\}$ and  $h_R \in \{25, 30, \dots, 95, 100\}$ (Km), which overall amounts at estimating 64 models. Table \ref{tab:0819} only reports the result of the best combination, that is the combination with the lowest AICc.


SARD model in the first column of Table \ref{tab:0819} is able to explain almost 77\% of the cross-sectional variance of the income change of Italian municipalities over the period 2008-2019.
The coefficient $\widehat{\tilde{\phi}}$, measuring the impact of the initial level of income on its variation, is significant and equal to 0.013. From the comparison of $\widehat{\tilde{\phi}}$ and the long-run aggregate growth rate calculated by Eq. \eqref{eq:aggregateRateOfGrowth},  is it possible to calculate the value of $\widehat{\rho}_\phi$ and, consequently, 
the scale factor (equal to 0.8282) needed to retrieve the parameters of Eq.  \eqref{eq:dataGenerationProcessCellTotalIncome} from the estimated coefficients of Model \eqref{eq:SARD}.\footnote{For example, $\widehat{\gamma}_{S}$ is equal to $1.08 \text{e-}05 \times 0.8282 = 8.94\text{e-}06$.}

The estimated coefficient of topography $\widehat{\tilde{\gamma}}_S$ is positive and significant, i.e. a nonuniform topography hurts economic activity. This finding is in line with the well-known phenomenon of depopulation characterising the mountain areas of Italy \citep{formelivelli}. Although the estimated $\widehat{\tilde{\gamma}}_S$ appears very low in absolute value, the analysis in Section \ref{subsec:effectOfTheCoefficients} shows its substantial impact on the spatial distribution of Italian income density.

The estimated coefficients of aggregation and repulsion ($\widehat{\tilde{\gamma}}_A$ and  $\widehat{\tilde{\gamma}}_R$, respectively) display the signs suggested by Eq.  \eqref{eq:dataGenerationProcessCellTotalIncome}, and are both statistically significant at the usual levels. In absolute value, $\widehat{\tilde{\gamma}}_A$ is marginally but significantly higher than $\widehat{\tilde{\gamma}}_R$. The aggregation is therefore prevailing on repulsion although aggregation forces work within a smaller radius given that the best specification for the spatial structure is given by $h_A=10$ km and $h_R=45$ km. Overall, this implies the emergence of urban agglomerates, but whose radius should be limited within 10 km \citep{rosenthal2008attenuation}. At the same time, the crowding/congestion effects, working in a radius of 45 km, explain the extension of suburban areas and commuting zones \citep{alonso1964location,duranton2020economics,glaeser2001job,kneebone2009job,travisi2010impacts}.  

The estimated coefficient $\widehat{\tilde{\gamma}}_D$ has the expected positive and significant sign, highlighting the presence of diffusive forces in the spatial distribution of income density which tend to equalise it.  A detailed quantitative assessment of these forces is left to Section \ref{subsec:effectOfTheCoefficients}.
The coefficients for the corrections of time discretization are all highly significant.\footnote{In particular, the estimated five coefficients are $\widehat{\tilde{\rho}}_S=-8.27\text{e-}04$, $\widehat{\tilde{\rho}}_A=1.99\text{e-}06$, $\widehat{\tilde{\rho}}_R=-1.96\text{e-}06$ and $\widehat{\tilde{\rho}}_D=-5.97\text{e-}05$.}

Finally, the spatial dependence in the error, as measured by the coefficient $\hat{\lambda}$, is positive and significant, and is close to one.  Given that all elements of $W_\epsilon$ are non-negative (see Appendix \ref{app:spatialMatrixErrorsRealData}), in our specific application the error exhibits positive spatial correlation. This means that neighbouring municipalities' dynamics are still positively correlated even after controlling for topographical, aggregative, repulsive, and diffusive forces. This positive correlation could also result from a (unaccounted) place-based policy that affects multiple neighbouring municipalities, leading agents (individuals/firms) to coordinate their spatial movements.

The SARD model lacking spatial dependence in the errors exhibits inferior performance in terms of AICc and $R^2$-N, as demonstrated by the WN-SARD model estimated through either maximum likelihood (ML) or instrumental variables (IV). The estimated IV coefficients have marginally different magnitudes, yet they retain their respective signs and levels of significance. Instead, omitting the corrections for time discretization (NAIVE SARD) results in a significant decrease in model performance and discrepancies in the estimated coefficients, despite retaining their signs and significance. 

\begin{landscape}
	\begin{table}
		\tiny 
		\hskip-1.0cm\begin{tabular}{lccccc|ccc|ccc}
			\\[-1.8ex]\hline 
			\hline \\[-1.8ex]   
			& \multicolumn{10}{c}{\textit{Dependent variable: $\Delta_{11} y$}} \\ 
			\cline{5-8} 
			\\[-1.8ex] & \multicolumn{9}{c}{} \\
			\\[-1.8ex]		        & SARD                & WN SARD   &  WN SARD   & WN SARD &  NAIVE SARD         & INCOME LAG       & ALT INCOME LAG   &  S INCOME LAG    & SLX               & SPATIAL LAG         & SPATIAL DURBIN      \\ 
			& \multicolumn{5}{c|}{}& \multicolumn{3}{c|}{}  \\
			&    (ML)            &  (ML)                 & (IV)                 &(OLS)             &      (OLS)     &     (OLS)           &  (OLS)            &   (OLS)               &   (OLS)        &         (ML) & (ML)       \\  \hline & \multicolumn{5}{c|}{}& \multicolumn{3}{c|}{}  \\ 
			& (1)                & (2)                 & (3)                 & (4)                 & (5)              & (6)              & (7)              & (8)               & (9)                 & (10)        & (11)         \\  \hline & \multicolumn{5}{c|}{}& \multicolumn{3}{c|}{}  \\ 
			
			$\tilde{\alpha}$ 		& 1e-03$^{**}$   & 5.91e-04$^{**}$   & 1.27e-03$^{*}$    &   1.01e-03$^{*}$   & 1.66e-03$^{**}$          & 6.31e-03$^{***}$ & 0.012$^{***}$    & 6.41e-03$^{***}$ & $-$3.33e-03$^{**}$ & $-$3.08e-03$^{***}$ & $-$2.87e-03$^{**}$  \\
			& (6.87e-04)          & (3.18e-04)          & (6.64e-04)         &(5.93e-04) & (7.85e-04)          & (8e-04)          & (1.21e-03)       & (7.94e-04)       & (1.37e-03)        & (1.14e-03)          & (1.28e-03)          \\
			&                     &                     &                     &              &       &                  &                  &                  &                   &                     &                     \\
			$\tilde{\phi}$   		&0.013$^{***}$       & 0.013$^{***}$       & 0.013$^{***}$  &   0.013$^{***}$   & 0.013$^{***}$       & 0.011$^{***}$    & 0.011$^{***}$    & 0.011$^{***}$    & 7.58e-03$^{***}$  & 7.48e-03$^{***}$    & 7.90e-03$^{***}$    \\
			& (5.98e-05)          & (4.14e-05)          & (9.13e-05)    &   (7.79e-05)   & (1.02e-04)          & (8.76e-05)       & (9e-05)          & (8.75e-05)       & (1.45e-04)        & (1.36e-04)          & (1.36e-04)          \\ 
			&                     &                     &                     &                &     &                  &                  &                  &                   &                     &                     \\
			$\gamma_{ALT}$  &                     &                     &                     &            &         &                  & $-$0.011$^{***}$ &                  & $-$0.017$^{***}$  & $-$2.49e-03         & $-$0.015$^{***}$    \\
			&                     &                     &                     &                     &                  && (1.74e-03)       &                  & (4.95e-03)        & (1.56e-03)          & (4.62e-03)          \\ 
			&                     &                     &                     &                     &&                  &                  &                  &                   &                     &                     \\
			$\theta$        &                     &                     &                     &                    & &                  &                  &                  & 6.32e-03$^{***}$  &                     & $-$2.24e-03$^{***}$ \\
			&                     &                     &                     &                     &&                  &                  &                  & (2.09e-04)        &                     & (3.36e-04)          \\
			&                     &                     &                     &                     &&                  &                  &                  &                   &                     &                     \\
			$\theta_{ALT}$  &                     &                     &                     &                  &   &                  &                  &                  & 0.015             &                     & 0.015               \\
			&                     &                     &                     &                     &&                  &                  &                  & (5.93e-03)        &                     & (5.54e-03)          \\
			&                     &                     &                     &                    & &                  &                  &                  &                   &                     &                     \\
			$\tilde{\gamma}_S$      & 1.08e-05$^{***}$    & 1.08e-05$^{***}$    & 8.88e-06$^{***}$   & 1.15e-05$^{***}$    & 7.67e-06$^{***}$    &                  &                  & 5.4e-06$^{***}$  &                   &                     &                     \\
			& (2.17e-07)          & (1.98e-07)          & (6.65e-07)        & (4.61e-07) & (4.88e-07)          &                  &                  & (4.89e-07)       &                   &                     &                     \\
			&                     &                     &                     &                    & &                  &                  &                  &                   &                     &                     \\
			$\tilde{\gamma}_A$      & $-$3.55e-01$^{***}$ & $-$4.3e-01$^{***}$ & $-$3.4e-01$^{***}$& $-$4.81e-01$^{***}$  & $-$1.7e-01$^{***}$ &                  &                  &                  &                   &                     &                     \\
			& (2.93e-03)          & (2.99e-03)          & (0.017)       & (0.012)  & (7.39e-03)          &                  &                  &                  &                   &                     &                     \\
			&                     &                     &                     &                &     &                  &                  &                  &                   &                     &                     \\
			$\tilde{\gamma}_R$      & 3.397$^{***}$    &3.043$^{***}$    & 3.616$^{***}$   &4.588$^{***}$  & 0.751$^{***}$    &                  &                  &                  &                   &                     &                     \\
			& (0.049)          & (0.044)          & (0.529)        & (0.21) & (0.109)          &                  &                  &                  &                   &                     &                     \\
			&                     &                     &                     &           &          &                  &                  &                  &                   &                     &                     \\
			$\tilde{\gamma}_D$      & 2.38e-06$^{***}$    & 2.19e-06$^{***}$    & 2.48e-06$^{***}$  & 3.7e-06$^{***}$  & 1.61e-06$^{***}$    &                  &                  &                  &                   &                     &                     \\
			& (3.61-08)          & (3.69e-08)          & (1.36e-07)     & (7.45e-08)    & (9.11e-08)          &                  &                  &                  &                   &                     &                     \\
			&                     &                     &                     &                     &&                  &                  &                  &                   &                     &                     \\ \cline{2-12}  \\[-1.8ex]
			${\rho}$          &                     &                     &                     &                    & &                  &                  &                  &                   & 0.494$^{***}$       & 0.607$^{***}$       \\
			&                     &                     &                     &                    & &                  &                  &                  &                   & (0.011)             & (0.019)             \\
			&                     &                     &                     &                    & &                  &                  &                  &                   &                     &                     \\
			$\lambda$               & 0.682$^{***}$       &                     &                  &   &                     &                  &                  &                  &                   &                     &                     \\
			& (0.018)             &                     &                     &           &          &                  &                  &                  &                   &                     &                     \\ \hline \\[-1.8ex]
			Correction      & YES & YES  & YES &YES&NO & & & & & &\\ 
			for time      &  &   &  & & & & & & &&\\
			discretization     &  &   &  & & & & && & &\\ \hline
			AICc                    & -23220.5            & -22169.6         & -21881.4         & -21436.3 &   -21434.3          & -20678.3         & -20716.8         & -20797.5         & -21582.8          & -22318.4            & -22370.4            \\
			
			MSE                & 0.0022             & 0.0029            & 0.0026             &0.0021 &       0.0037      & 0.0041 		   & 0.0041 		  & 0.0040           & 0.0036 		     & 0.0032 			   & 0.0032			     \\ 
			
			$R^2$-N                 & 0.7694             & 0.7360            & 0.7261             &0.7096 &       0.7096      & 0.6802 		   & 0.6818 		  & 0.6851           & 0.7153 		     & 0.7410 			   & 0.7427			     \\ \hline \\[-1.8ex]
			$D$                     &                     &                     &                     &      &               &                  &                  &                  & 25                & 25                  & 25                  \\
			$h_{A}$                 & 10                  & 10                  & 10             & 10         & 10                  &                  &                  &                  &                   &                     &                     \\
			$h_{R}$                 & 45                  & 45                  & 45              & 45     & 45                  &                  &                  &                  &                   &                     &                     \\ \hline \\[-1.8ex] 
			\textit{Note:}         	& \multicolumn{11}{r}{ AICc: corrected Akaike Information Criteria.  MSE: Mean Squared Error. $R^2$-N: Nagelkerke Pseudo-$R^2$. Significance levels: $^{*}$p$<$0.1; $^{**}$p$<$0.05; $^{***}$p$<$0.01} \\ 
		\end{tabular} 
		\caption{\footnotesize The estimated coefficients of different econometric models for Italian municipality incomes per $km^2$ for the period 2008-2019: ML estimation of SARD model (1); ML estimation of SARD model without spatially correlated errors (2); IV estimation of SARD model without spatially correlated errors (3); OLS estimation of SARD model without spatially correlated errors (4); OLS estimation of SARD model without spatially correlated errors and corrections for time discretization (5); estimation of a benchmark model with only lagged income (6); estimation of a benchmark model also including altimetry (7); estimation of a benchmark model also including our proxy for topography (8); estimation of SLX model with lagged income and altimetry (9); estimation of SPATIAL LAG model with lagged income and altimetry (10); estimation of SPATIAL DURBIN model with lagged income and altimetry (11). Source: our estimations on data from the Italian Ministry of Economy and Finance (Agenzia delle Entrate).}
		\label{tab:0819}
	\end{table}\vspace{1cm}
\end{landscape}
\noindent 

\subsection{A comparison with other (spatial) econometric models \label{sec:comparisonSpatialEconometricModels}}
The comparison between the SARD model and the simpler model that solely encompasses the starting income level (INCOME LAG in Table \ref{tab:0819}, absolute convergence in growth rate) indicates that the $\widehat{\phi}$ coefficient is consistently stable across both models, but the explained variance reduces to approximately 0.68 in the latter. The incorporation of altimetry in the ALT INCOME LAG model (convergence conditioned to altimetry), despite being negative and significant, does not significantly improve the fitness of the model. A notable improvement is instead reached in the S INCOME LAG model (convergence conditioned to $\mathbf{x}_{S}$) where we consider a more complex measure of topography such as the $\mathbf{x}_S$ previously employed in the SARD model. However, the fitness of S INCOME LAG model still lags behind that of the SARD model.

Finally, to compare the performance of Model \eqref{eq:SARD} with respect to models typically used in spatial econometrics, we estimate SLX, SPATIAL LAG and SPATIAL DURBIN models, which consider both the starting income level and altimetry and use a spatial weight matrix $W$ whose elements are defined as: 
\begin{equation}
	w_{ij} = \left\{ \begin{array}{ll}
		\frac{1}{d_{ij}^2} & \textrm{if $d_{ij} \leq \bar{d} \; \forall i \neq j$}\\
		0 & \textrm{otherwise},\\
	\end{array} \right.
	\label{eq:weigthMatrix}
\end{equation}
where $\bar{d}$ is chosen in the set $\{5, 10, \dots, 95,100\}$ kilometres on the base of the lowest AICc. 
Spatial dependence is present in all the models, with the estimated coefficient for the spatially lagged municipal income change $\hat{\rho}$ being positive and significant (ranging from 0.49 to 0.61), as well as the coefficient of the spatially lagged municipal income $\hat{\theta}$ (ranging from $-2.24\text{e-}03$ to $6.32\text{e-}03$). The SPATIAL DURBIN model exhibits the best fit, although it remains lower than the SARD model.


\subsection{Decomposing the local growth}\label{subsec:effectOfTheCoefficients}

The impact on the spatial dynamic of the Italian municipal income of each component of the SARD model in Table \ref{tab:0819} is calculated by a counterfactual methodology. In particular, for each reallocative component of the SARD model, we calculate a counterfactual in sample average annual growth rate of municipality setting to zero the coefficient related to the component of interest, i.e. $\widehat{\mathbf{g}}_{j}^{CF} = \left[\left(\widehat{\mathbf{y}}^{2019}|\gamma_{j}=0\right) /{\mathbf{y}}^{2008}\right]^{1/11}-1$, where $\widehat{\mathbf{y}}^{2019}|\gamma_{j}=0$ is the vector of forecasted municipal incomes in 2019 setting $\gamma_{j}=0$, with $j \in \left\{S,A,R;D\right\}$. The contribution of component $j$ is therefore measured by $\widehat{\mathbf{g}}_{j} = \widehat{\mathbf{g}} - \widehat{\mathbf{g}}_{j}^{CF}$.

\begin{figure}[!htbp]	
	\begin{subfigure}[b]{0.49\textwidth}
		\centering
		\includegraphics[width=\textwidth]{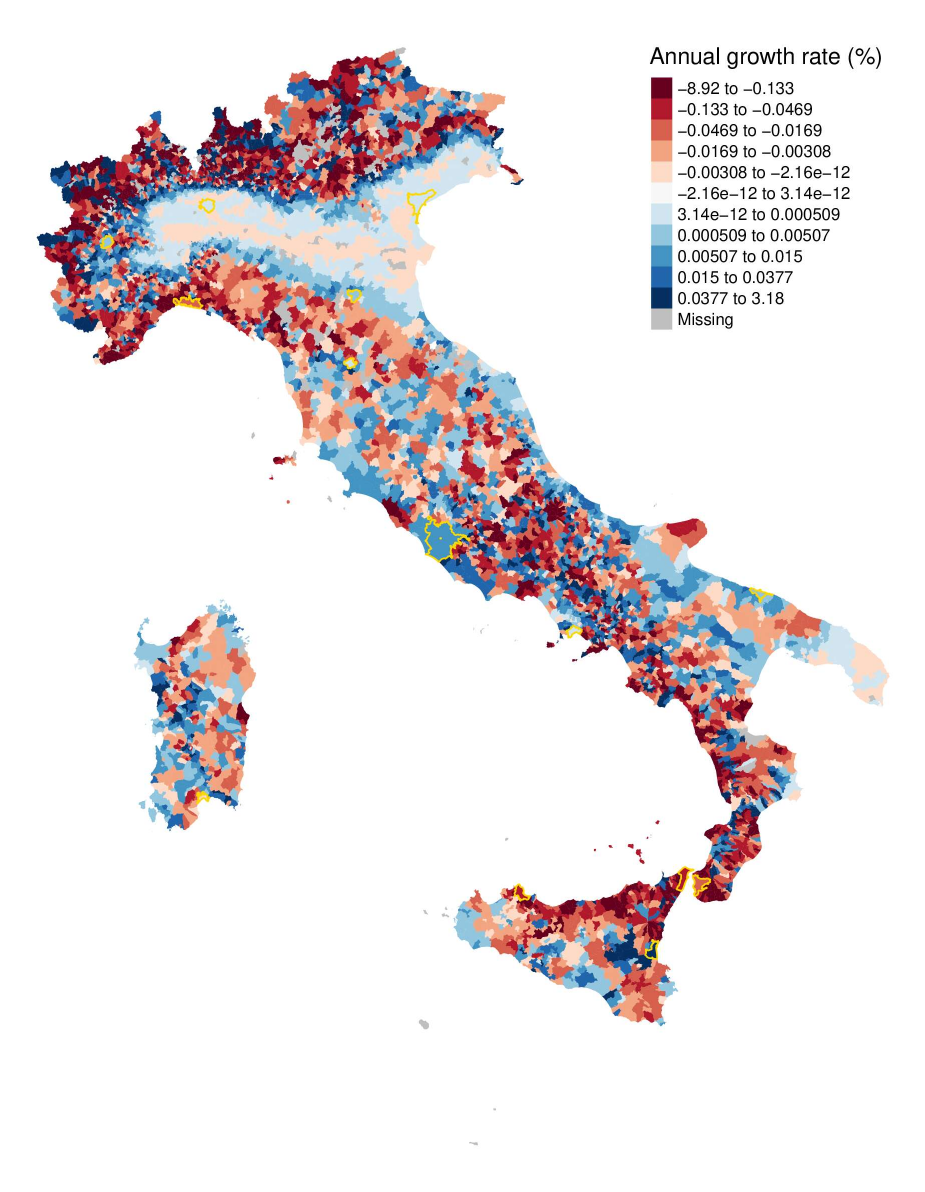}
		\caption{Contribution of topography $S$}
		\label{fig:deltaSISample}
	\end{subfigure}
	\begin{subfigure}[b]{0.49\textwidth}
		\centering
		\includegraphics[width=\textwidth]{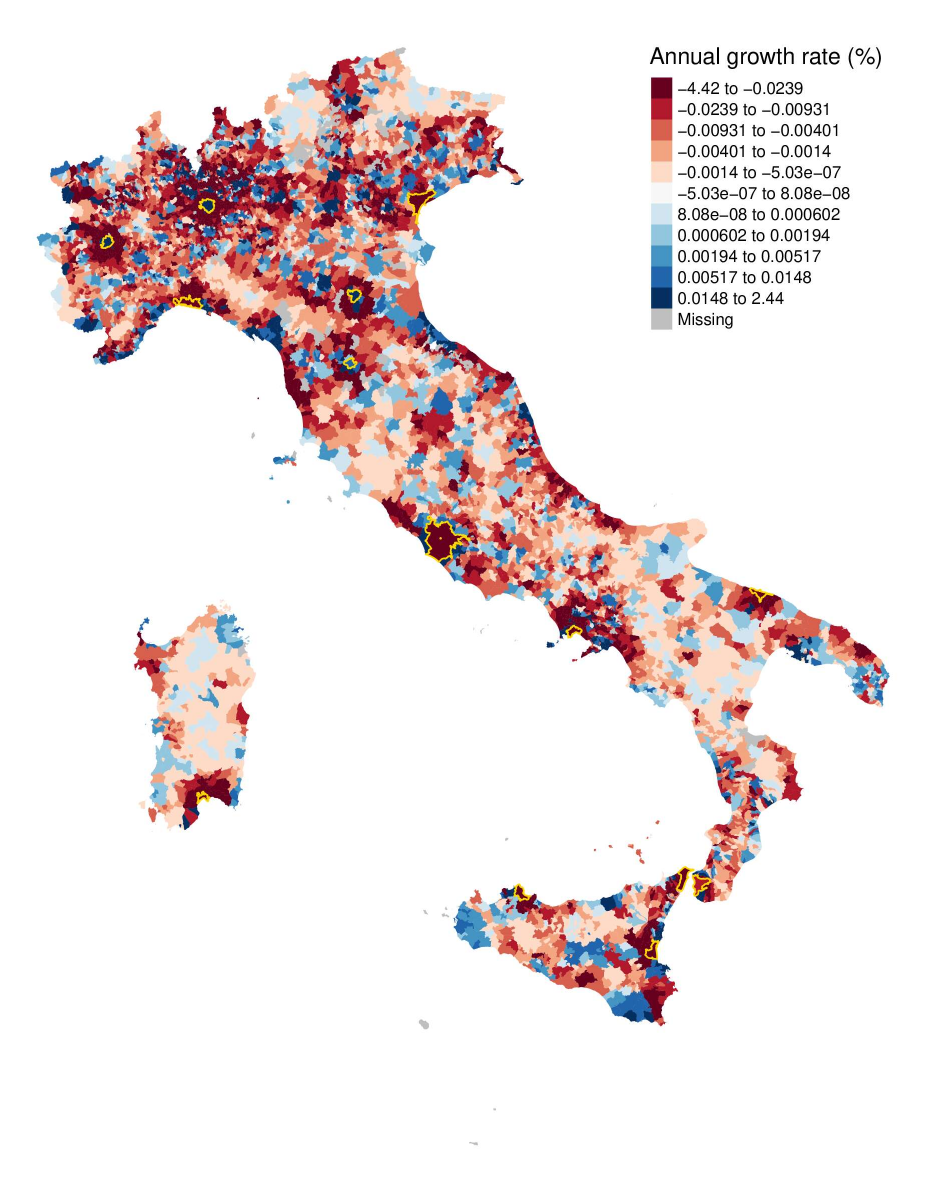}
		\caption{Contribution of aggregation $A$}
		\label{fig:deltaAInSample}
	\end{subfigure}
	\begin{subfigure}[b]{0.49\textwidth}
		\centering
		\includegraphics[width=\textwidth]{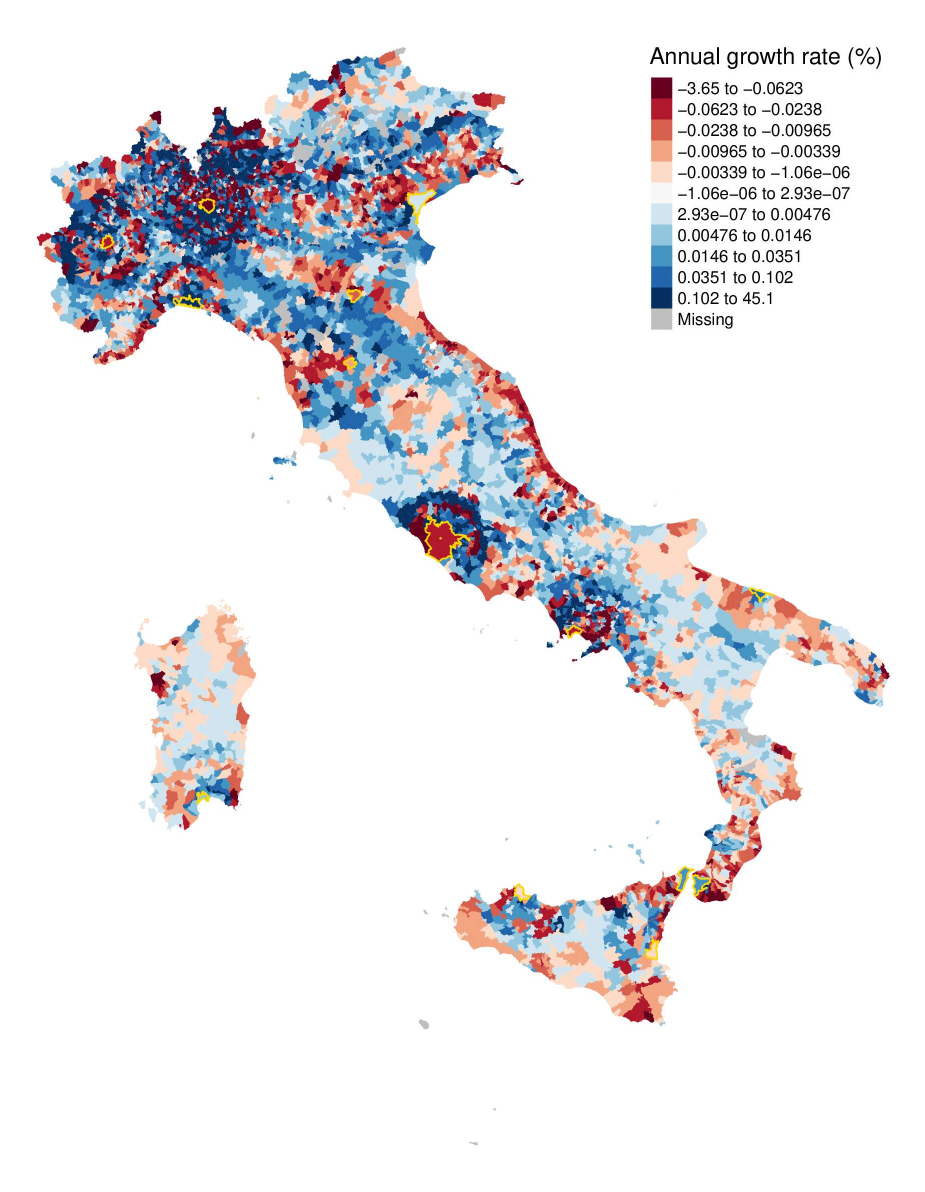}
		\caption{Contribution of repulsion $R$}
		\label{fig:deltaRInSample}
	\end{subfigure}
	\begin{subfigure}[b]{0.49\textwidth}
		\centering
		\includegraphics[width=\textwidth]{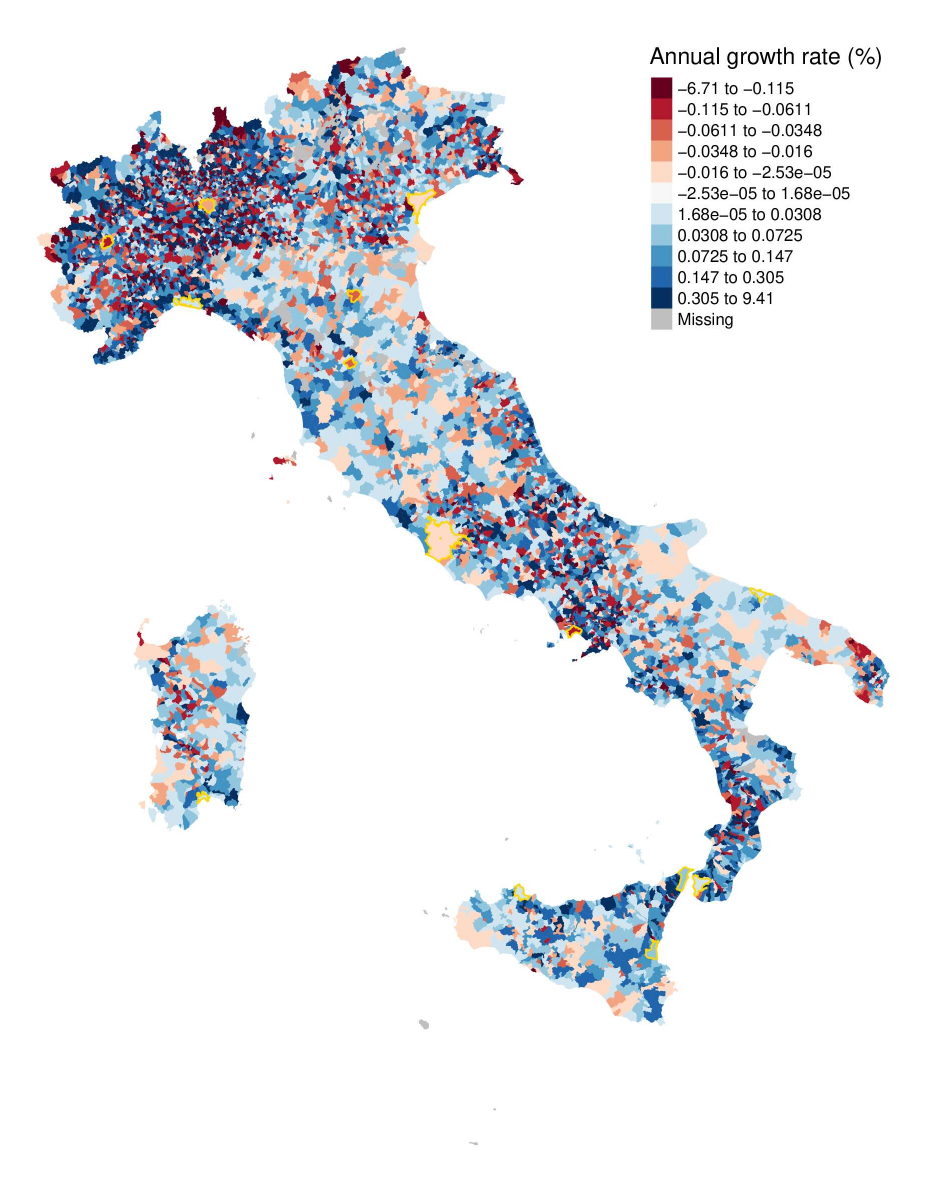}
		\caption{Contribution of diffusion $D$}
		\label{fig:deltaDInSample}
	\end{subfigure}
	\caption{The individual contribution of components $\{S,A,R,D\}$ to the forecasted in sample average annual growth rate 2008-2019 based on the estimate of SARD model reported in Table \ref{tab:0819}. Yellow borders indicate the 14 Italian metropolitan municipalities.}
	\label{fig:couterfactual}
\end{figure}

Figure \ref{fig:deltaSISample} shows that the areas at the bottom of mountains have benefited at most by the topography $S$, with mountain areas having been mainly hurt, and all flat areas having no effect. Topography appears a driving force for the emergence of medium cities on the borders of the Po Valley and in the central part of the Adriatic coast. Large alpine valleys, such as the ones in Valle d'Aosta, Val di Susa, Trento and Bolzano, also display a positive effect. Finally, the estimated benefit for the municipalities just around Volcanos Vesuvio and Etna's areas is further evidence of the importance of topography.
The effect of the aggregation component is spatially heterogeneous, with a prevalent positive contribution for municipalities in coastal areas and the Po Valley (see Figure \ref{fig:deltaAInSample}). However, being a metropolitan municipality does not imply gaining from agglomeration forces because, in the final outcome, a key role is played by the income of neighbouring municipalities acting as a competitive force: Milan and Naples, where the second has very rich neighbouring municipalities with respect to the first, have a gain of 0.31\% and a loss of 0.25\% from the aggregation component, respectively (see column $W_{10}\mathbf{y}^{2008}$ versus column $\mathbf{y}^{2008}$ and column $\hat{\mathbf{g}}_A$ in Table \ref{tab:metropolitanCities}). The extension of Rome cannot allow for a precise identification of an aggregation effect with respect to its centre.
The effect of the repulsion component is instead negative for the six richest Metropolitan cities and, in general for the richest municipalities (see Figure \ref{fig:deltaRInSample} and column $\hat{\mathbf{g}}_R$ in Table \ref{tab:metropolitanCities}). 
Finally, the effect of the diffusion component appears random across Italian municipalities (Figure \ref{fig:deltaDInSample}), but negative for the greatest Metropolitan municipalities (Milan, Turin, Naples, Florence, Bologna, Rome, see column $\hat{\mathbf{g}}_D$ in Table \ref{tab:metropolitanCities}).
Overall, aggregation and repulsion components seem to play the major role for most of the Metropolitan municipalities (Table \ref{tab:metropolitanCities}).

\begin{table}[!htbp]
	\hspace{-1.5cm}
	\footnotesize{
		\begin{tabular}{lrrrrrrrrrr}
			\hline
			\hline\\[-1.8ex]
		   Metropolitan city & $\mathbf{y}^{2008}$ & $W_{10}\mathbf{y}^{2008}$ & $W_{45}\mathbf{y}^{2008}$ & $\dfrac{W_{10}\mathbf{y}^{2008}}{\mathbf{y}^{2008}}$ & $\dfrac{W_{45}\mathbf{y}^{2008}}{\mathbf{y}^{2008}}$ & $\hat{\mathbf{g}}$ & $\hat{\mathbf{g}}_{S}$ & $\hat{\mathbf{g}}_{A}$ & $\hat{\mathbf{g}}_{R}$ & $\hat{\mathbf{g}}_{D}$ \\[1.8ex] 
			\hline
 Milan              & 164 & 1006 & 9502 & 6.13 & 57.87 & 1.22 & 0.00  & 0.31  & -0.12 & -0.02 \\
Turin              & 115 & 236  & 1259 & 2.06 & 10.98 & 1.04 & 0.01  & 0.08  & -0.03 & -0.07 \\
Naples             & 85  & 807  & 3036 & 9.51 & 35.76 & 0.7  & 0.00  & -0.25 & -0.04 & -0.09 \\
Florence           & 65  & 46   & 392  & 0.71 & 6.04  & 1.01 & 0.00  & 0.01  & -0.01 & -0.06 \\
Bologna            & 53  & 76   & 398  & 1.41 & 7.4   & 1.02 & 0.00  & 0.02  & -0.02 & -0.05 \\
Palermo            & 46  & 32   & 175  & 0.7  & 3.83  & 0.96 & -0.06 & -0.05 & -0.01 & 0.00  \\
Genoa              & 43  & 10   & 446  & 0.24 & 10.33 & 0.88 & -0.03 & -0.34 & 0.16  & 0.02  \\
Rome               & 37  & 0    & 459  & 0    & 12.33 & 0.98 & 0.00  & -0.03 & -0.05 & -0.02 \\
Bari               & 37  & 52   & 154  & 1.41 & 4.18  & 0.98 & 0.00  & -0.16 & 0.05  & 0.02  \\
Cagliari           & 29  & 58   & 117  & 1.98 & 3.97  & 1.03 & 0.00  & -0.07 & 0.01  & 0.02  \\
Catania            & 17  & 11   & 458  & 0.67 & 27.05 & 1.13 & 0.08  & -0.06 & -0.01 & 0.03  \\
Messina            & 13  & 10   & 218  & 0.75 & 17.08 & 0.98 & -0.08 & -0.07 & 0.04  & 0.04  \\
Venice             & 11  & 9    & 902  & 0.83 & 84.18 & 1.01 & 0.00  & -0.06 & 0.01  & 0.00  \\
Reggio di Calabria & 8   & 2    & 220  & 0.24 & 26.16 & 1.04 & -0.04 & -0.02 & 0.01  & 0.00 \\  \hline
		  \end{tabular}
		  }
	  \caption{Income per Km$^2$ in 2008 ($\mathbf{y}^{2008}$), cumulative income per Km$^2$ of municipalities within 10 Kms in 2008 ($W_{10}\mathbf{y}^{2008}$), cumulative income per Km$^2$ of municipalities within 45 Kms in 2008 ($W_{45}\mathbf{y}^{2008}$), the fitted growth rate of income per Km$^2$ (in \%) in the period 2008-2019 ($\hat{\mathbf{g}}$), the counterfactual growth rate of income per Km$^2$ (in \%) in the period 2008-2019 due to component $j$, with $j \in \{S,A,R,D\}$ ($\hat{\mathbf{g}}_j$) for the fourteen Italian Metropolitan cities.
	  }
  \label{tab:metropolitanCities}
	\end{table}

\subsection{Convergence in local income}

\begin{figure}[!htbp]
	\centering
	\includegraphics[width=0.6\textwidth]{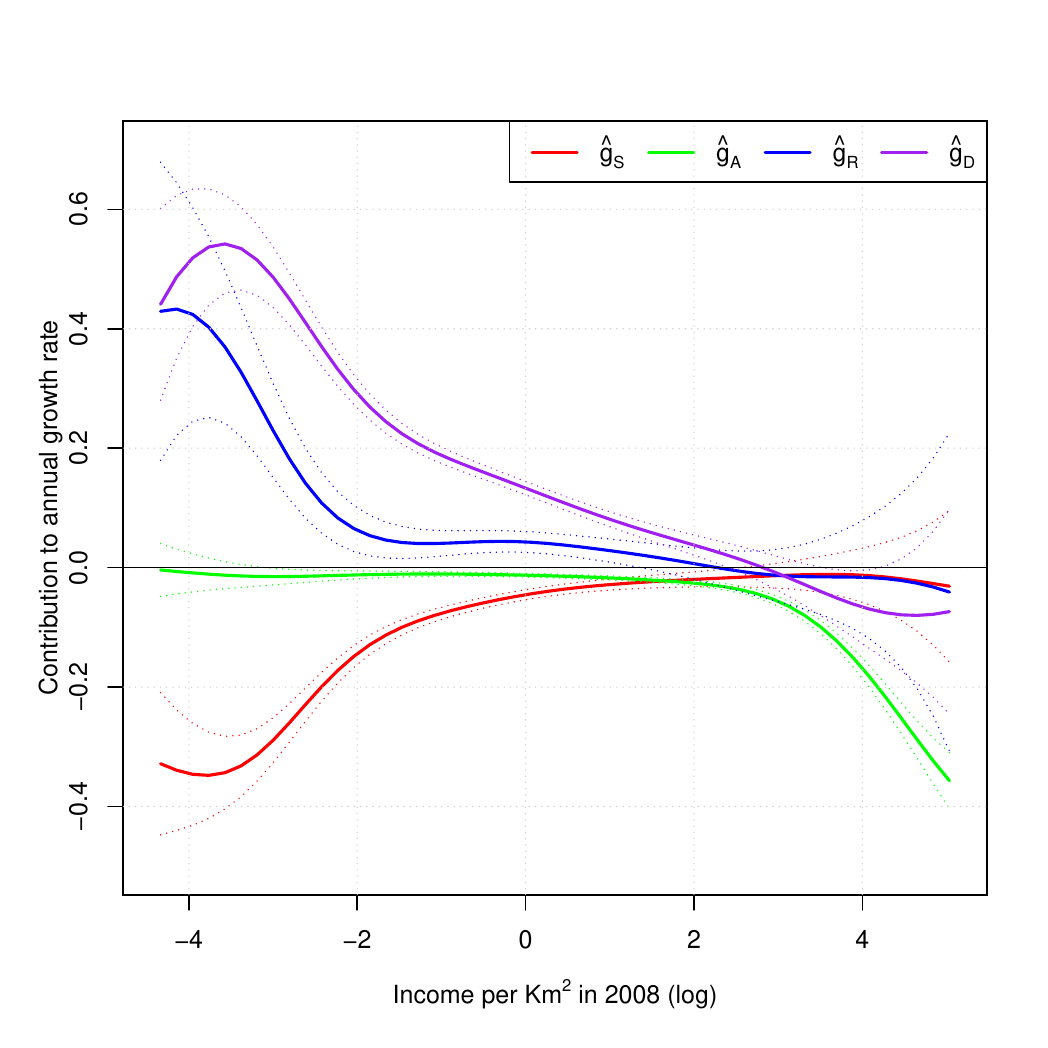}
	\caption{The nonparametric estimate (kernel regression) of the relation between each reallocative component (S, A, R, D) of local growth and the initial level of income (dotted lines represent 95\% confidence bands).}
	\label{fig:gjCrossSectional}
\end{figure}

Our model allows to account for the sources of (di)convergence in local income. Figure \ref{fig:gjCrossSectional} reports the nonparametric estimate (kernel regression) of the relationship between each one of the four components of local growth and the initial level of local income. A positive slope points to that component as a source of divergence, for example component S (topography) represented by the red line in Figure \ref{fig:gjCrossSectional} decreases up to a 0.4\% in annual growth rate for municipalities with a very low income density in 2008 (e.g. the mountain areas).
On the contrary, components R (repulsion) and D (diffusion) (blue and violet lines, respectively) display a negative slope, i.e. they are sources of convergence. In particular, diffusion exerts such convergent impact on the whole range of income, while repulsion only benefits low-income municipalities. Finally, component A (aggregation) (green line) is a source of convergence only for very high-income municipalities, i.e. for an income per Km$^2$ in 2008 greater than $\exp(3)\approx 20$ (current thousands of  euros per squared kilometre), in agreement with the figures in Table \ref{tab:metropolitanCities}, where the most of metropolitan municipalities showed a negative contribution from the aggregative component.

\subsection{Forecasting the local income \label{sec:forecasting}}

\begin{figure}[!htbp]
	\centering
	\begin{subfigure}[b]{0.4\textwidth}
		\centering
		\includegraphics[width=\textwidth]{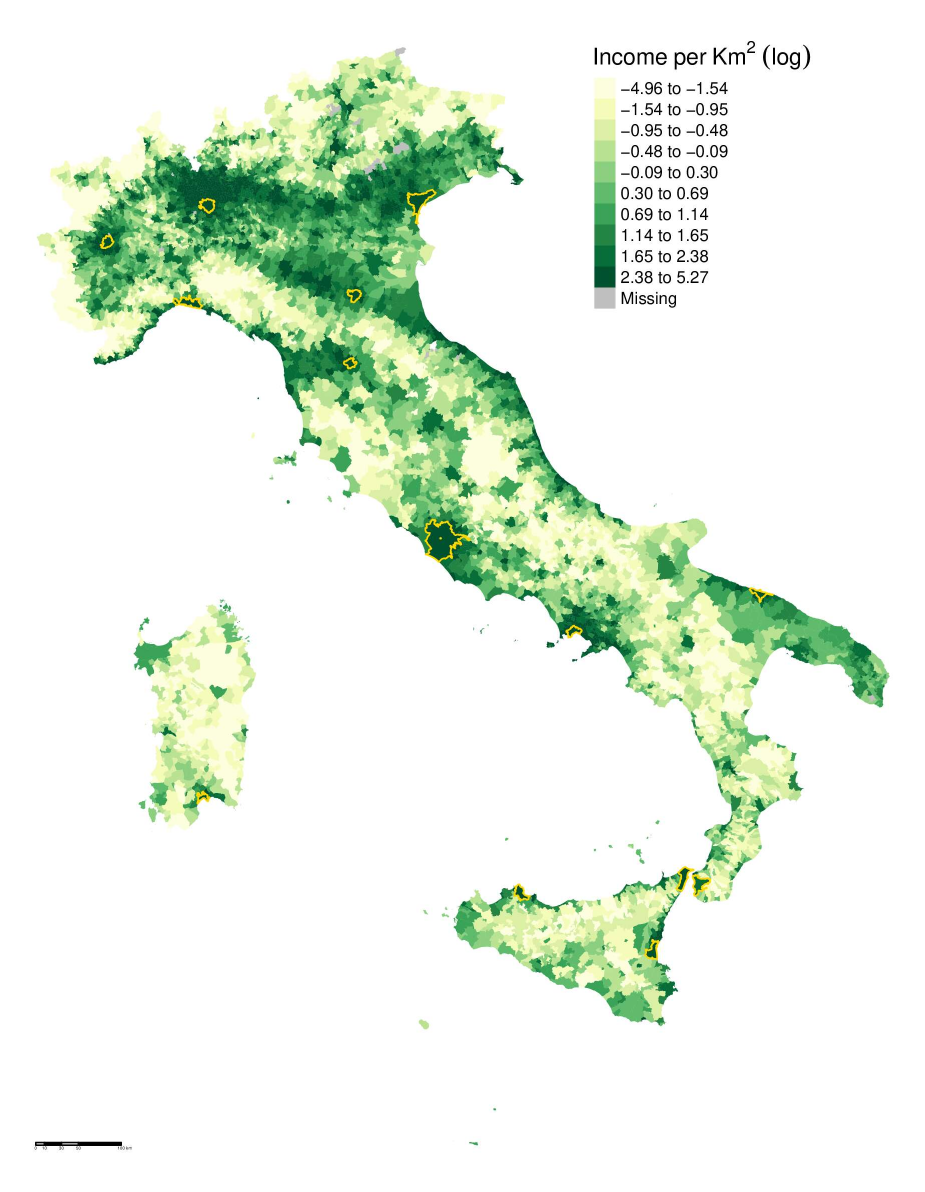}
		\caption{Income per Km$^2$ in 2019.}
		\label{fig:incomeDistribution2019}
	\end{subfigure}
	\begin{subfigure}[b]{0.4\textwidth}
		\centering
		\includegraphics[width=\textwidth]{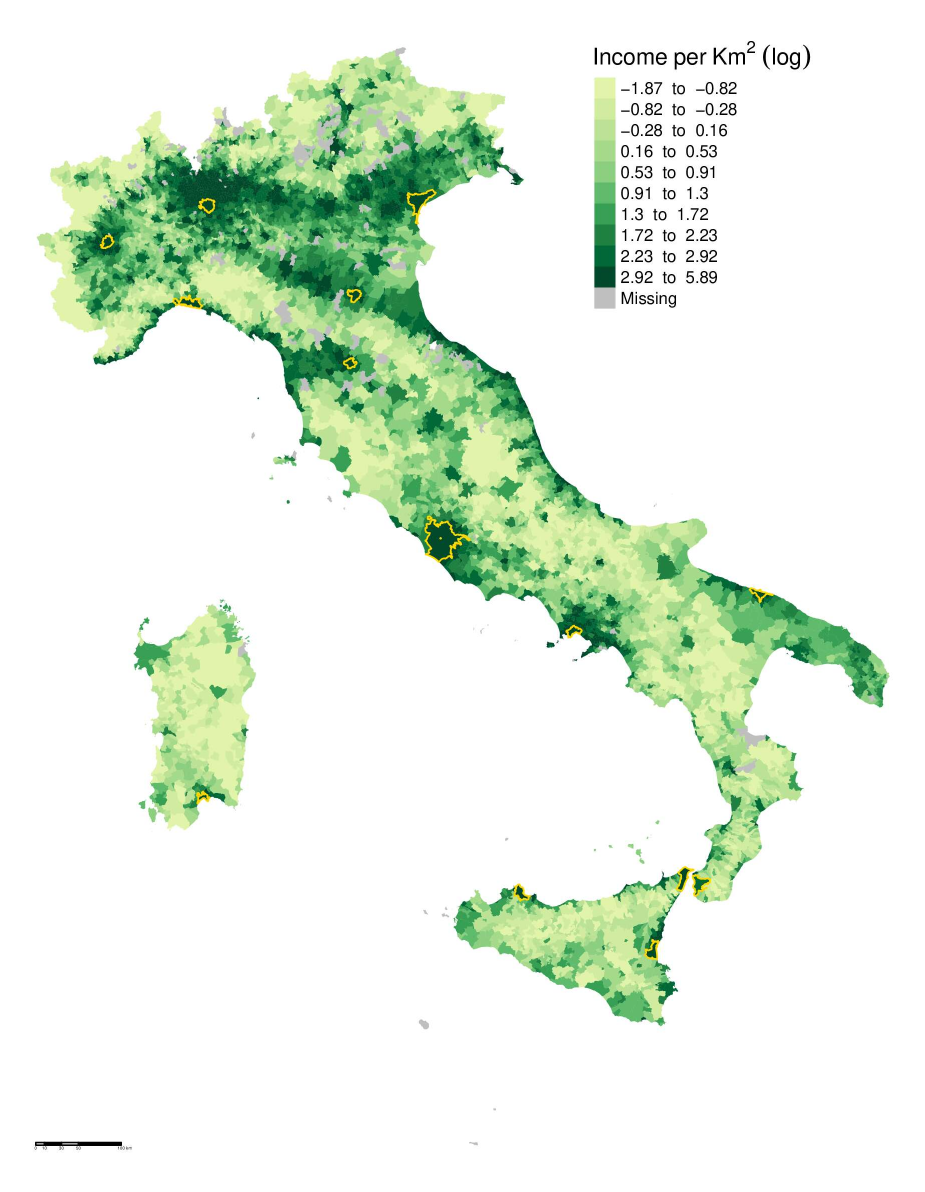}
		\caption{Forecasted income per Km$^2$ in 2069.}
		\label{fig:incomeSARD}
	\end{subfigure}
	\caption{The comparison between the income per Km$^2$ of Italian municipalities in 2019 versus the forecasted income per Km$^2$ in 2069 based on the estimated SARD model in Table \ref{tab:0819}. Yellow borders indicate the 14 Italian metropolitan municipalities.}
\label{fig:forecastedIncome}
\end{figure}

A potential application of the SARD model is forecasting local income. In Figure \ref{fig:incomeSARD} we map the forecasted income per Km$^2$ in 2069 using the SARD model, taking 2019 as the starting year.
In the forecast, most of the municipalities are anticipated to grow, with a median growth rate amounting to 1.17\% (municipalities with a forecasted negative growth are only 74). A graphical comparison of Figures \ref{fig:incomeDistribution2019} and \ref{fig:incomeSARD} suggests that (relative) persistence is the dominant feature for spatial distribution dynamics: in other words, the spatial pattern of income density is already consolidated in Italy, which implies that the spatial distribution in 2019 is close to its long-run equilibrium. 

\section{Concluding remarks \label{sec:concludingRemarks}}

We have proposed a novel family of spatial econometric models, derived from continuous time-space models defined by the formalism of PDEs. This new family offers an alternative to traditional spatial econometrics \citep{lesage2009introduction}, and the burgeoning field of quantitative spatial economics \citep{redding2017quantitative}. 

Our models allow to disentangle the mechanism of \textit{accumulation} from the one of \textit{reallocation}, and enable the identification of four different classes of spatial effects, i.e. \textit{topography}, \textit{aggregation}, \textit{repulsion} and \textit{diffusion}. These effects are identified through spatial matrices derived from our theoretical framework and computed based on the spatial distribution of observations. Consequently, our methodology overcomes the traditional identification issues faced by other spatial econometric models \citep{gibbons2012mostly}.
Furthermore, our approach allows for a more sophisticated incorporation of topographical features, such as roads, rivers, and railways, compared to existing literature (e.g., \citealp{allen2014trade}), by constructing additional exogenous regressors informed by the theoretical framework. In the context of quantitative spatial economics, our method requires less information for estimation, representing a significant advantage.
Additionally, the proposed methodology demonstrates a strong capability for producing accurate forecasts by accounting for the endogeneity of the spatial dynamics of the variable of interest. This feature makes it particularly useful for forecasting local dynamics of climate change and pollution, as illustrated by \citet{cruz2024economic}.

Our methodology, unlike other spatial econometric models, performs exceptionally well as the frequency and resolution of data increase. Its full potential will be realized with the advent of high-resolution satellite imagery and the involvement of major companies in academic research, which will expand the availability of large, high-resolution georeferenced datasets at higher frequencies (e.g., nightlights, population, climate, and pollution data).\footnote{For instance, see data from the \textit{Visible Infrared Imaging Radiometer Suite} (VIIRS) on the Suomi satellite launched in 2011 by NASA and the National Oceanic and Atmospheric Administration (NOAA) at \url{https://eogdata.mines.edu/products/vnl/}, the Copernicus Programme at \url{https://www.copernicus.eu/en}, and \textit{Data For Good}'s High-Resolution Population Density Maps by META at \url{https://dataforgood.facebook.com/dfg/tools}.}

\medskip

\noindent \textbf{Acknowledgements}: The authors have been supported by the Italian Ministry of University and Research (MIUR), in the framework of PRIN project 2017FKHBA8 001 (The Time-Space Evolution of Economic Activities: Mathematical Models and Empirical Applications).

\bibliographystyle{apalike}

\bibliography{biblio}

\clearpage

\appendix

\textbf{\Large Appendix}

\section{The matrices for the correction of discretization over time \label{sec:appendixMatricesdiscretizationOverTime}}
In this appendix, we describe how to compute the matrices $M_S$, $M_A$, $M_R$ and $M_D$ for the discretization in time of Eq. \eqref{eq:dataGenerationProcessCellTotalIncome}. In Section \ref{subsec:discretizationTime} we saw the case where $\gamma_S,\gamma_A,\gamma_R = 0$. We now deal with the full case. In the general case Eq. \eqref{eq:partialtty} becomes:
\begin{eqnarray}\label{eq:partialttyApp}
	\partial_{tt}^{2}y(t',z^i) &=& 	\partial_{t}\left[\partial_{t}y(t,z^i)\right]\big\vert_{t=t'}= \nonumber \\ \nonumber
	&=& \partial_t a\left(t,z^i\right)\big\vert_{t=t'} + \phi\partial_{t}y(t,z^i)\big\vert_{t=t'} + \\ \nonumber
	&+& \gamma_S \bigg[\partial_{z_1} \left( \partial_{t}y(t',z) \partial_{z_1} S(z)\right) + \partial_{z_2} \left( \partial_{t}y(t',z) \partial_{z_2} S(z)\right)\bigg]\bigg\vert_{z=z^i} +  \\ \nonumber
	&+& \gamma_A \bigg[\partial_{z_1} \left(\partial_{t}y(t',z) \partial_{z_1} \left(K_{h_A}  * y\right) (t',z) + y(t',z) \partial_{z_1}\left(K_{h_A}  * \partial_{t}y\right) (t',z)\right)  +  \\ \nonumber
	&+& \partial_{z_2} \left(\partial_{t}y(t',z) \partial_{z_2} \left(K_{h_A}  * y\right) (t',z) + y(t',z) \partial_{z_2}\left(K_{h_A}  * \partial_{t}y\right) (t',z)\right)\bigg]\bigg\vert_{z=z^i}  + \\ \nonumber
	&+& \gamma_R \bigg[\partial_{z_1} \left(\partial_{t}y(t',z) \partial_{z_1} \left(K_{h_R}  * y\right) (t',z) + y(t',z) \partial_{z_1}\left(K_{h_R}  * \partial_{t}y\right) (t',z)\right)  +  \\ \nonumber
	&+& \partial_{z_2} \left(\partial_{t}y(t',z) \partial_{z_2} \left(K_{h_R}  * y\right) (t',z) + y(t',z) \partial_{z_2}\left(K_{h_R}  * \partial_{t}y\right) (t',z)\right)\bigg]\bigg\vert_{z=z^i}  + \\ 
	&+& \gamma_D \bigg[\partial_{z_1 z_1} \partial_t y(t',z)+ \partial_{z_2 z_2} \partial_t y(t',z)  \bigg]\bigg\vert_{z=z^i}. 
\end{eqnarray}
Applying the discretization also in space and assuming $\partial_t y(t,z^i)\big\vert_{t=t'} \approx \Delta_\tau y_i$ as in Section \ref{subsec:discretizationTime}, we have 
\begin{eqnarray}\label{eq:partialttyMatrixPartialApp}
	\partial_{tt}^{2}y(t',z^i) &\approx& \partial_t a\left(t',z^i\right) + {\rho_\phi} \Delta_\tau y_i + \nonumber \\ \nonumber
	&+& \rho_S \bigg[ M_{z_1} \left( \Delta_{\tau} \mathbf{y} \odot  M_{z_1} \mathbf{s}\right) +  M_{z_2} \left( \Delta_{\tau} \mathbf{y} \odot  M_{z_2} \mathbf{s}\right)\bigg]_i +  \\ \nonumber
	&+& \rho_A \bigg[ M_{z_1} \left(\Delta_{\tau} \mathbf{y} \odot M_{z_1} W_{h_A} \mathbf{y} + \mathbf{y} \odot  M_{z_1}W_{h_A} \Delta_{\tau}\mathbf{y}\right)  +  \\ \nonumber
	&+&  M_{z_2} \left(\Delta_{\tau} \mathbf{y} \odot M_{z_2} W_{h_A} \mathbf{y} + \mathbf{y}\odot M_{z_2}W_{h_A} \Delta_{\tau}\mathbf{y}\right)\bigg]_i  + \\ \nonumber
	&+& \rho_R \bigg[ M_{z_1} \left(\Delta_{\tau} \mathbf{y}\odot  M_{z_1} W_{h_R} \mathbf{y} +\mathbf{y}  \odot M_{z_1}W_{h_R} \Delta_{\tau}\mathbf{y}\right)  +  \\ \nonumber
	&+&  M_{z_2} \left(\Delta_{\tau} \mathbf{y} \odot M_{z_2} W_{h_R} \mathbf{y} + \mathbf{y} \odot M_{z_2}W_{h_R} \Delta_{\tau}\mathbf{y}\right)\bigg]_i  + \\ 
	&+& \rho_D \left[( M_{z_1 z_1} +  M_{z_2 z_2}) \Delta_{\tau} \mathbf{y}  \right]_i, 
\end{eqnarray}
for some $\rho_\phi \approx \phi$, $\rho_S \approx \gamma_S$, $\rho_A \approx \gamma_A$, $\rho_R \approx \gamma_R$ and $\rho_D \approx \gamma_D$.

The expression related to the coefficient $\rho_D$ can be immediately interpreted as 
\begin{equation*}
	\rho_D \left[( M_{z_1 z_1} +  M_{z_2 z_2}) \Delta_{\tau} \mathbf{y}  \right]_i = \rho_D (M_D \Delta_{\tau} \mathbf{y}) _i
\end{equation*}
by introducing $M_D \equiv  M_{z_1 z_1} +  M_{z_2 z_2}$. The terms related to the coefficients $\rho_S$, $\rho_A$ and $\rho_R$, even though cannot be immediately expressed in matrix form, are linear with respect to $\Delta_{\tau} \mathbf{y}$. Therefore 
it is possible to construct matrices $M_S,M_A$ and $M_R$ depending on $\mathbf{y}$
such that Eq. \eqref{eq:partialttyMatrixPartialApp} becomes  
\begin{eqnarray}\label{eq:partialttyMatrixApp}
	\partial_{tt}^{2}y(t',z^i) &\approx& \partial_t a\left(t',z^i\right) + {\rho_\phi} \Delta_\tau y_i + \nonumber \\ \nonumber
	&+& {\rho}_S (M_S \Delta_\tau \mathbf{y})_i + {\rho}_A (M_A \Delta_\tau \mathbf{y})_i + 
	{\rho}_R (M_R \Delta_\tau \mathbf{y})_i + {\rho}_D (M_D \Delta_\tau \mathbf{y})_i.
\end{eqnarray}
To construct such matrices recall that given any linear function $\psi:\RR^N \to \RR^N$, this can be represented in matrix form with respect to the canonical basis of $\RR^N$ $\{\mathbf{e}_i\}_{i=1,\dots,N}$ by the matrix which has as the $i$-th row the vector $\psi(\mathbf{e}_i)$.

Eq. \eqref{eq:partialttyMatrixApp}  is the analogous of Eq. \eqref{eq:partialttyMatrix} when all the terms of Eq. \eqref{eq:dataGenerationProcessCellTotalIncome} are present. Repeating the same computation of Section \ref{subsec:discretizationTime} we end up with the analogous of Eq. \eqref{eq:discretizationSpaceTimeSimplifiedModel} with all the terms, which reads
\begin{eqnarray}\label{eq:discretizationSpaceTimeSimplifiedModelApp}
	\Delta_\tau \mathbf{y} &\approx&  \left[ \left(1 - \frac{\tau{\rho_\phi}}{2}\right)\mathbf{I} - \frac{\tau{\rho}_S}{2} M_S - \frac{\tau{\rho}_A}{2} M_A - \frac{\tau{\rho}_R}{2} M_R - \frac{\tau{\rho}_D}{2} M_D\right]^{-1} \times \nonumber \\ \nonumber 
	&\times& \bigg[ \mathbf{a} +  \frac{\tau}{2}\Delta_\tau \mathbf{a} + \phi \mathbf{y} + \\ \nonumber 
	&+& {\gamma}_{S} \left[ M_{z_1} \left(\mathbf{y} \odot M_{z_1} \mathbf{s} \right)  +  M_{z_2} \left(\mathbf{y} \odot M_{z_2} \mathbf{s} \right)  \right] +   \\ \nonumber
	&+& {\gamma}_A \left[ M_{z_1} \left(\mathbf{y} \odot M_{z_1} W_{h_A} \mathbf{y} \right)  +  M_{z_2} \left(\mathbf{y} \odot M_{z_2} W_{h_A} \mathbf{y} \right)  \right] + \\  \nonumber
	&+& {\gamma}_R \left[ M_{z_1} \left(\mathbf{y} \odot M_{z_1} W_{h_R} \mathbf{y} \right)  +  M_{z_2} \left(\mathbf{y} \odot M_{z_2} W_{h_R} \mathbf{y} \right)  \right] +  \\  
	&+& {\gamma}_D (M_{z_1 z_1} + M_{z_2 z_2})\mathbf{y} \bigg].
\end{eqnarray}
From Eq. \eqref{eq:discretizationSpaceTimeSimplifiedModelApp} by rearranging the terms we arrive at Eq. \eqref{eq:modelDiscretizedOvertime}.

\section{Estimation procedure of spatial matrix $W_\epsilon$  \label{app:spatialMatrixErrors}}

In this section, we describe the data-driven procedure for the calculation of $W_\epsilon$.
Let $\hat{\mathbf{e}}$ be the $1 \times N$ vector of residuals of Model \eqref{eq:SARD} estimated by ML without controlling for spatial correlation in the errors, and $W_q$ the spatial matrix based on the $q$-th order of contiguity across cells and zero diagonal. As the first step, estimate by OLS the model:
\begin{equation} 
	\hat{\mathbf{e}} = \left[\ell_1 W_1 + \ell_2 \left(W_2 -W_1\right) + \ell_3 \left(W_3 -W_2\right) +  \cdots + \ell_Q \left(W_Q -W_{Q-1}\right)\right]\hat{\mathbf{e}} + \mathbf{u},
	\label{eq:spatialWeightMatrix}
\end{equation}
where $\mathbf{u}$ is a vector of random components. Then, calculate the spatial weight matrix $W_\epsilon$ used in Model \eqref{eq:SARD} with the specification of error in Eq. \eqref{eq:SARDspatialError} as:
\begin{equation}
	W_\epsilon = \hat{\ell}_1 W_1 + \hat{\ell}_2 \left(W_2 -W_1\right) + \hat{\ell}_3 \left(W_3 -W_2\right) +  \cdots + \hat{\ell}_{\hat{Q}} \left(W_{\hat{Q}} -W_{\hat{Q}-1}\right),
	\label{eq:estimatedSpatialWeightMatrix}
\end{equation}
where $\hat{Q}$ is the appropriate maximum order of contiguity suggested by the statistical significance of $\hat{\ell}$'s. As regards the desired properties, $W_\epsilon$ has zero diagonal, is symmetric, and allows for nonlinear diffusion of errors over space; in particular, the intensity of diffusion of the first-contiguous cells is equal to $\hat{\ell}_1$, of the only second -contiguous cells is equal to $\hat{\ell}_2$, etc...
The use of $W_\epsilon$ of Eq. \ref{eq:estimatedSpatialWeightMatrix} implies the estimated $\lambda$ should be very close to one. However, this could not be the case because, in principle, OLS estimates of Model \eqref{eq:spatialWeightMatrix} are biased for the presence of endogeneity. More sophisticated methods of estimate, such as ML, could overcome such bias but also imply not trivial numerical difficulties which we leave to future research \citep{lesage2011pitfalls}.

\clearpage

\textbf{\Large Online Appendix}

\section{The micro-foundation of the spatial growth model\label{sec:microfoundation}}
This appendix summarizes the main ideas about the microfoundation of Eq. \eqref{eq:dataGenerationProcessCellTotalIncome}. Consider a set of $N_a$ rational agents, which can be understood as units of production/consumers. For the sake of simplicity, we briefly describe here the case where $\phi = 0$ and $a\left(t,z\right)=0 \:\forall (t,z)$, i.e. there is no change in the total number of agents but they only relocate across space. The case $\phi \neq 0$ can be treated similarly, see for example \cite{catellier2021mean}. Each agent is identified by index $i$, with $i = 1,\dots, N_a$, and is characterised by its location in the domain $\Omega\subseteq \RR^2$, labelled by $X^{i,N_a}_t$. At $t = 0$ agents are independently distributed at random in the domain $\Omega$ following a common probability density distribution on $\Omega$ denoted by $y_0(z)$. For $t > 0$ each agent's location evolves by obeying the following Stochastic Differential Equation (SDE):
\begin{eqnarray}  \nonumber
	dX^{i,N_a}_t = &-& \gamma_S \nabla_z S\left(X^{i,N_a}_t\right)\,dt + \\ \nonumber
	&-& \gamma_A \frac{1}{N_a}\sum_{j=1}^{N_a}\nabla_z K_{h_A}\left(X^{i,N_a}_t - X^{j,N_a}_t\right)\,dt+\\ \nonumber
	&-& \gamma_R \frac{1}{N_a}\sum_{j=1}^{N_a}\nabla_z K_{h_R}\left(X^{i,N_a}_t - X^{j,N_a}_t\right)\,dt+\\
	&+& \sqrt{2\gamma_D}\,dB^i_t,
	\label{eq:agentMovement}
\end{eqnarray}
where $(B^i_t)_{i \in \NN}$  is a sequence of independent Brownian motions, and $\gamma_A < 0$, while $\gamma_S \text{,} \gamma_R \text{ and } \gamma_D >0$, to respect the coherence with the phenomena (spatial non-uniformity, aggregation, repulsion, diffusion) we are interested to model.

According to Eq. \eqref{eq:agentMovement} the spatial position of each agent is evolving by keeping into account its relative position with respect to all other agents. In particular, the first term on the right-hand side of Eq. \eqref{eq:agentMovement} expresses the tendency of agents to move where the function $S$ is lower, which can be decided by the agents' spatial distribution and/or the particular characteristics of different locations. The second and third terms reflect the interactions among agents. In particular, by the linearity of the derivative: 
\begin{eqnarray*}  
	\frac{1}{N_a}\sum_{j=1}^{N_a}\nabla_z K_{h_A}\left(X^{i,N_a}_t - X^{j,N_a}_t\right) &=& \nabla_z\left(\frac{1}{N_a}\sum_{j=1}^{N_a} K_{h_A}\left(\cdot - X^{j,N_a}_t\right) \right)(X^{i,N_a}_t); \text{ and} \\
	\frac{1}{N_a}\sum_{j=1}^{N_a}\nabla_z K_{h_R}\left(X^{i,N_a}_t - X^{j,N_a}_t\right) &=& \nabla_z\left(\frac{1}{N_a}\sum_{j=1}^{N_a} K_{h_R}\left(\cdot - X^{j,N_a}_t\right) \right)(X^{i,N_a}_t),
\end{eqnarray*}
that is, each agent is moving along the direction of the gradient of a local average (the concept of locality is defined by the functions $K_{h_A}$ and $K_{h_R}$) of the other agents' location. Finally, the fourth term of Eq. \eqref{eq:agentMovement} represents the agents' \textit{idiosyncratic and independent random movements} and is expressed through the independent additive noise $B^i_t$.

The \textit{empirical distribution} of agent's location is: 
\begin{equation}
	E^{N_a}_t := \frac{1}{N_a}\sum_{i=1}^{N_a} \delta_{X^{i,{N_a}}_t},
\end{equation}
where $\delta_z$ is the random variable on $\RR^2$ with unitary mass in the point $z$. $E^{N_a}_t$ is a continuous set of random variables on $\RR^2$ depending on time. For any given $N_a \in \NN$ and $t>0$, this distribution is singular, in the sense that is a distribution over $\RR^2$ which does not admit a probability density function, since it has a positive probability only on a finite set (corresponding to the location of the $N_a$ agents). However, when $N_a$ goes to infinity the family of random variable $E^{N_a}_t$ becomes diffuse and converges (in distribution) to a continuous family of random variables over $\RR^2$, labelled by $E_t$ for any $t > 0$. The distribution of $E_t$ is now regular and admits a probability density function for every $t$, called $y(t,z)$. An explicit expression for $y(t,z)$ for every $t$ is not available. However one can prove that the probability density function $y(t,z)$ is the unique solution to Eq.  \eqref{eq:dataGenerationProcessCellTotalIncome} (see \citealp[Section 1.1]{sznitman1991topics}).

\clearpage

\section{A numerical investigation of the properties of the spatial growth model \label{sec:propertiesModel}}

An explicit solution for Eq. \eqref{eq:dataGenerationProcessCellTotalIncome} is not available in general. We will then proceed to discuss some of the basic properties of Eq. \eqref{eq:dataGenerationProcessCellTotalIncome} by numerical simulations. In all examples, we make some simplifications to better discern the peculiar features of the model from the exogenous effects related to the particular case considered in Section \ref{sec:empiricalApplication}. In particular, we take the total cross-sectional amount of the variable $y$ of the system constant and equal to 1, i.e. we take $\phi = 0$. Moreover, we take the exogenous function $S(z)$ to be identically zero. We also neglect the repulsive effect, i.e. set $\gamma_R = 0$. Summarising, we consider only the \textit{aggregation} ($\gamma_A$) and \textit{diffusion} ($\gamma_D$) effects. We study the problem on a squared domain $\Omega = [0,4] \times [0,4]$ with a discrete set of locations uniformly spaced with a distance between contiguous locations of $\Delta z = 10^{-2}$ (i.e. 160,000 total points). 

\begin{figure}[!htbp]
	\centering
	\begin{subfigure}[b]{0.3\textwidth}
		\centering
		\includegraphics[width=\textwidth]{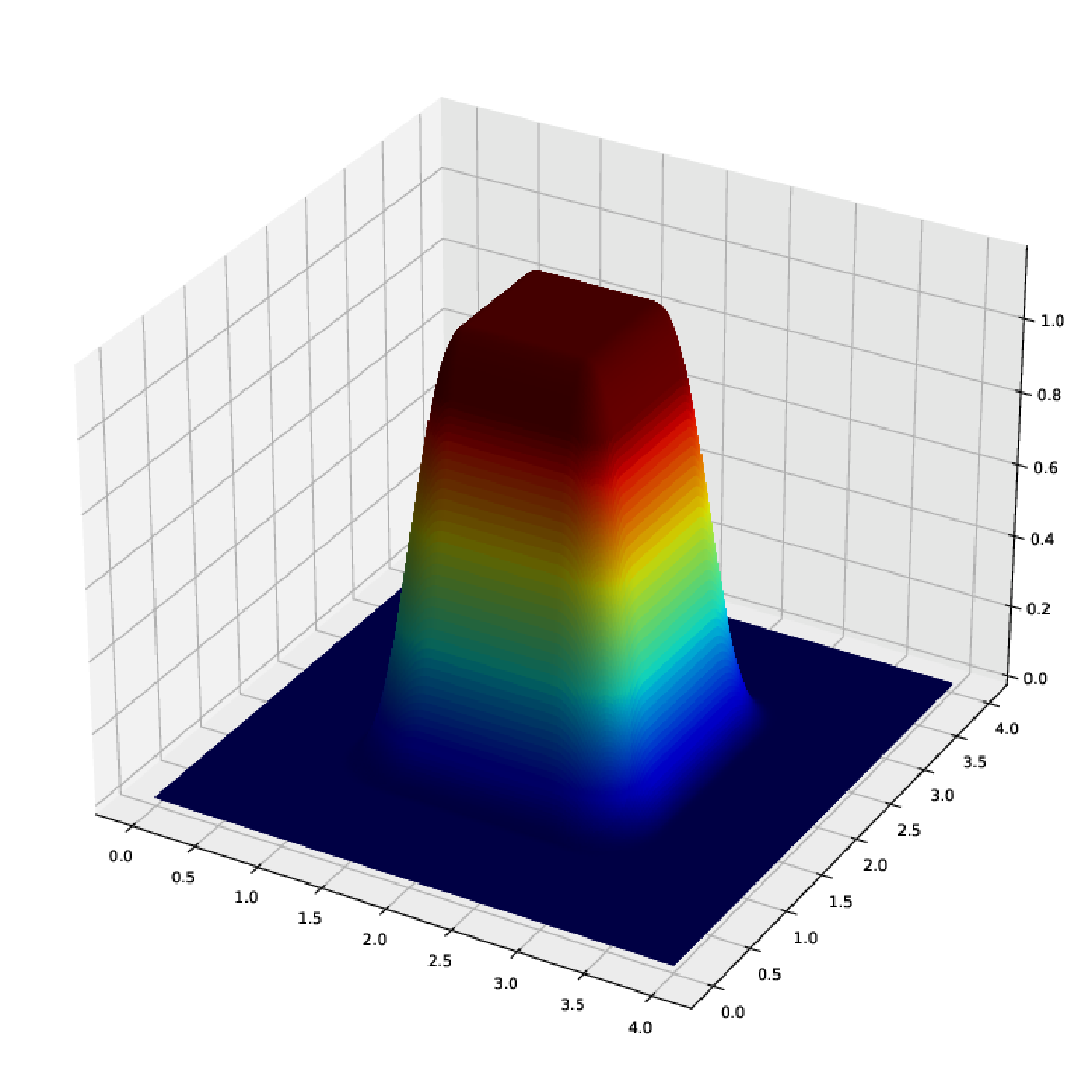}
		\caption{$t=0$}
	\end{subfigure}
	\begin{subfigure}[b]{0.30\textwidth}
		\centering
		\includegraphics[width=\textwidth]{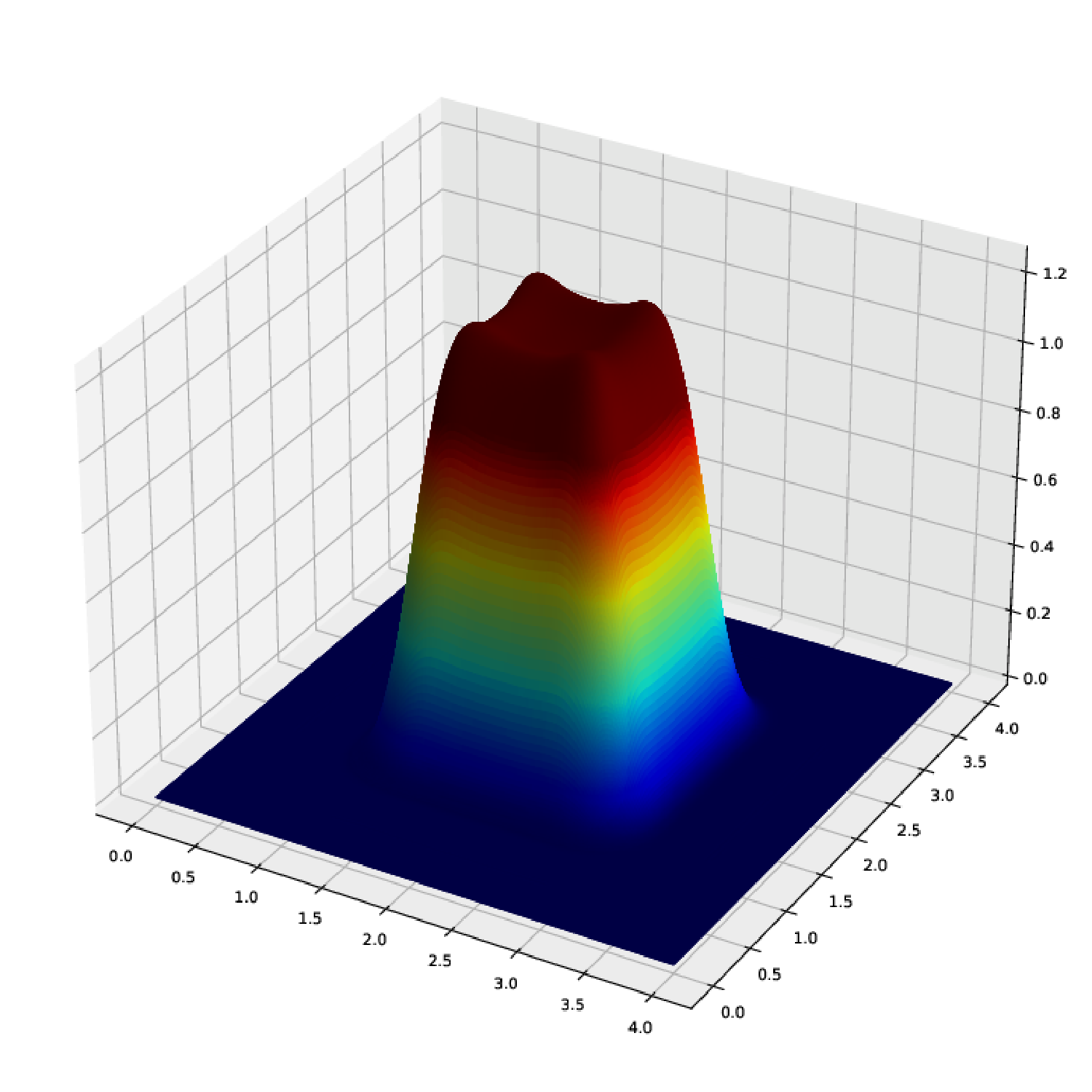}
		\caption{$t=0.5$}
	\end{subfigure}
	\begin{subfigure}[b]{0.30\textwidth}
		\centering
		\includegraphics[width=\textwidth]{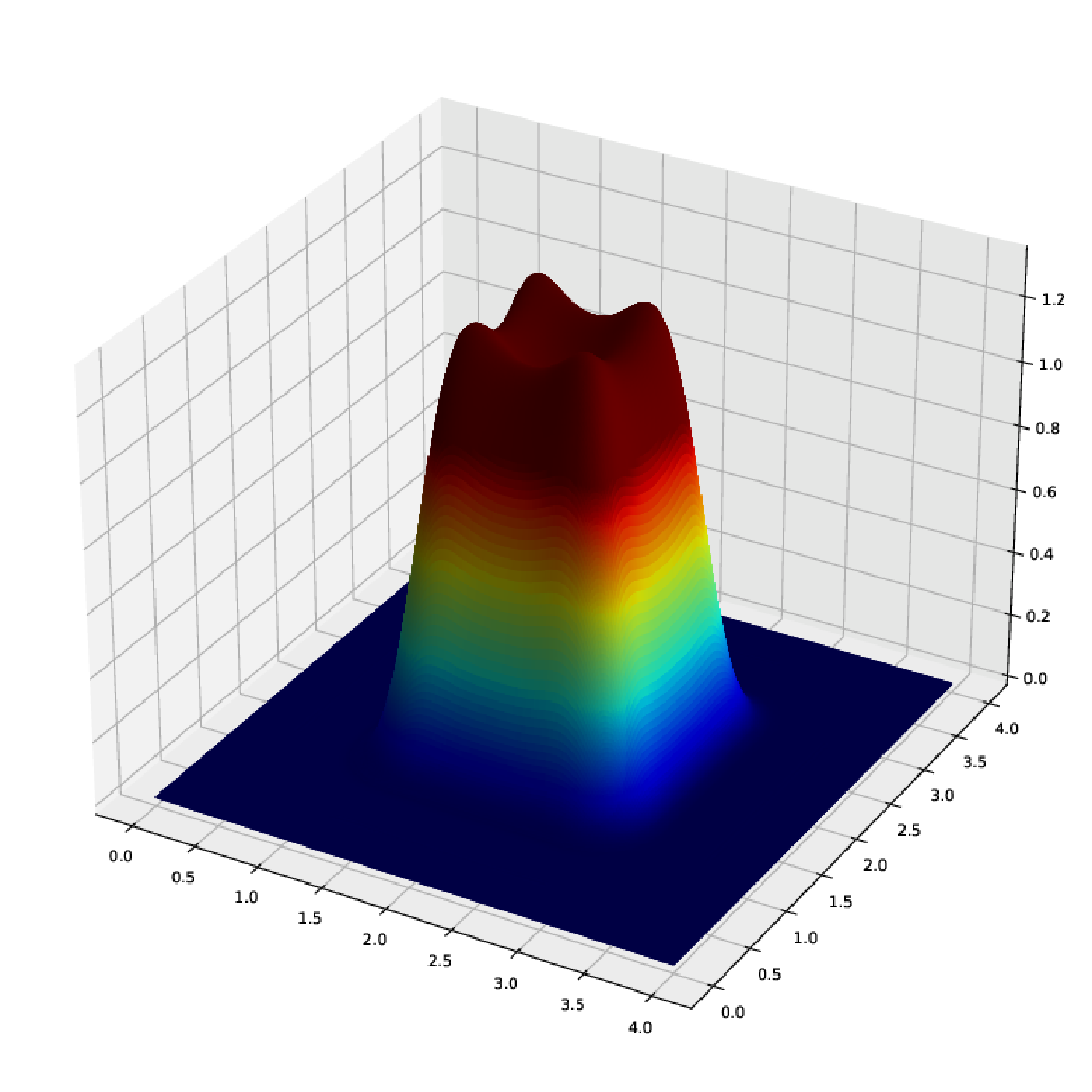}
		\caption{$t=1$}
	\end{subfigure}
	\begin{subfigure}[b]{0.3\textwidth}
		\centering
		\includegraphics[width=\textwidth]{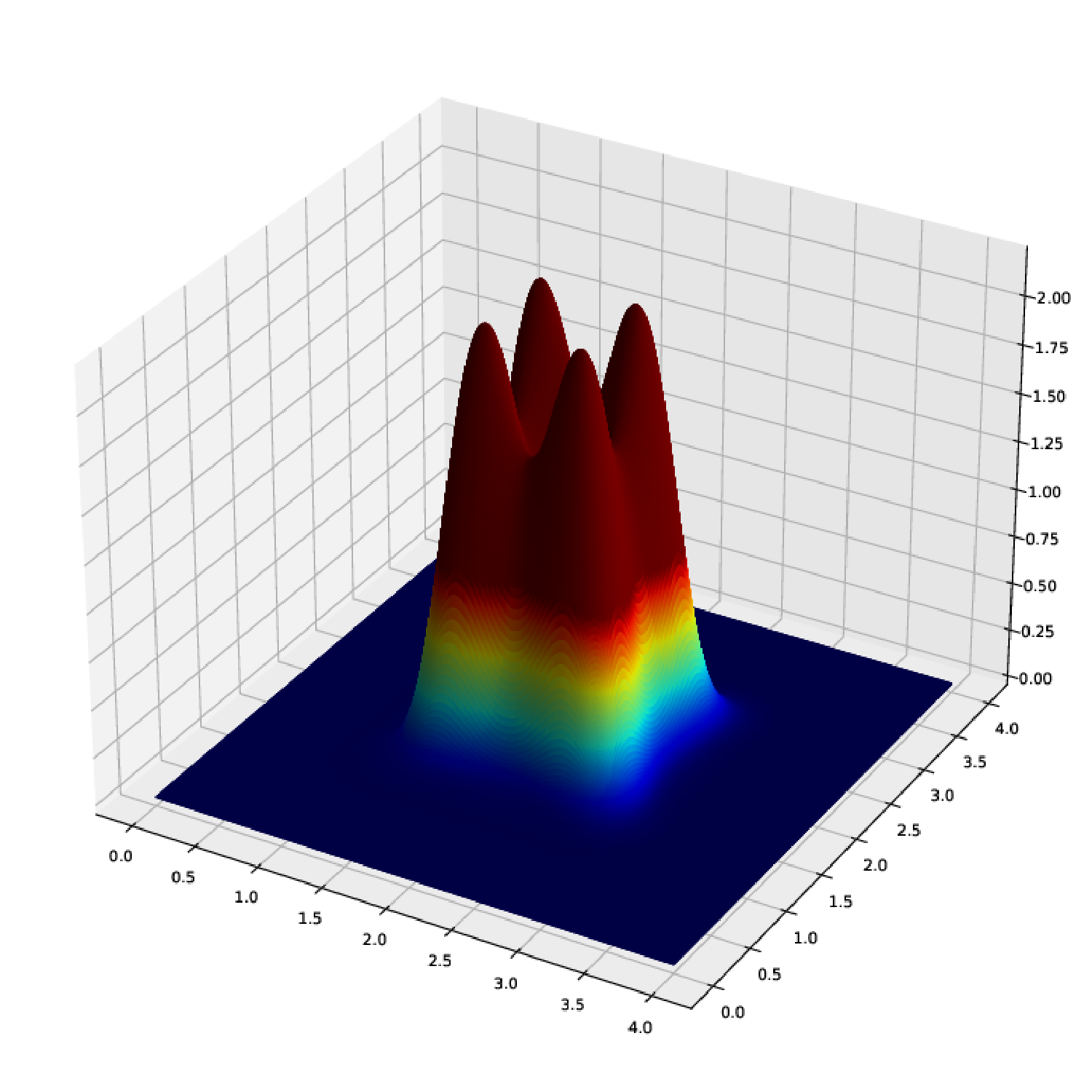}
		\caption{$t=5$}
	\end{subfigure}
	\begin{subfigure}[b]{0.30\textwidth}
		\centering
		\includegraphics[width=\textwidth]{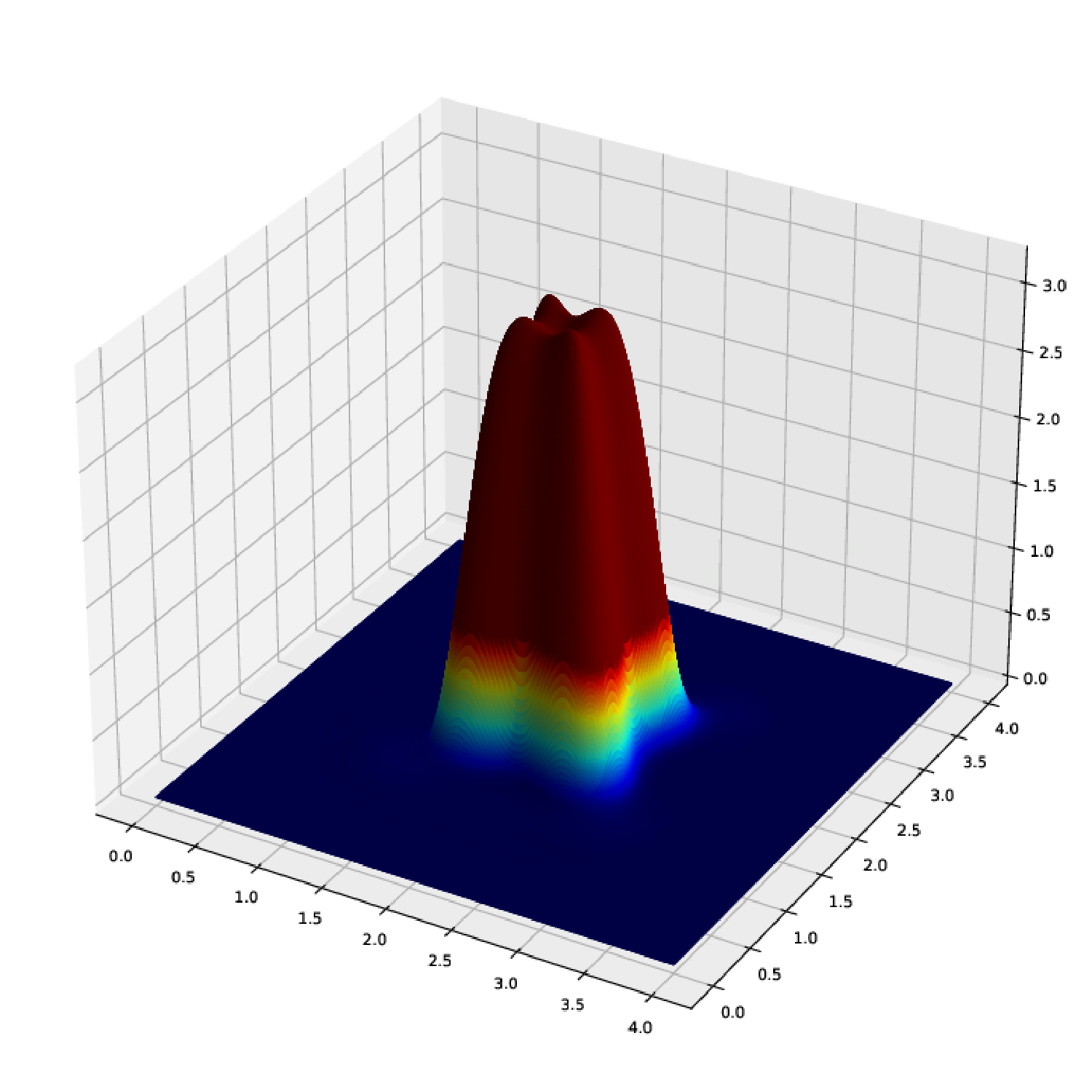}
		\caption{$t=10$}
	\end{subfigure}
	\begin{subfigure}[b]{0.30\textwidth}
		\centering
		\includegraphics[width=\textwidth]{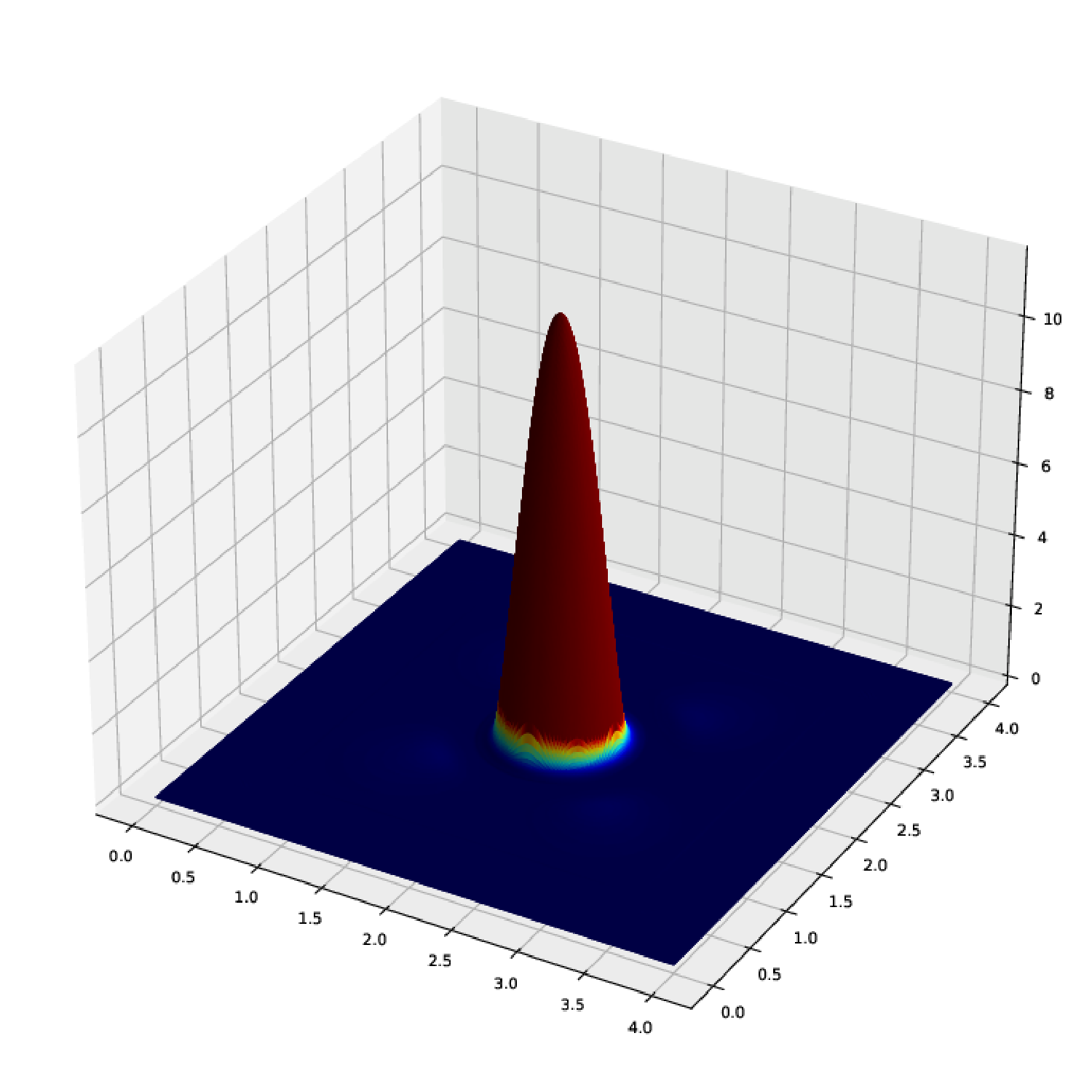}
		\caption{$t=20$}
	\end{subfigure}
	\caption{The distribution dynamics of $y(t,z)$ over space and time for the baseline case with $\gamma_S = 0,\gamma_A=-0.01,\gamma_R = 0$, $\gamma_D=0.005$, $h_A=0.4$, $\Omega=[0,4]\times[0,4]$.}
	\label{fig:numericalInvestigationh=0.4}
\end{figure}

\begin{figure}[!htbp]
	\centering
	\begin{subfigure}[b]{0.3\textwidth}
		\centering
		\includegraphics[width=\textwidth]{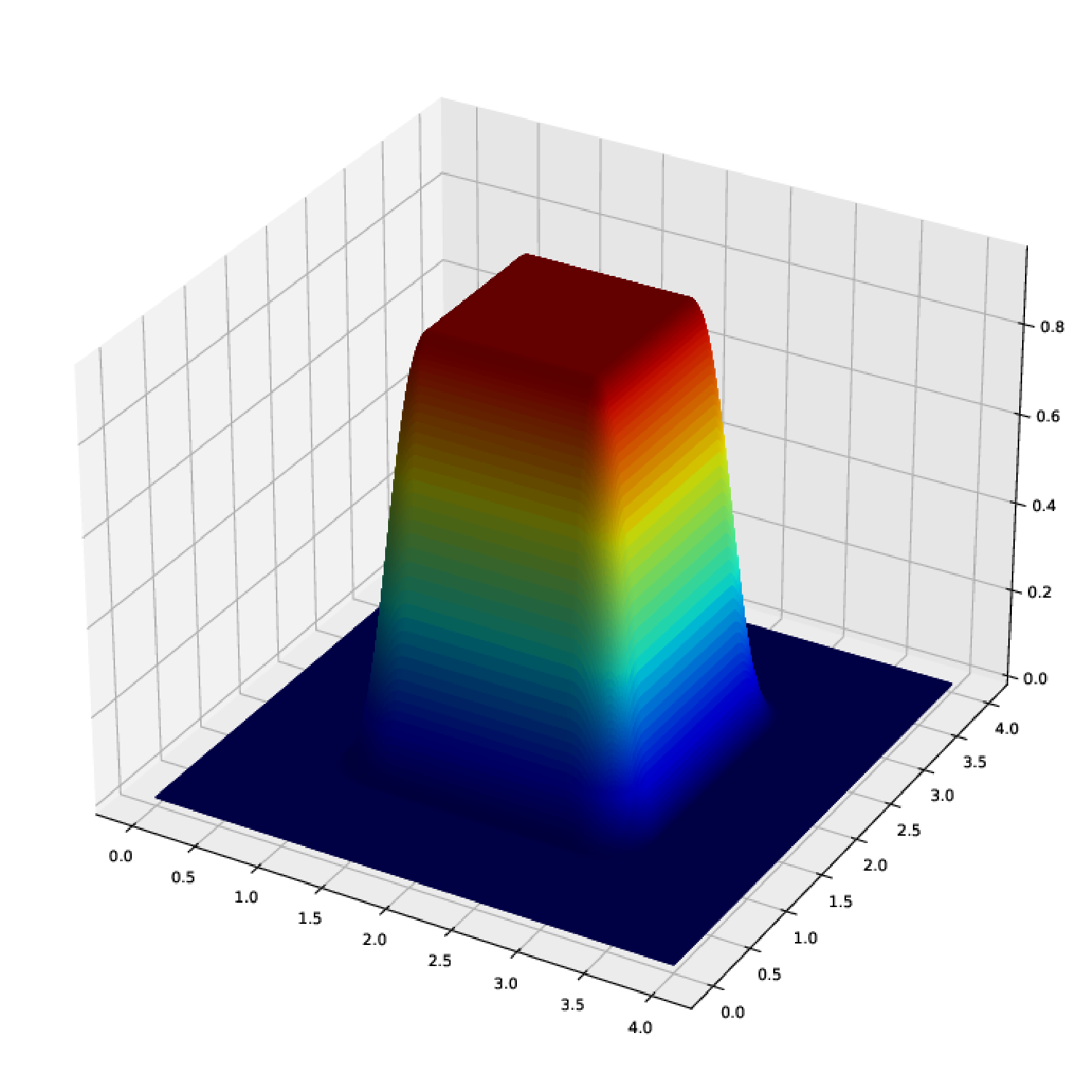}
		\caption{$t=0$}
	\end{subfigure}
	\begin{subfigure}[b]{0.30\textwidth}
		\centering
		\includegraphics[width=\textwidth]{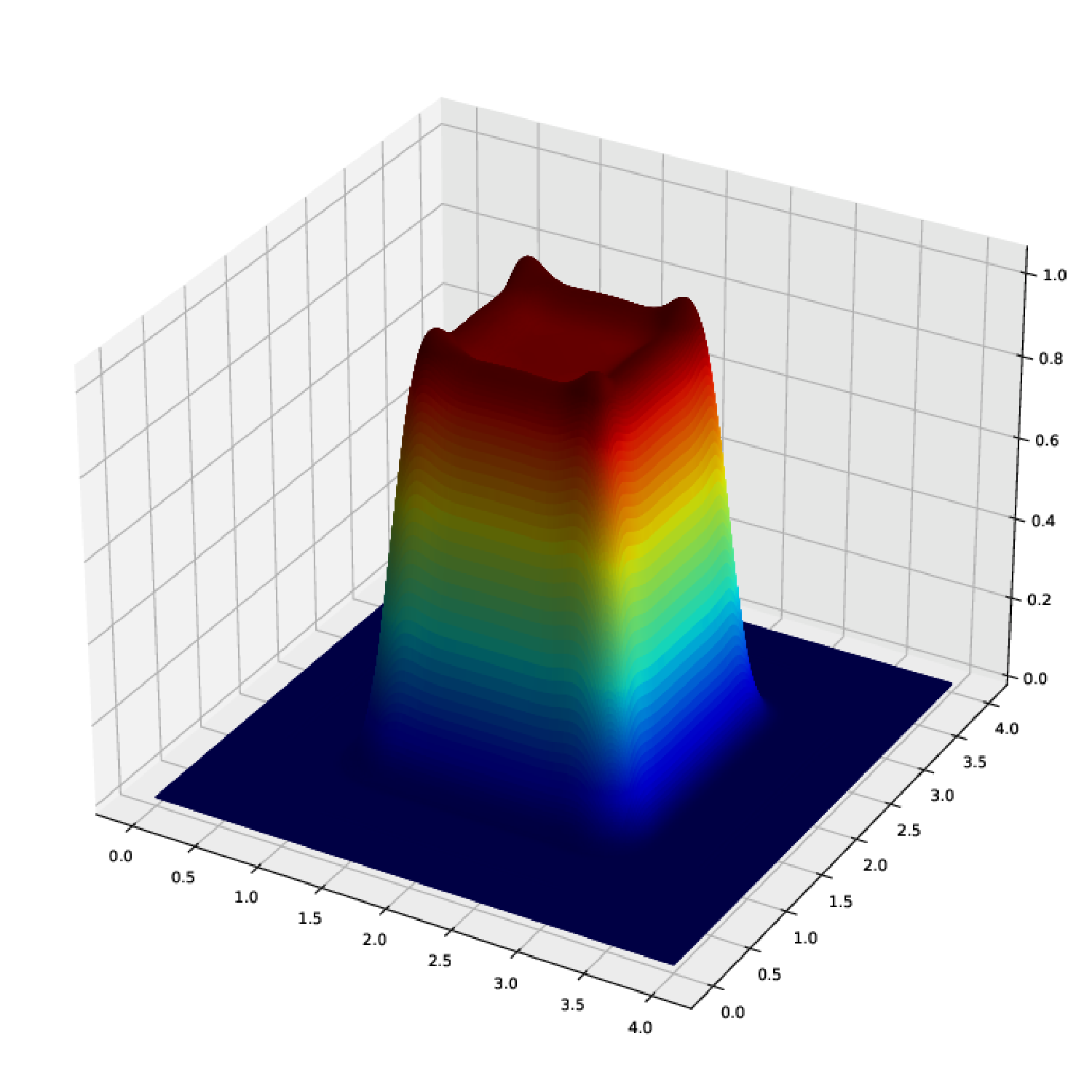}
		\caption{$t=0.5$}
	\end{subfigure}
	\begin{subfigure}[b]{0.30\textwidth}
		\centering
		\includegraphics[width=\textwidth]{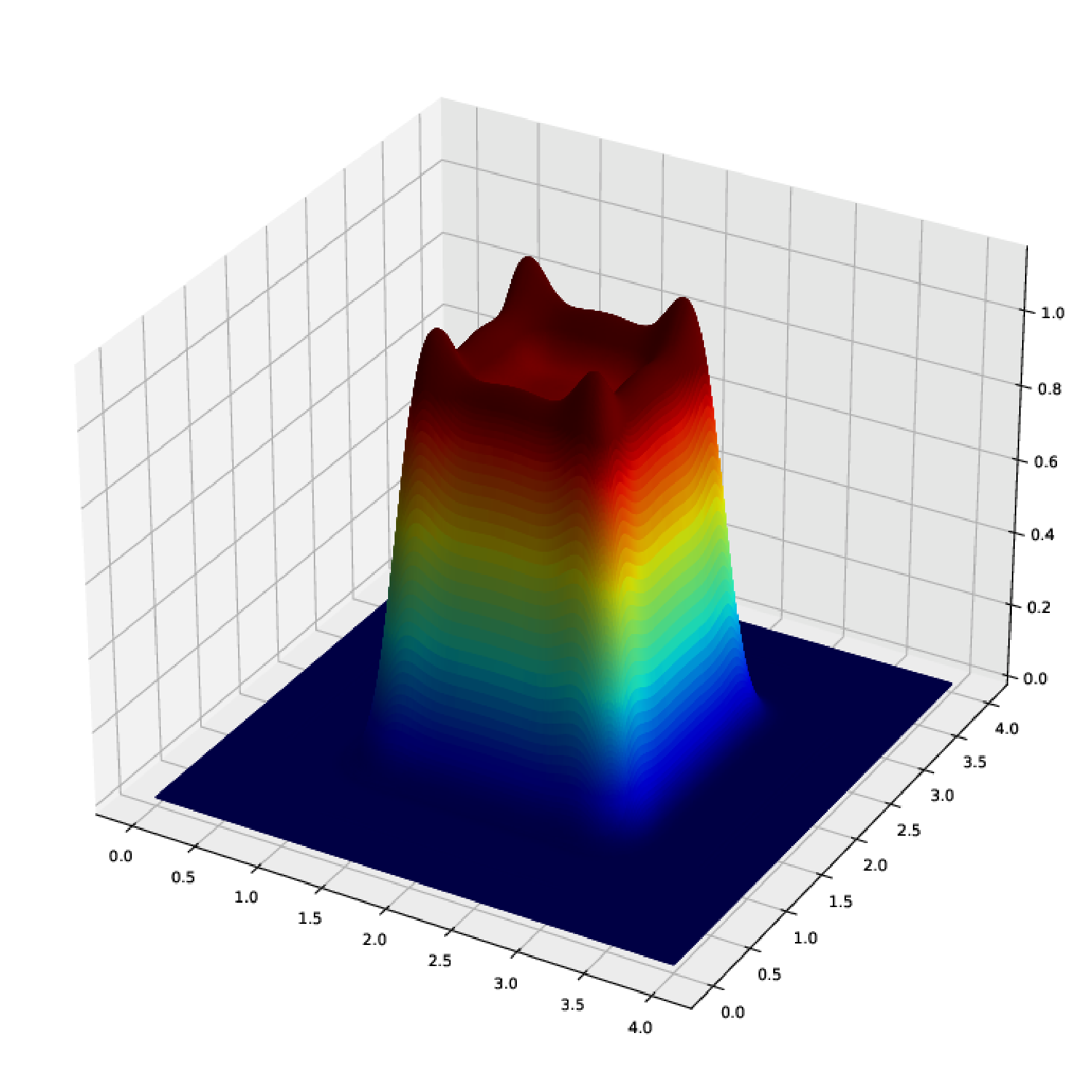}
		\caption{$t=1$}
	\end{subfigure}
	\begin{subfigure}[b]{0.3\textwidth}
		\centering
		\includegraphics[width=\textwidth]{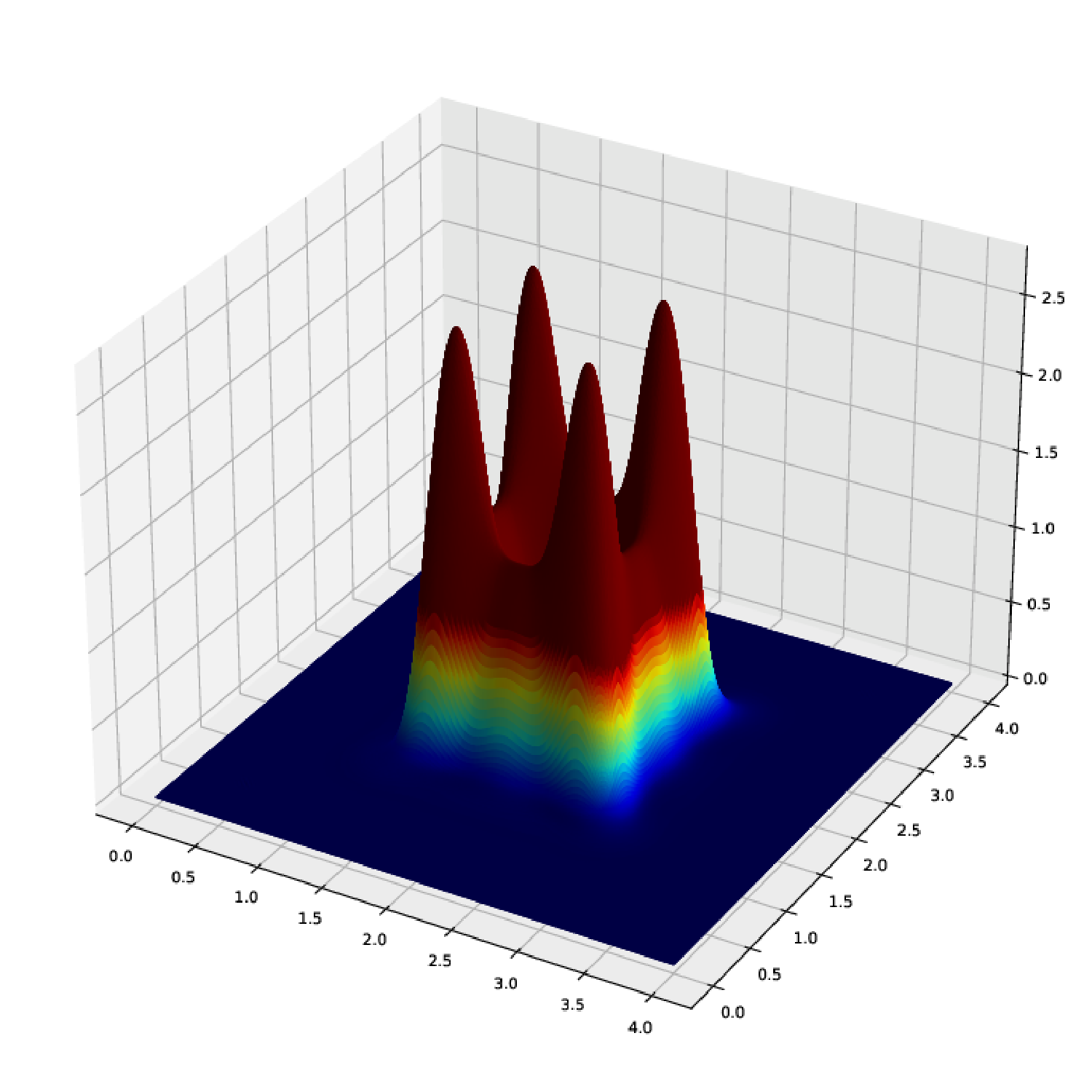}
		\caption{$t=5$}
	\end{subfigure}
	\begin{subfigure}[b]{0.30\textwidth}
		\centering
		\includegraphics[width=\textwidth]{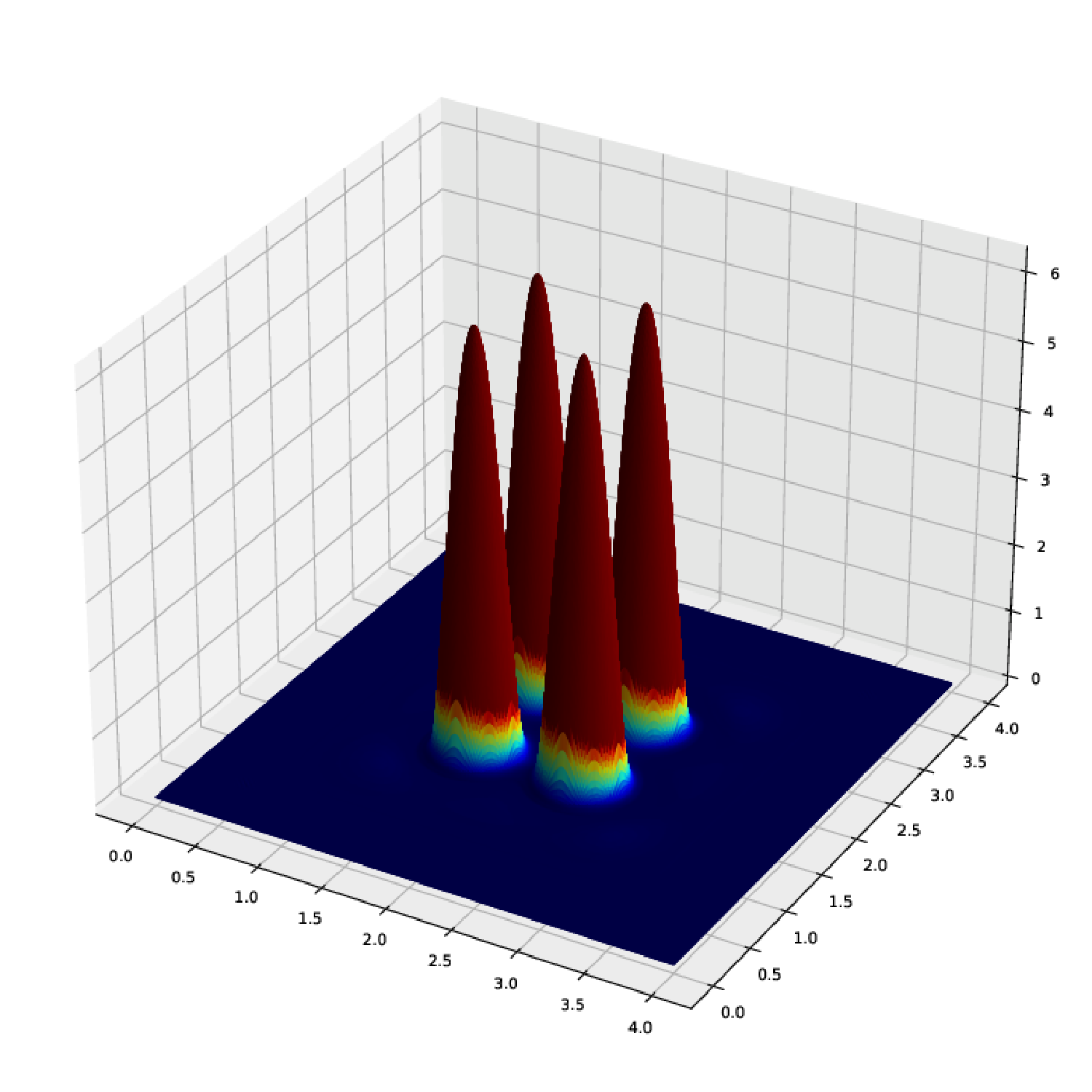}
		\caption{$t=10$}
	\end{subfigure}
	\begin{subfigure}[b]{0.30\textwidth}
		\centering
		\includegraphics[width=\textwidth]{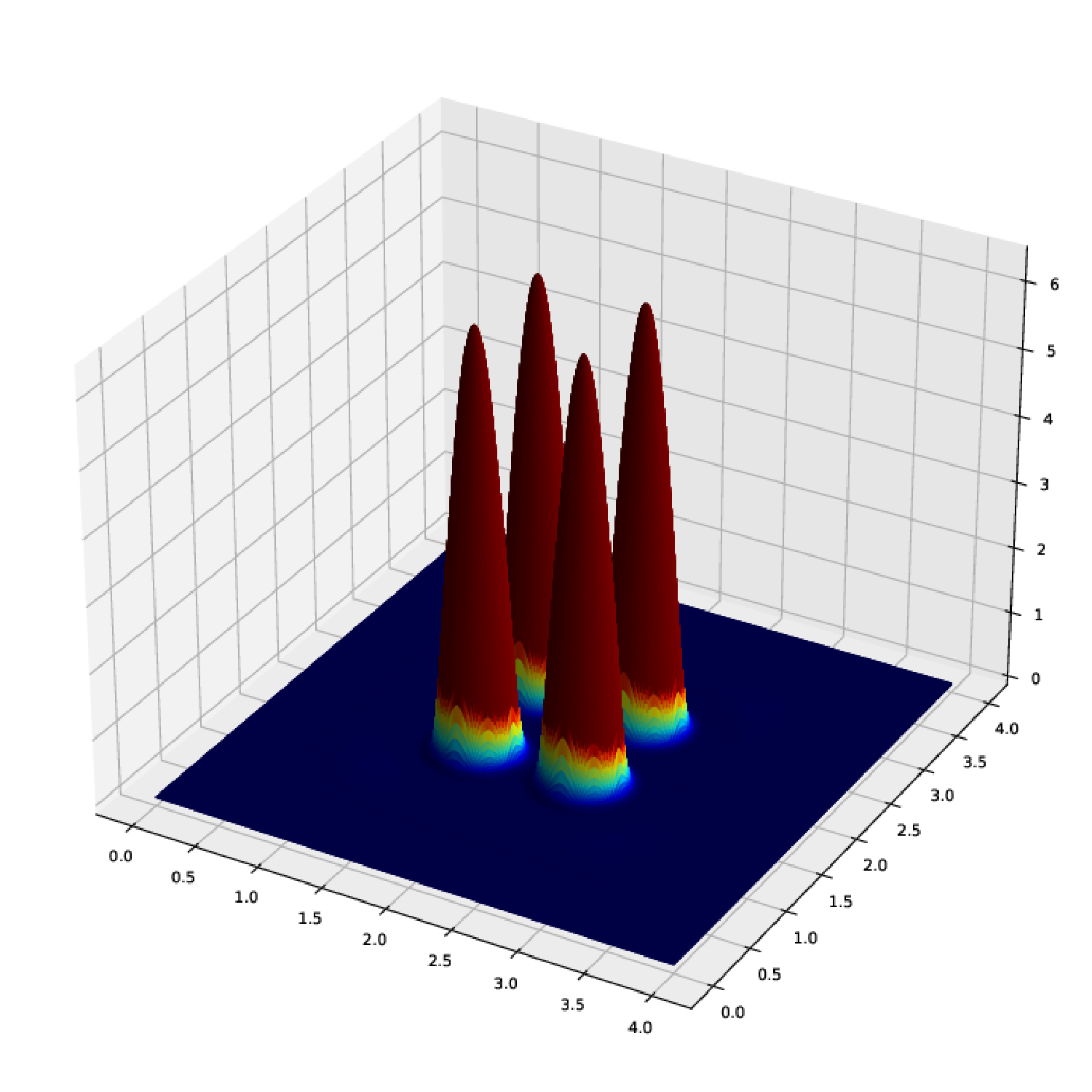}
		\caption{$t=20$}
	\end{subfigure}
	\caption{The distribution dynamics of $y(t,z)$ over space and time for the baseline case with $\gamma_S = 0,\gamma_A=-0.01,\gamma_R = 0$, $\gamma_D=0.005$, $h_A=0.3$, $\Omega=[0,4]\times[0,4]$.}
	\label{fig:numericalInvestigationh=0.3}
\end{figure}

In figures \ref{fig:numericalInvestigationh=0.4} and \ref{fig:numericalInvestigationh=0.3}  we set an initial condition that is flat at the centre of the domain and zero close to the boundary, with some intermediate values in between so that the initial profile is a continuous function.  
We then set $\gamma_A=-0.01, \gamma_D = 0.005$ and explore the impacts of changes in $h_A$ since the distance at which aggregation takes place is one of the main characteristics of the system which determines the evolution of the spatial pattern. 
Figure \ref{fig:numericalInvestigationh=0.4} reports the distribution dynamics of the baseline model which highlights the dynamics of aggregation of income, i.e. the emergence of a city, in a central location at around $t = 20$. We notice that there is an initial temporary formation of four smaller clusters around $t = 5$, which then agglomerates together to contribute to the formation of the single large cluster in the centre of the domain. 
In Figure \ref{fig:numericalInvestigationh=0.3} we keep the initial condition unchanged and only reduce the distance of aggregation from $0.4$ to $0.3$. The spatial pattern in the first few periods is roughly the same, exhibiting the formation of the four temporary clusters around period $t = 5$. However, since the distance at which aggregation takes place is smaller, the four clusters are now so far apart that they are not able to merge anymore. Therefore we observe at the final period $t = 20$ a configuration which is made of four separate clusters instead of a single larger one.  
The spatial pattern strongly resembles the ones discussed in \cite{krugman1994complex} and in \citet[cap. 8]{barthelemy2016structure}.

\clearpage 

\section{The computation of the matrices of spatial partial derivative \label{sec:computationGradientLaplacian}}

This appendix introduces to \emph{generalized finite difference method} for calculating the spatial partial derivatives when observations are not equally distributed across the space, as originally introduced in \cite{jensen1972finite} (for a detailed overview of the method see, e.g., \citealp{benito2001influence}).

Let $y(z)$ be the value of a function at location $z \in \RR^2$; define a set of neighbouring areas $z^j$ whose function values are indicated by $y(z^j)$, for $j = 1,\cdots,n_s$, where $n_s$ is the number of neighbouring locations. Define the following function:
\begin{eqnarray}\nonumber
	B(z) &=& \sum_{j=1}^{n_s} \left\{ \left[ y(z) - y(z^j) + h_j\dfrac{\partial y(z)}{\partial z_1} + k_j\dfrac{\partial y(z)}{\partial z_2} +  \right. \right.\\
		&+& \left. \left. \dfrac{1}{2} \left( h_j^2 \dfrac{\partial^2 y(z)}{\partial z_1^2} + k_j^2\dfrac{\partial^2 y(z)}{\partial z^2_2} + 2 h_j k_j \dfrac{\partial^2 y(z)}{\partial z_1 \partial z_2} \right)
		\right] w(h_j,k_j) \right\}^2,
	\label{eq:secondOrderTaylorApproximationFunction}
\end{eqnarray}
where $h_j = z_1-z^j_1$, $k_j = z_2-z^j_2$ and $w\left(\cdot,\cdot\right)$ is a weighting function decreasing in both arguments and always not negative. The function $B(z)$ is a weighted linear combination of squares of the error that one has by approximating the function $y(z^j)$ with its second-order linear approximation in the point $y(z)$, for every $j=1,\dots,n_s$. Therefore, if this approximation is sufficiently accurate, $B(z)$ is close to zero. Given $y(z)$, $y(z^j)$,  $h_j$, $k_j$ and $w_j$, for $j = 1,2,\cdots,n_s$, Eq. \eqref{eq:secondOrderTaylorApproximationFunction} allows to calculate the values of $\dfrac{\partial y(z)}{\partial z_1}$, $\dfrac{\partial y(z)}{\partial z_2}$, $ \dfrac{\partial^2 y(z)}{\partial z_1^2}$, $ \dfrac{\partial^2 y(z)}{\partial z_2^2}$ and $\dfrac{\partial^2 y(z)}{\partial z_1 \partial z_2}$ under the condition that $B(z)$ is minimized. In particular, due to the quadratic shape of the function $B$, this minimization amounts to solving the following system of linear equation (see \citealp[p. 6]{benito2001influence}):
\begin{eqnarray}
	D_z &= \begin{bmatrix}
		\dfrac{\partial y(z)}{\partial z_1}     \\
		\dfrac{\partial y(z)}{\partial z_2}     \\
		\dfrac{\partial^2 y(z)}{\partial z_1^2} \\
		\dfrac{\partial^2 y(z)}{\partial z_2^2} \\
		\dfrac{\partial^2 y(z)}{\partial z_1 \partial z_2}
	\end{bmatrix} = \mathbf{D} \begin{bmatrix}
		y(z)   \\
		y(z^1) \\
		y(z^2) \\
		\cdots \\
		y(z^{n_s})
	\end{bmatrix},
	\label{eq:matrixCalculationDerivative}
\end{eqnarray}
with $\mathbf{D} \equiv \mathbf{A}^{-1} \mathbf{B}$ is a $(5 \times (n_s+1))$ matrix, $w_j = w(h_j,k_j)$ and where $\mathbf{A}$ is a symmetric $(5 \times 5)$ matrix defined as:
\begin{eqnarray}
	A &= \begin{bmatrix*}[l]
		\sum_{j=1}^{n_s} h_j^2 w_j^2 &  \sum_{j=1}^{n_s} h_jk_j w_j^2 &   \frac{1}{2}\sum_{j=1}^{n_s} h_j^3 w_j^2  & \frac{1}{2}\sum_{j=1}^{n_s} h_jk_j^2 w_j^2 & \sum_{j=1}^{n_s} h_j^2 k_j w_j^2  \\
		&   \sum_{j=1}^{n_s} k_j^2 w_j^2 &  \frac{1}{2}\sum_{j=1}^{n_s} h_j^2 k_j w_j^2 & \frac{1}{2}\sum_{j=1}^{n_s}  k_j^3 w_j^2  & \sum_{j=1}^{n_s} h_j k_j^2 w_j^2  \\
		& & \frac{1}{4}\sum_{j=1}^{n_s} h_j^4 w_j^2 &  \frac{1}{4}\sum_{j=1}^{n_s} h_j^2 k_j^2 w_j^2 &  \frac{1}{2}\sum_{j=1}^{n_s} h_j^3 k_j w_j^2 \\
		& & & \frac{1}{4}\sum_{j=1}^{n_s} k_j^4 w_j^2 & \frac{1}{2}\sum_{j=1}^{n_s} h_j k_j^3 w_j^2 \\
		& & & & \sum_{j=1}^{n_s} h_j^2 k_j^2 w_j^2
	\end{bmatrix*},
	\label{eq:matrixA}
\end{eqnarray}
and $\mathbf{B}$ is a $(5 \times (n_s+1)) $ matrix defined as:
\begin{eqnarray}
	B & =\begin{bmatrix}
		- \sum_{j=1}^{n_s} h_j w_j^2                   & h_1 w_1^2                 & h_2 w_2^2              & \cdots & h_{n_s} w_{n_s}^2              \\
		- \sum_{j=1}^{n_s} k_j w_j^2                   & k_1 w_1^2                 & k_2 w_2^2      & \cdots & k_{n_s} w_{n_s}^2              \\
		- \frac{1}{2}\sum_{j=1}^{n_s} h_j^2 w_j^2 & \frac{1}{2}h_1^2{2} w_1^2 & \frac{1}{2}k_1^2 w_2^2 & \cdots & \frac{1}{2}h_{n_s}^2 w_{n_s}^2 \\
		- \frac{1}{2}\sum_{j=1}^{n_s} k_j^2 w_j^2      & \frac{1}{2}h_2^2 w_1^2    & \frac{1}{2}k_2^2 w_2^2 & \cdots & \frac{1}{2}k_{n_s}^2 w_{n_s}^2 \\
		- \sum_{j=1}^{n_s} h_j k_j w_j^2               & h_1 k_1 w_1^2             & h_2 k_2 w_2^2          & \cdots & h_{n_s} k_{n_s} w_{n_s}^2
	\end{bmatrix}.
	\label{eq:matrixB}
\end{eqnarray}
\cite{benito2001influence} suggests using as a weighting function:
\begin{equation}
	w(d_j) = 1 - 6 \left( \dfrac{d_j}{dm_j} \right)^2 + 8 \left( \dfrac{d_j}{dm_j} \right)^3 - 3 \left( \dfrac{d_j}{dm_j} \right)^4,
	\label{eq:weightingFunction}
\end{equation}
where $d_j^2 = h_j^2 + k_j^2$ is the squared distance between $z$ and $z_j$ and $dm_j$ is the maximum distandiffusivece on all possible neighbouring locations.

The set of all neighbouring locations used to approximate the partial derivatives in location $z$ is called the \emph{star} of the location; the choice of the elements of the \textit{star} is an important factor to ensure the accuracy of the method. We will adopt the \emph{closest neighbourhood criterion}, that is, fixed the number of elements of the \textit{star} $n_s$, we select the $n_s$ closest locations to each given location.

For every given location $z$, it is, therefore, possible to build the matrix $D$ and all the partial derivatives in location $z$ up to the second order. To carry out these operations set $(z^i)_{i=1,\dots,N}$ a finite number of locations; for each of these locations denote by $\mathbf{J^i} = (\mathbf{J^i}_k)_{k=1,\dots,n_s}$ the vector of indices of locations in the \textit{star} of location $z^i$ (by definition of star, $i$ is not an element of $\mathbf{J^i}$), and by $\mathbf{D_i}$ the matrix $\mathbf{D}$ associated to the location $z^i$; thus
\begin{eqnarray}
	\begin{bmatrix}
		\dfrac{\partial y(z^i)}{\partial z_1}     \\
		\dfrac{\partial y(z^i)}{\partial z_2}     \\
		\dfrac{\partial^2 y(z^i)}{\partial z_1^2} \\
		\dfrac{\partial^2 y(z^i)}{\partial z_2^2} \\
		\dfrac{\partial^2 y(z^i)}{\partial z_1 \partial z_2}
	\end{bmatrix} = \mathbf{D_i} \underbrace{\begin{bmatrix}
			y(z^i)                \\
			y(z^{\mathbf{J^i}_1}) \\
			y(z^{\mathbf{J^i}_2}) \\
			\vdots                \\
			y(z^{\mathbf{J^i}_{n_s}})
		\end{bmatrix}}_{\overline{\mathbf{y}}_i} = \mathbf{D_i}\overline{\mathbf{y}}_i.
\end{eqnarray}
Therefore, denoting by $(D_i)_{k,}$ the $k$-th row of the matrix $D_i$
\begin{eqnarray*}
	\dfrac{\partial y(z^i)}{\partial z_1} &= \ang{(\mathbf{D_i})_{1,},\overline{\mathbf{y}}_i};\\
	\dfrac{\partial y(z^i)}{\partial z_2} &= \ang{(\mathbf{D_i})_{2,},\overline{\mathbf{y}}_i};\\
	\dfrac{\partial^2 y(z^i)}{\partial z_1^2} &= \ang{(\mathbf{D_i})_{3,},\overline{\mathbf{y}}_i}; \text{ and}\\
	\dfrac{\partial^2 y(z^i)}{\partial z_2^2} &= \ang{(\mathbf{D_i})_{4,},\overline{\mathbf{y}}_i},\\
\end{eqnarray*}
where $\ang{\cdot,\cdot}$ is the standard scalar product of $\RR^{n_s+1}$.

To compute all the partial derivatives in all the locations via single matrix multiplications, set
\begin{equation*}
	\mathbf{y} = (y(z^1),y(z^2),\dots, y(z^N))^t
\end{equation*}
and the vectors of partial derivatives in all the locations as
\begin{eqnarray*}
	\dfrac{\partial y}{\partial z_1} = \begin{bmatrix}
		\dfrac{\partial y(z^1)}{\partial z_1} \\
		\dfrac{\partial y(z^2)}{\partial z_1} \\
		\vdots                                \\
		\dfrac{\partial y(z^N)}{\partial z_1} \\
	\end{bmatrix},
	\dfrac{\partial y}{\partial z_2} = \begin{bmatrix}
		\dfrac{\partial y(z^1)}{\partial z_2} \\ 
		\dfrac{\partial y(z^2)}{\partial z_2} \\
		\vdots                                \\
		\dfrac{\partial y(z^N)}{\partial z_2} \\
	\end{bmatrix},
	\dfrac{\partial^2 y}{\partial z_1^2} = \begin{bmatrix}
		\dfrac{\partial^2 y(z^1)}{\partial z_1^2} \\
		\dfrac{\partial^2 y(z^2)}{\partial z_1^2} \\
		\vdots                                    \\
		\dfrac{\partial^2 y(z^N)}{\partial z_1^2} \\
	\end{bmatrix},
	\dfrac{\partial^2 y}{\partial z_2^2} = \begin{bmatrix}
		\dfrac{\partial^2 y(z^1)}{\partial z_2^2} \\
		\dfrac{\partial^2 y(z^2)}{\partial z_2^2} \\
		\vdots                                    \\
		\dfrac{\partial^2 y(z^N)}{\partial z_2^2} \\
	\end{bmatrix}.
\end{eqnarray*}
Consider the $(N \times N)$ matrices ${M_{z_1}}$, ${M_{z_2}}$, ${M_{z_{1}z_{1}}}$ and ${M_{z_{2}z_{2}}}$ defined by
\begin{equation*}
	({M_{z_1}})_{i,j} = \begin{cases}
		(\mathbf{D_i})_{1,1}     & \text{ if } j = i,                                                        \\
		(\mathbf{D_i})_{1,k + 1} & \text{ if } j = \mathbf{J^i}_k \text{ for some } k \in  \{1,\dots, n_s\}, \\
		0                        & \text{otherwise},
	\end{cases}
\end{equation*}
\begin{equation*}
	({M_{z_2}})_{i,j} = \begin{cases}
		(\mathbf{D_i})_{2,2}     & \text{ if } j = i,                                                        \\
		(\mathbf{D_i})_{2,k + 1} & \text{ if } j = \mathbf{J^i}_k \text{ for some } k \in  \{1,\dots, n_s\}, \\
		0                        & \text{otherwise},
	\end{cases}
\end{equation*}
\begin{equation*}
	({M_{z_{1}z_{1}}})_{i,j} = \begin{cases}
		(\mathbf{D_i})_{3,3}     & \text{ if } j = i,                                                        \\
		(\mathbf{D_i})_{3,k + 1} & \text{ if } j = \mathbf{J^i}_k \text{ for some } k \in  \{1,\dots, n_s\}, \\
		0                        & \text{otherwise},
	\end{cases}
\end{equation*}
\begin{equation*}
	({M_{z_{2}z_{2}}})_{i,j} = \begin{cases}
		(\mathbf{D_i})_{4,4}     & \text{ if } j = i,                                                        \\
		(\mathbf{D_i})_{4,k + 1} & \text{ if } j = \mathbf{J^i}_k \text{ for some } k \in  \{1,\dots, n_s\}, \\
		0                		 & \text{otherwise},
	\end{cases}
\end{equation*}
then, for all $i=1, \dots, N$,
\begin{equation*}
	\dfrac{\partial y(z^i)}{\partial z_1}  \approx \left({M_{z_1}} \mathbf{y}\right)_i,\quad
	\dfrac{\partial y(z^i)}{\partial z_2} \approx \left({M_{z_2}} \mathbf{y}\right)_i,\quad
	\dfrac{\partial^2 y(z^i)}{\partial z_1^2} \approx \left({M_{z_{1}z_{1}}} \mathbf{y}\right)_i,\quad
	\dfrac{\partial^2 y(z^i)}{\partial z_2^2} \approx \left({M_{z_{2}z_{2}}} \mathbf{y}\right)_i.
\end{equation*}

\clearpage

\section{Parameters' estimate for all levels of discretizations\label{app:estimatedParametersNumericalSimulation}}

\begin{figure}[!htbp]
	\centering
	\begin{minipage}{0.44\textwidth}
		\centering
		\begin{subfigure}[b]{0.56\textwidth}
			\includegraphics[width=\textwidth]{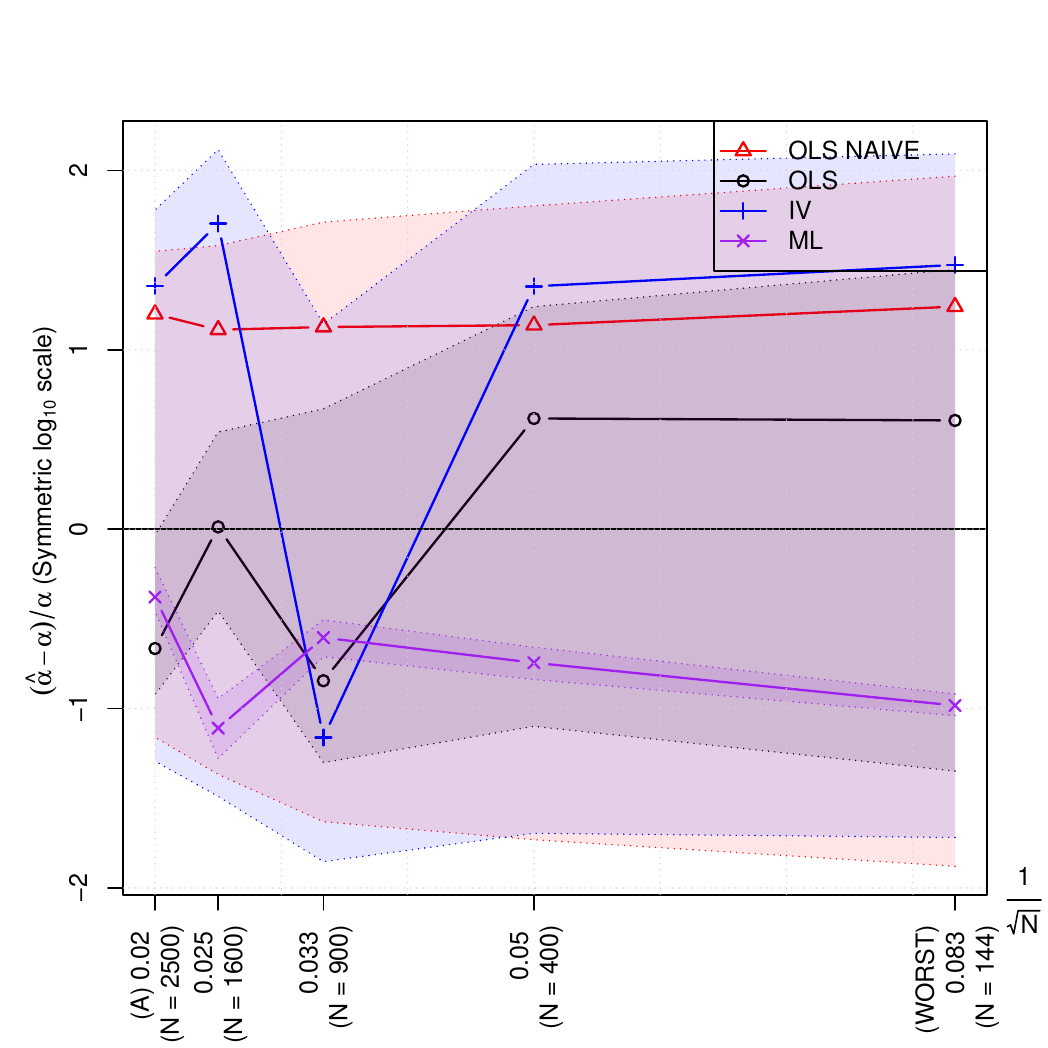}
		\end{subfigure}
		\begin{subfigure}[b]{0.56\textwidth}
			\includegraphics[width=\textwidth]{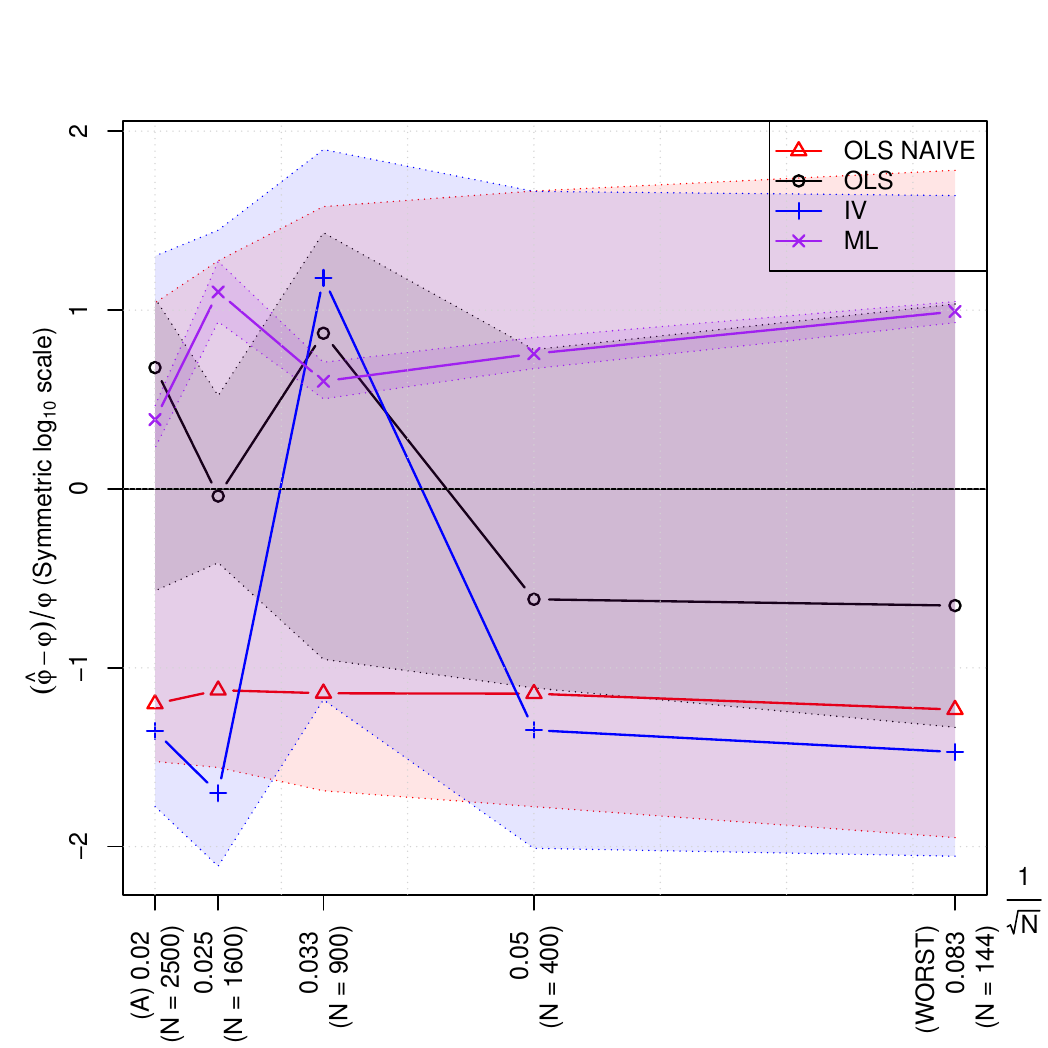}
		\end{subfigure}
		\begin{subfigure}[b]{0.56\textwidth}
			\includegraphics[width=\textwidth]{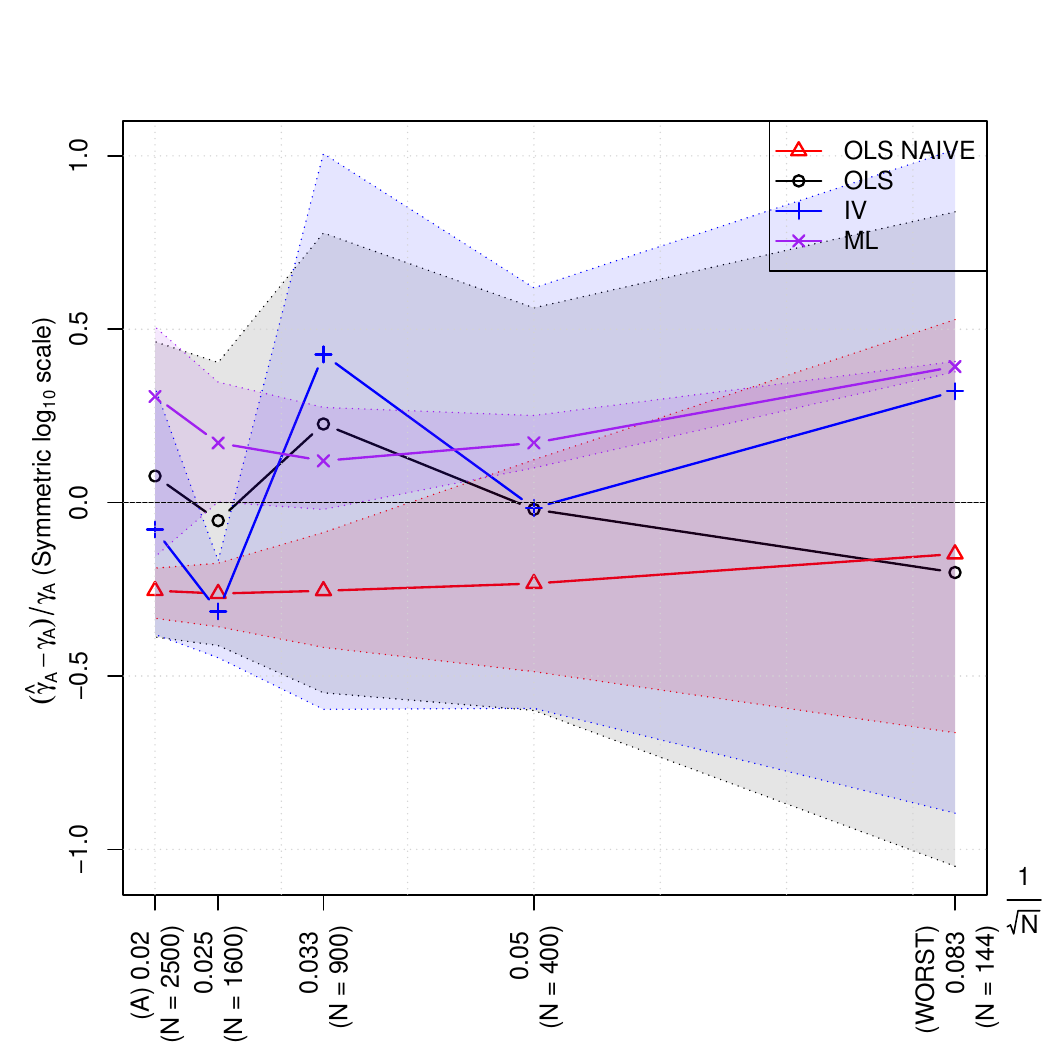}
		\end{subfigure}
		\begin{subfigure}[b]{0.56\textwidth}
			\includegraphics[width=\textwidth]{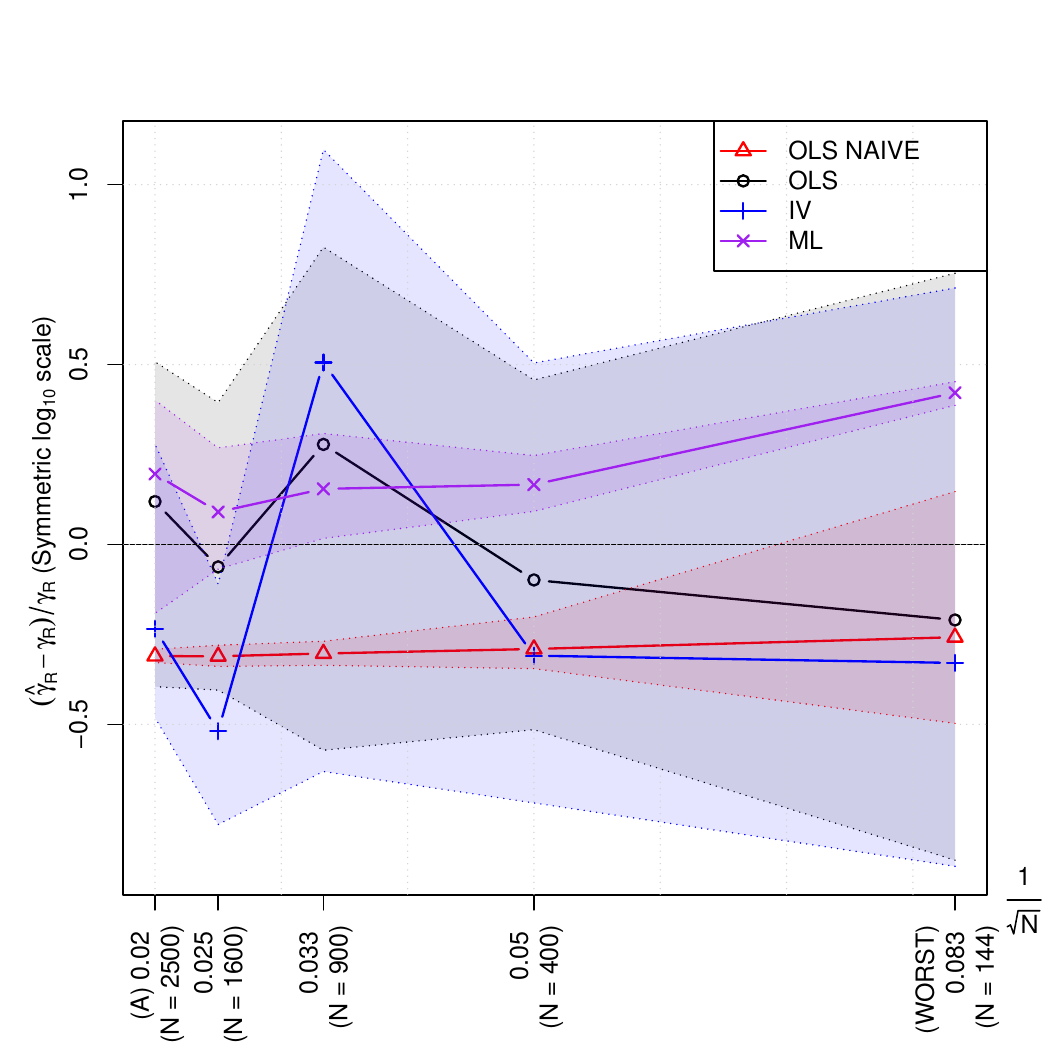}
		\end{subfigure}
		\begin{subfigure}[b]{0.56\textwidth}
			\includegraphics[width=\textwidth]{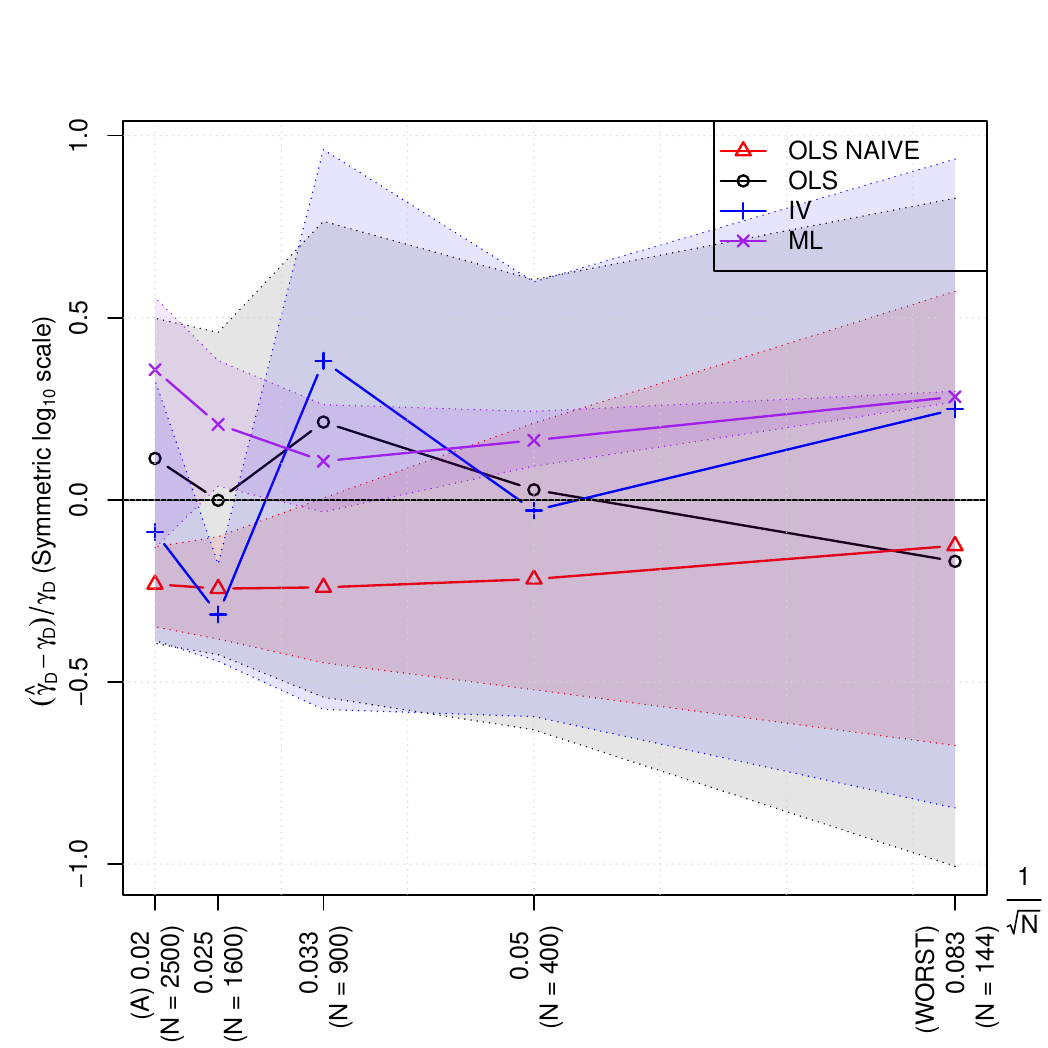}
		\end{subfigure}
	\end{minipage}
	\begin{minipage}{0.44\textwidth}
		\centering
		\begin{subfigure}[b]{0.56\textwidth}
			\includegraphics[width=\textwidth]{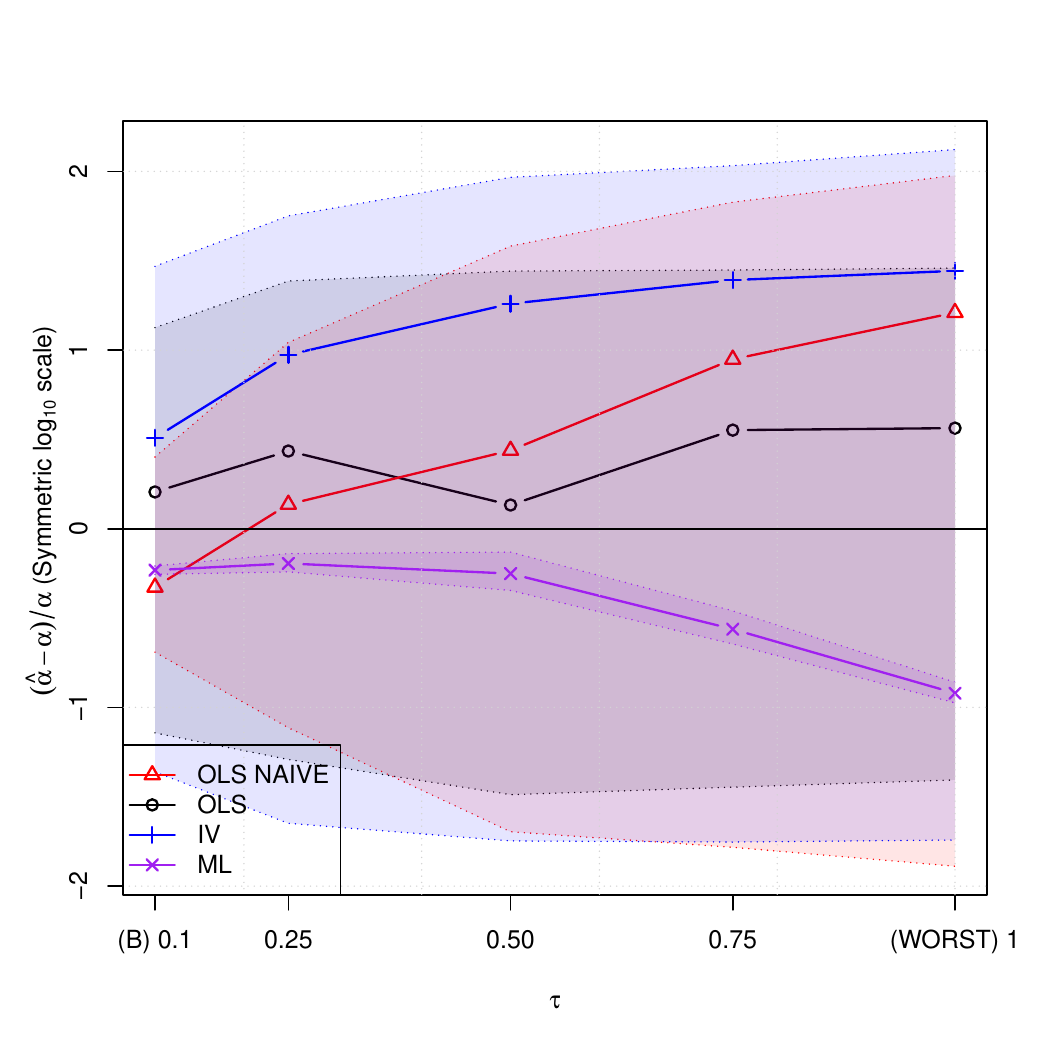}
		\end{subfigure}
		\begin{subfigure}[b]{0.56\textwidth}
			\includegraphics[width=\textwidth]{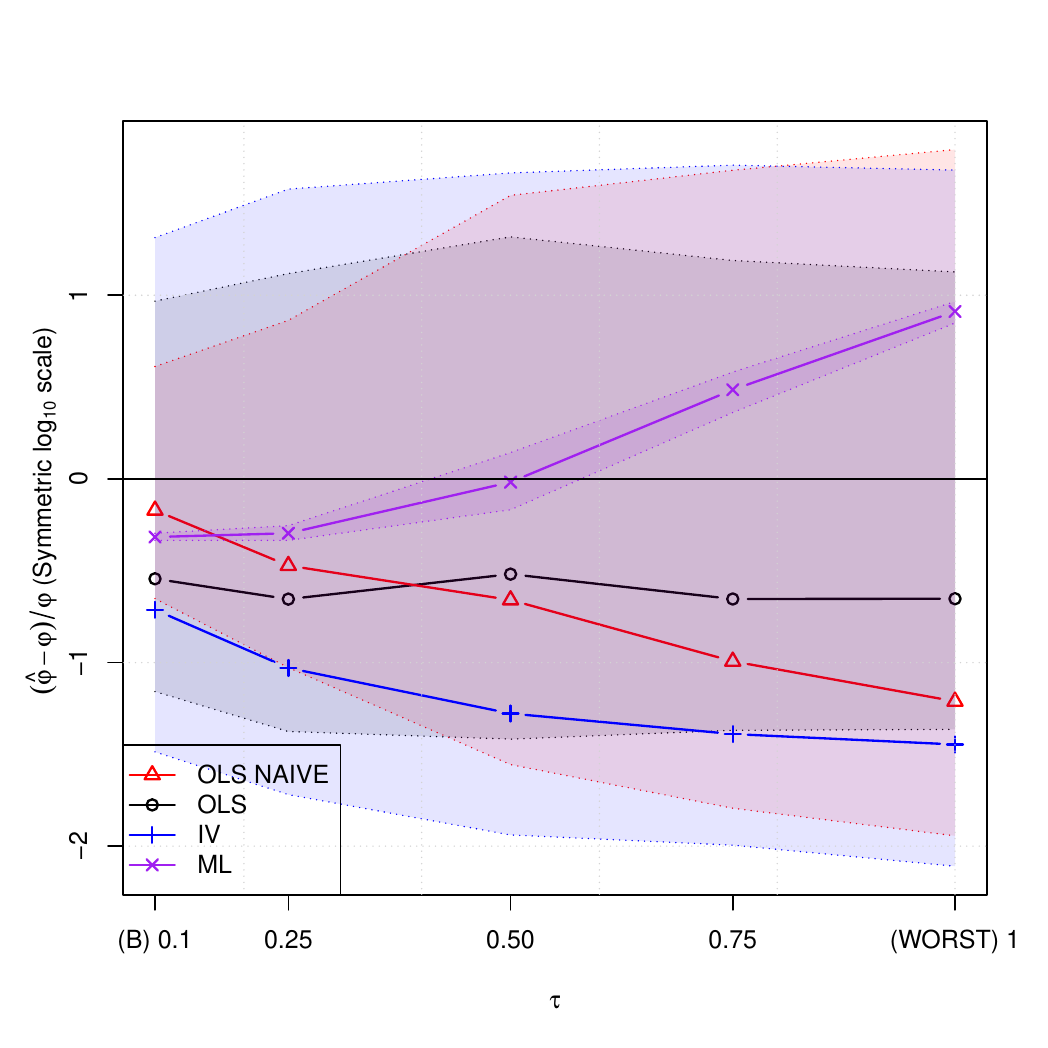}
		\end{subfigure}
		\begin{subfigure}[b]{0.56\textwidth}
			\includegraphics[width=\textwidth]{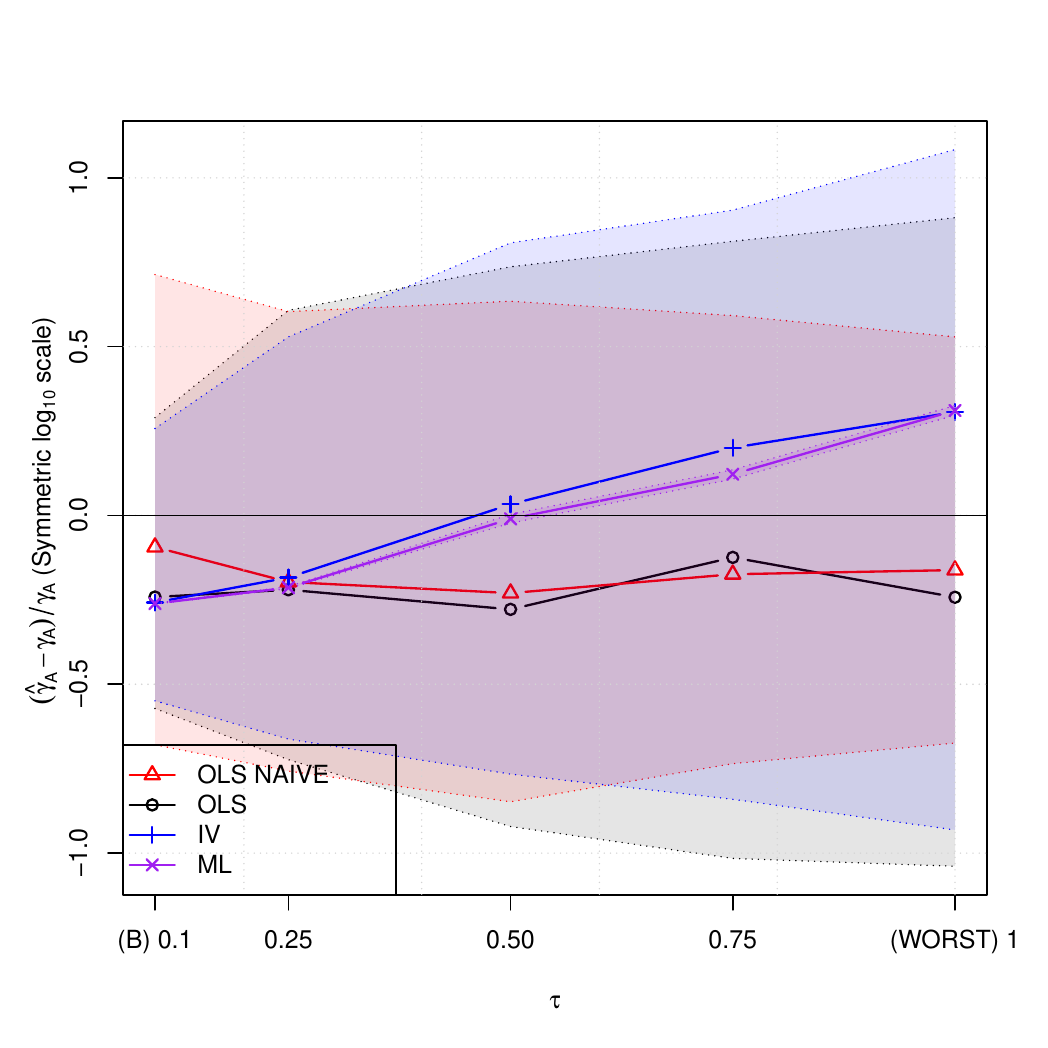}
		\end{subfigure}
		\begin{subfigure}[b]{0.56\textwidth}
			\includegraphics[width=\textwidth]{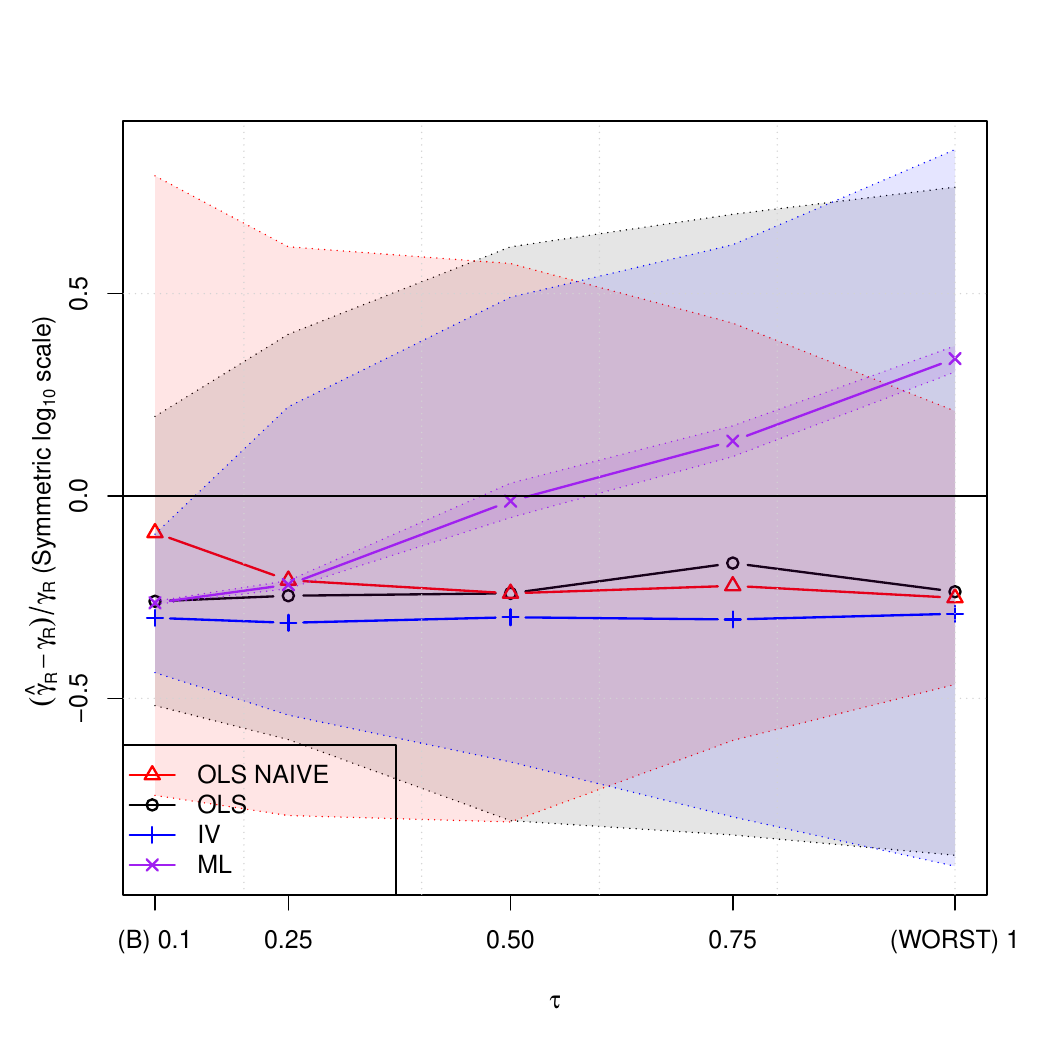}
		\end{subfigure}
		\begin{subfigure}[b]{0.56\textwidth}
			\includegraphics[width=\textwidth]{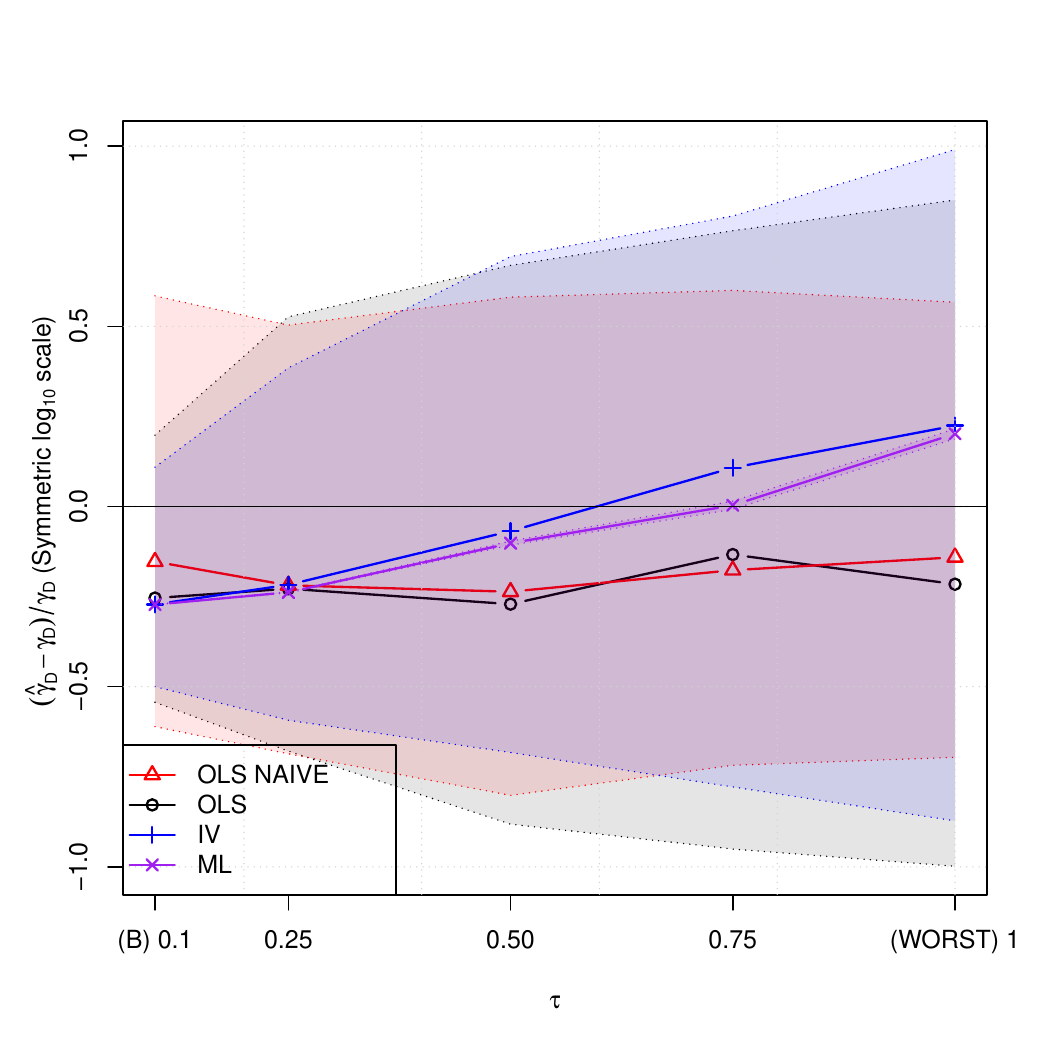}
		\end{subfigure}
		\label{fig:MonteCarloVariandoNpTau}
	\end{minipage}
	\caption{The estimated parameters and their standard errors (reported as bands of different colours) under alternative time and space discretizations for the four types of estimation of Figure \ref{fig:AICcDifferentModels}.}
	\label{fig:estimatedParametersNUmericalSImulation}
\end{figure}

\clearpage

\section{The estimated $W_{\epsilon}$ for Italian municipalities \label{app:spatialMatrixErrorsRealData}}

Table \ref{tab:estimatedSpatialWeightedMatrix} reports the estimated $\hat{\ell}$'s of Model \eqref{eq:spatialWeightMatrix} in Appendix \ref{app:spatialMatrixErrors} for Italian municipalities. The most appropriate maximum order of contiguity is $\hat{Q}=5$, decided based on the statistical significance of the estimated parameters. Notice that spatial correlation appears to suddenly drop after the first order of contiguity.
\begin{table}[!htbp]
	\centering
	\begin{tabular}{rrrrr}
		\hline
		\hline
		& Estimate & Std. Error & t value & Pr($>$$|$t$|$) \\ 
		\hline
		$\ell_{1 }$ & 0.1284 & 0.0035 & 36.56 & 0.0000 \\ 
		$\ell_{2 }$ & 0.0043 & 0.0022 & 2.01 & 0.0448 \\ 
		$\ell_{3 }$ & 0.0070 & 0.0017 & 4.01 & 0.0001 \\ 
		$\ell_{4 }$ & 0.0076 & 0.0016 & 4.88 & 0.0000 \\ 
		$\ell_{5 }$ & 0.0045 & 0.0014 & 3.14 & 0.0017 \\ 
		$\ell_{6 }$ & -0.0004 & 0.0013 & -0.27 & 0.7872 \\ 
		$\ell_{7 }$ & 0.0015 & 0.0012 & 1.26 & 0.2059 \\ 
		$\ell_{8 }$ & 0.0016 & 0.0011 & 1.37 & 0.1696 \\ 
		$\ell_{9 }$ & 0.0006 & 0.0011 & 0.52 & 0.6003 \\ 
		$\ell_{10}$ & 0.0015 & 0.0011 & 1.36 & 0.1748 \\ 
		\hline
		\hline
	\end{tabular}
	\caption{The estimate of Model \eqref{eq:spatialWeightMatrix}'s coefficients in Appendix \ref{app:spatialMatrixErrors} with $Q=10$. The calculation of $W_\epsilon$ is based on the first five coefficients, i.e. $\hat{Q}=5$, decided based on their statistical significance.}
	\label{tab:estimatedSpatialWeightedMatrix}
\end{table}

\clearpage

\clearpage

\section{The analysis of residuals for Italian municipalities}

\begin{figure}[!htbp]
	\centering
	\begin{subfigure}[b]{0.47\textwidth}
		\centering
		\includegraphics[width=\textwidth]{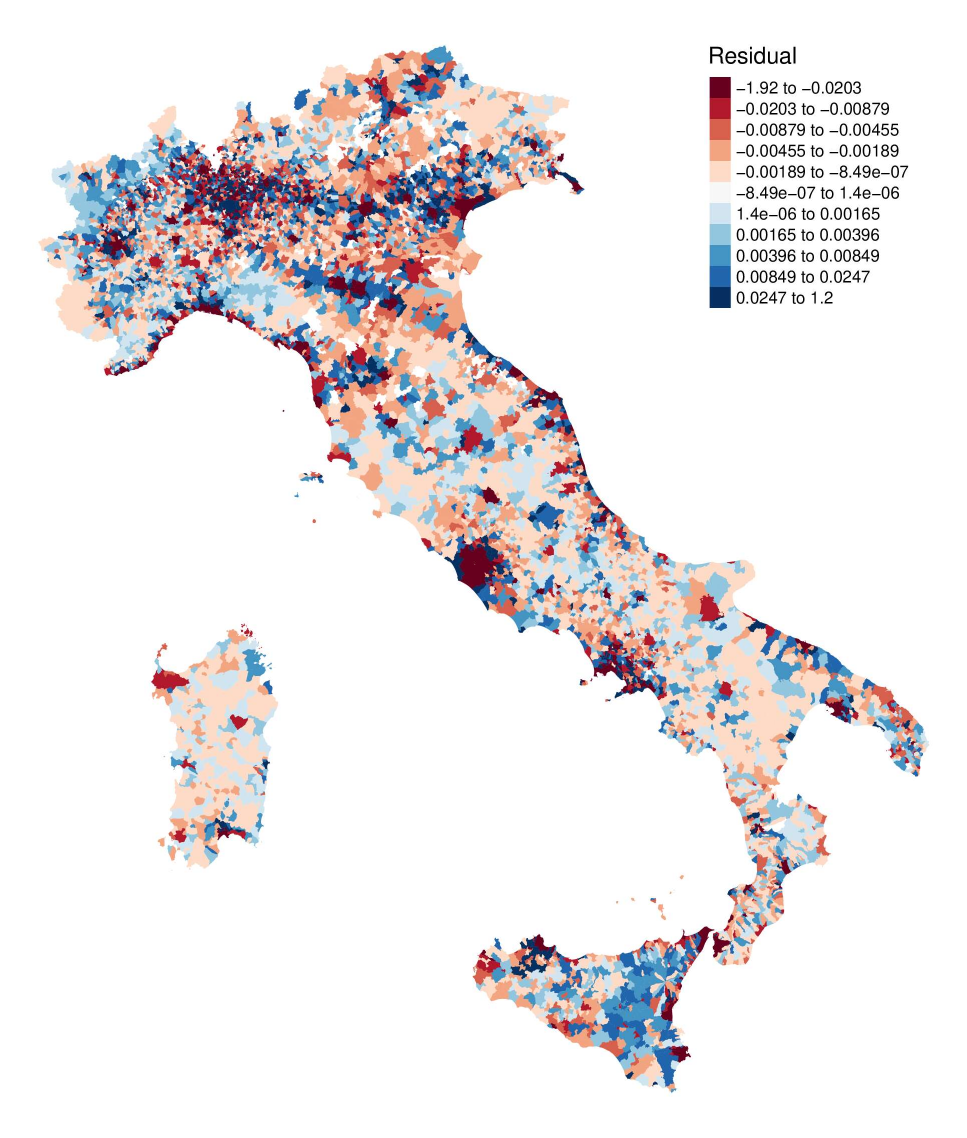}
		\caption{SARD}
	\end{subfigure}
	\begin{subfigure}[b]{0.47\textwidth}
		\centering
		\includegraphics[width=\textwidth]{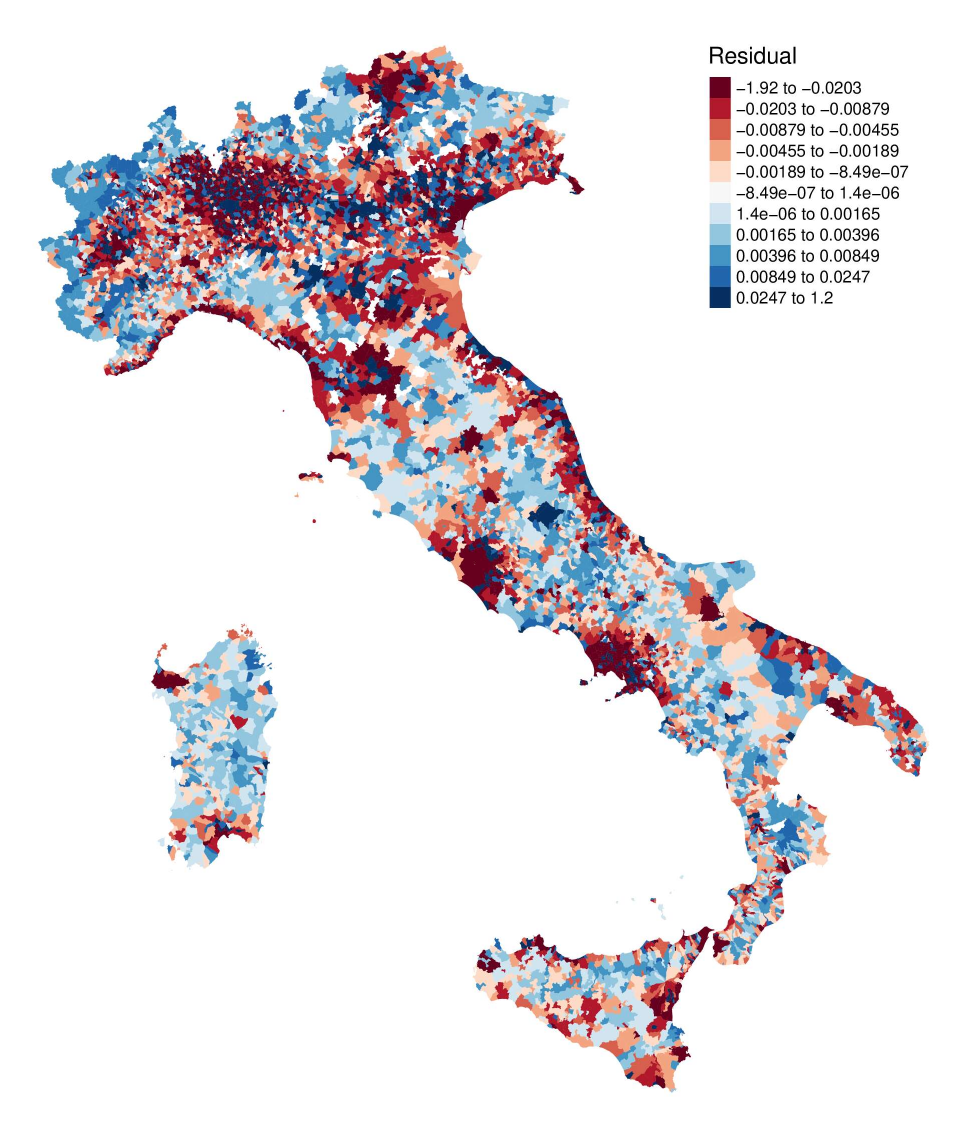}
		\caption{Durbin}
	\end{subfigure}
	\caption{The map of residuals for Italian Municipalities for the period 2008-2019 for SARD and spatial Durbin.}
	\label{fig:mapsResidual19}
\end{figure}

\begin{figure}[!htbp]
	\centering
	\begin{subfigure}[b]{0.47\textwidth}
		\centering
		\includegraphics[width=\textwidth]{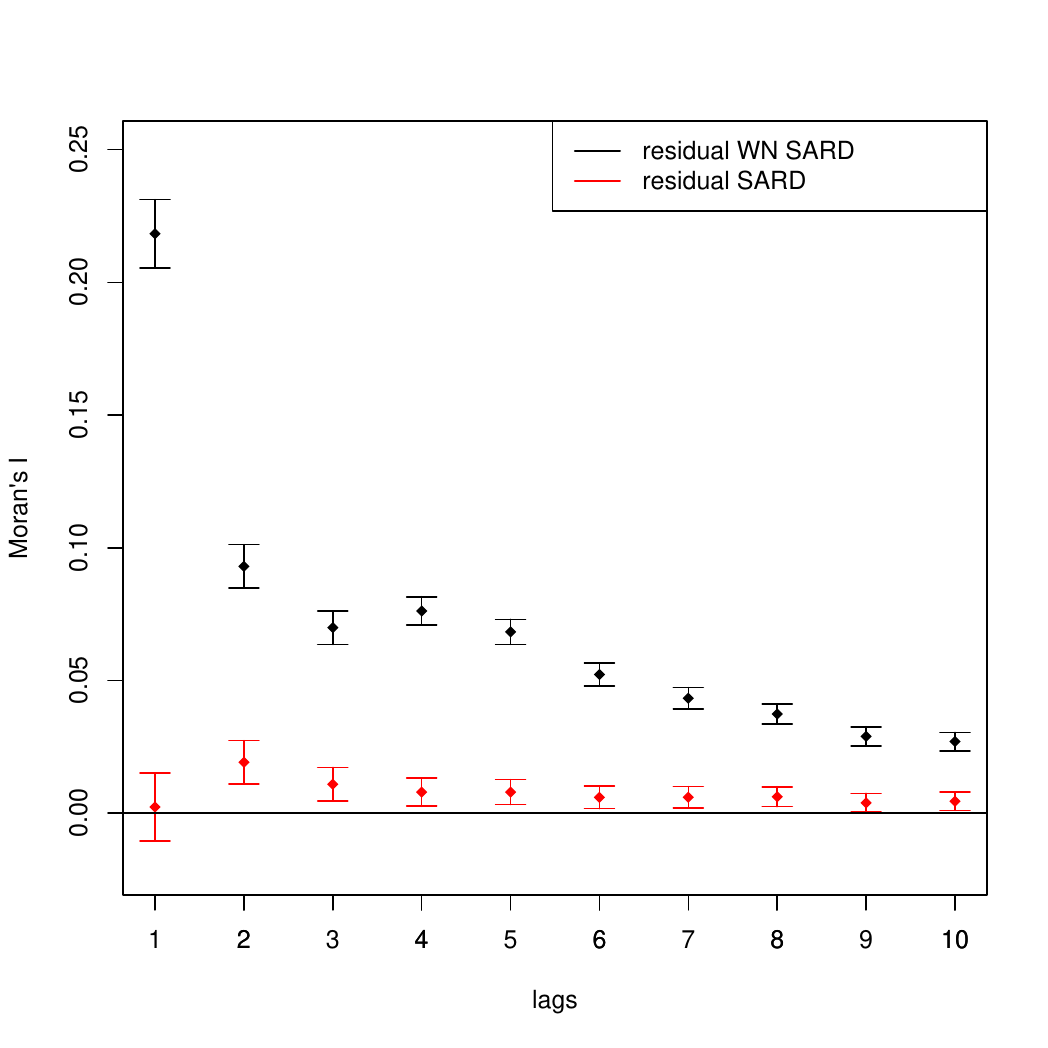}
		\caption{Moran's I: SARD}
	\end{subfigure}
	\begin{subfigure}[b]{0.47\textwidth}
		\centering
		\includegraphics[width=\textwidth]{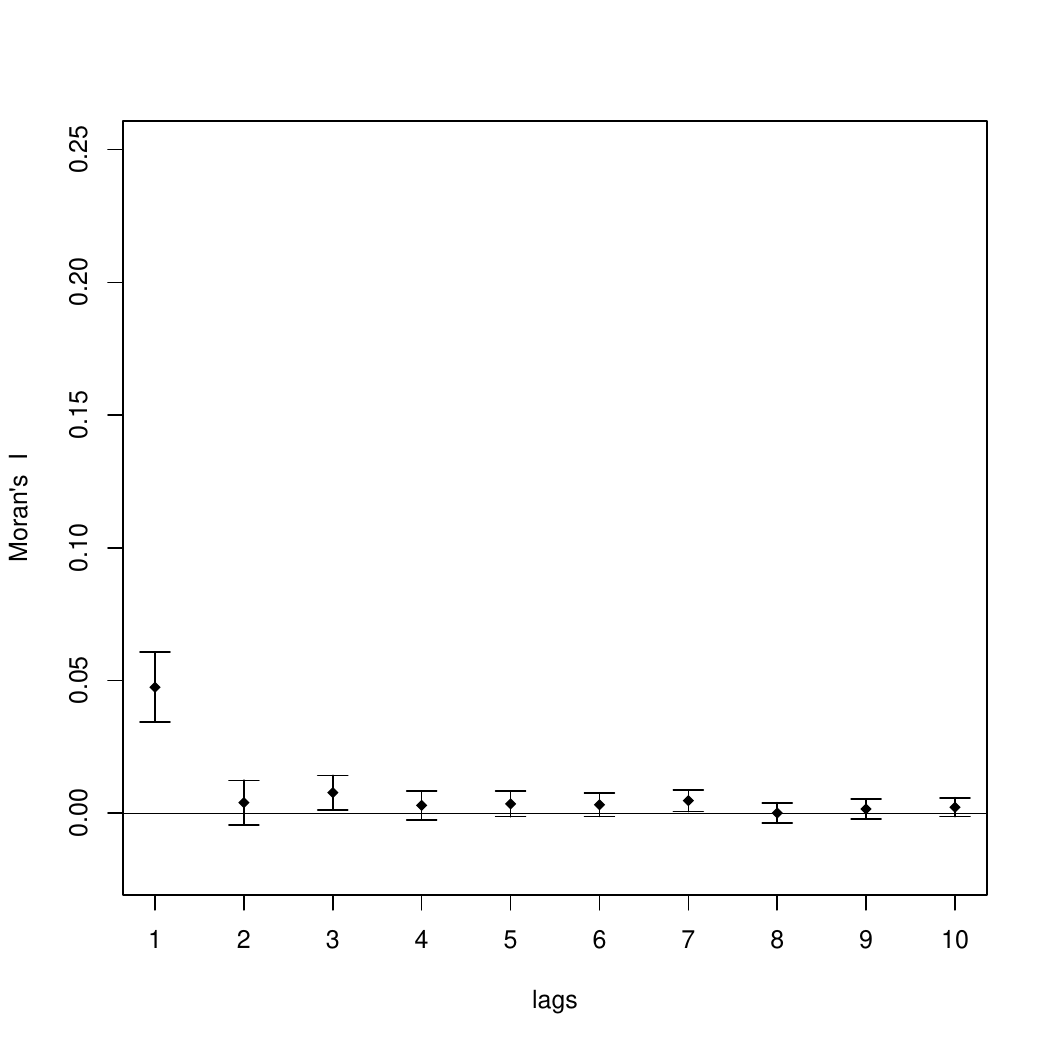}
		\caption{Moran's I: Durbin}
	\end{subfigure}
	\caption{Spatial correlograms of residuals for Moran's I for the period 2008-2019.}
	\label{fig:corrIG}
\end{figure}

Figure \ref{fig:mapsResidual19} reports the spatial distributions of residuals for SARD and SPATIAL DURBIN models. The SARD model can better explain the dynamics of the innermost regions, like the Apennines mountains and the centre of Sicily and Sardinia islands. This in particular highlights how our model is better able to exploit the information coming from the altimetry than the classical spatial econometrics models. Moreover, we also appreciate how the spatial pattern of residuals appears more spotted between regions of red and blue colours (corresponding to negative and positive residuals) in the case of the SARD model with respect to SPATIAL DURBIN. When looking at the spatial correlograms for Moran's I index (Figure \ref{fig:corrIG}), we find weak evidence of additional spatial dependence unaccounted for in both the SARD and SPATIAL DURBIN models.

\end{document}